\documentclass{aa}  

\usepackage{xcolor}
\usepackage{graphicx}
\usepackage{txfonts}
\usepackage{placeins}
\newcommand{\alphaco} {$\alpha_{\rm CO}$}

\newcommand{\alphacogal} {$\alpha_{\rm CO,Gal}$}
\newcommand{\ergs} {erg s$^{-1}$}

\newcommand{\faper} {\hbox{$f_{{\rm aper}}$}}
\newcommand{\fmol} {\hbox{$f_{{\rm mol}}$}}
\newcommand{\fhi} {\hbox{$f_{{\rm HI}}$}}
\newcommand{\fmolzcorr} {\hbox{$f_{{\rm mol, zcorr}}$}}

\newcommand{\halpha}{H$\alpha$}

\newcommand{\ico}{$I_{\rm CO}$}

\newcommand{\lsun}{$L_\odot$}
\newcommand{\msun}{$M_\odot$}
\newcommand{\mi}{$\mu$m}
\newcommand{\kms}{km~s$^{-1}$}

\newcommand{\htwo} {\hbox{H$_2$}}
\newcommand{\lco} {\hbox{$L^\prime_{\rm CO}$}}

\newcommand{\lwonesun} {$L_{\rm W1, \odot}$ }

\newcommand{\mmol} {\hbox{$M_{\rm mol}$}}

\newcommand{\mgas} {\hbox{$M_{{\rm gas}}$}}
\newcommand{\mstar} {\hbox{$M_{{\rm *}}$}}

\newcommand{\mhi}   {\hbox{$M_{\rm HI}$}}

\newcommand{\diso}{$d_{\rm 25}$}
\newcommand{\riso}{$r_{\rm 25}$}

\newcommand{\sfrbest}{SFR$_{\rm best}$}

\newcommand{\taudep}{$\tau_{\rm dep}$}

\newcommand{\figuredirpython}{figures}
\newcommand{\figuredirpythoncigale}{figures}

\newcommand{\figuredirspectra}{spectra}

\begin{document}

   \title{Molecular gas in super spiral galaxies
\thanks{Full Tables 1, 2, 4 and 5 are available at the CDS via anonymous ftp to cdsarc.u-strasbg.fr (130.79.128.5) or via http://cdsarc.u-strasbg.fr/viz-bin/ qcat?J/A+A/627/A107}}

   \author{Ute Lisenfeld
          \inst{1,2}      
    \and
    Patrick M. Ogle \inst{3}
    \and 
    Philip N. Appleton \inst{4}
    \and 
    Thomas H. Jarrett\inst{5}
    \and 
    Blanca M. Moncada-Cuadri \inst{6}
}

   \institute{Departamento de F\'isica Te\'orica y del Cosmos, Universidad de Granada, 18071 Granada, Spain, \email{ute@ugr.es}
   \and 
   Instituto Carlos I de F\'isica T\'eorica y Computacional, 
   Facultad de Ciencias, 18071 Granada, Spain
   \and 
   Space Telescope Science Institute, 3700 San Martin Drive, Baltimore, MD 21218, USA
   \and 
   Caltech/IPAC, 1200 E. California Blvd., Pasadena, CA 91125, USA
   \and
   Department of Astronomy, University of Cape Town, Private Bag X3, Rondebosch 7701, South Africa
   \and
   Department of Physics, University of Bath, Claverton Down, Bath, BA2 7AY, UK
}

   \date{Received September 15, 1996; accepted March 16, 1997}

 
  \abstract{
   At the highest stellar masses (log(\mstar) $\gtrsim$ 11.5 \msun), only  a small fraction of  galaxies  are disk-like and actively star-forming objects.  These so-called  `super spirals' 
   are ideal objects to better understand how galaxy evolution proceeds and to extend our knowledge about the relation between stars and gas to a higher stellar mass regime. We present new CO(1-0) data for a sample of 46 super spirals and  for 18 slightly lower-mass (log(\mstar) $>$ 11.0 \msun ) galaxies with  broad HI lines  -- HI fast-rotators (HI-FRs).  We analyze their molecular gas mass, derived from CO(1-0), in relation to their star formation rate (SFR) and stellar mass, and compare the results to values and scaling relations derived from lower-mass  galaxies. 
   We confirm that super spirals follow the same star-forming main sequence (SFMS) as lower-mass galaxies. We find that they possess abundant molecular gas  (mean redshift-corrected molecular gas mass fraction (log(\fmolzcorr)  = -1.36 $\pm$ 0.02), which lies above  the extrapolation of  the scaling relation with stellar mass derived from lower-mass galaxies, but within the relation between \fmol\ and the distance to the SFMS. The molecular gas depletion time, \taudep = \mmol/SFR, is  higher  than for lower-mass  galaxies on the SFMS (\taudep = 9.30 $\pm$ 0.03, compared to \taudep = 9.00 $\pm$ 0.02 for the comparison sample) and seems to continue an increasing trend with stellar mass. HI-FR galaxies have an atomic-to-molecular gas mass ratio that is in agreement with that of lower-mass galaxies, indicating that the conversion from  the atomic  to molecular gas proceeds in a similar way.  We conclude that the availability of molecular gas is a crucial factor to enable star formation to continue and that, if gas is present, quenching is not a necessary destiny for high-mass galaxies.  The difference in gas depletion time  suggests that the properties of the molecular gas at high stellar masses are less favorable for star formation.}


   \maketitle
%

\section{Introduction} 

Considerable progress has been made in recent years  in our understanding of how galaxies evolve.  For gas-rich disk galaxies, there exists a tight relation between star formation rate (SFR) and stellar mass, usually referred to as the  star-forming main sequence \citep[SFMS, e.g.,][]{brinchmann04, elbaz07}.  The slope of this relation is slightly less than unity (in a log-log representation), so that the specific SFR (sSFR = SFR/\mstar) decreases with stellar mass, \mstar. 
This relation indicates that spiral galaxies evolve a large fraction of their  lifetime along the SFMS by converting a relatively steady gas supply into stars. The molecular gas depletion time (\taudep = \mmol/SFR) is surprisingly constant as a function of redshift for galaxies close to the SFMS \citep{tacconi20}, indicating that the conditions under which star formation (SF) occurs are very uniform in normal disk galaxies.

At high stellar masses (log(\mstar) $\sim 10.5$ \msun),  the growth of disks seems to come to a halt and disk galaxies become more and more rare, whereas spheroidal galaxies become more frequent  \citep{kauffmann03}. SF seems to become quenched at these high stellar masses. There are several possible explanations for a drastic decrease in SF for high-mass galaxies:
Major galaxy mergers may disrupt disk galaxies and transform them rapidly into elliptical galaxies \citep{baldry04}.  Increasing feedback from a growing supermassive black hole in an active galactic nucleus (AGN) may shock or eject gas from the galaxy disk, reducing its capacity to form stars \citep{hopkins06a, ogle14}). Ram-pressure stripping of the interstellar medium by the intercluster medium in a galaxy cluster can also remove cold gas \citep{Sivanandam14}.  An available cold gas reservoir is furthermore fundamental to maintain SF. A lack of gas or a lack of molecular gas that formed from atomic gas or inefficient SF due to the properties of the molecular gas could all decrease the SFR in a galaxy.  The accretion of cold gas onto a galaxy may be stopped when the galaxy halo becomes so massive that accretion shocks develop, interrupting the cold streams of gas needed to replenish the disk \citep{dekel06}. The molecular-to-atomic gas mass ratio depends on properties like the  midplane pressure in galaxies \citep{wong02, blitz04, blitz06, leroy08}  and variations in these parameters among galaxies can affect the formation of molecular gas. And finally, the SF efficiency (SFE =SFR/\mmol, the inverse of \taudep) depends on the physical properties of the molecular gas, such as the density and temperature of the giant molecular clouds (GMCs) and the fraction of diffuse molecular gas, not bound to GMCs. 

Studies of large galaxy samples have demonstrated how the gas mass fraction (\mgas/\mstar), the molecular gas mass fraction, \mmol/\mstar,  and the molecular gas depletion time, \taudep,  depend on the position of the galaxy in the \mstar - SFR plane. Scaling relations have been derived for local galaxies  \citep[e.g.,][]{saintonge11a,saintonge11b, saintonge17, janowiecki20, casasola20} and for galaxies at high redshift \citep{genzel15, tacconi18}.  Results show that \taudep\ has only a weak dependence on stellar mass and on redshift, and it changes most significantly as a function of  the distance to the SFMS ($\triangle$SFMS), with longer times below the SFMS. The total gas fraction \mgas/\mstar,  and molecular gas fraction, \mmol/\mstar,  decrease with stellar mass and also with $\triangle$SFMS. Together, these relations  imply that a lack of gas is an important reason for the quenching of SF, but that changes in \taudep\ also play a role.

In contrast to what one would expect, even at very high stellar masses, about 6\% of galaxies have disks that have not quenched SF \citep{ogle16, ogle19b}.  \citet{ogle19b} selected  a catalog of 84 super spirals (SSs)  from the 1525 most optically luminous galaxies from the Sloan Digital Sky Survey. These spiral galaxies are extreme by many measures, with r-band luminosities of L = 8--14 L$^{\star}$, stellar masses of \mstar  = 0.2--1$\times$10$^{12}$ \msun, and giant isophotal diameters of \diso $=$ 55--134 kpc. Their sSFR puts them on the SFMS. They have redshifts of 0.1 $< z <$ 0.3 and appear uncommon in the local Universe.  Progenitors for SSs have not yet been identified at much higher redshifts.   Super spirals are very likely a remnant population of unquenched, massive disk galaxies. A large fraction (41\%) have double nuclei, double disks or other signatures of ongoing mergers. Presently, their high mass protects their disks from destruction in a merger because the majority of super spiral mergers are now minor mergers  \citep{ogle19b}. However, this leaves the open question of how SSs managed to become such massive disks in the first place. Possibly, super spirals have  remained star-forming disk galaxies compared to giant ellipticals because they reside in less massive dark halos than giant ellipticals of similar mass in stars. Alternatively, the super spirals with large bulge fractions may have formed more recently from a gas-rich spiral-elliptical minor merger \citep{jackson22}.

Super spirals are excellent objects to test galaxy evolution.  Their extreme properties (size, stellar mass) provide a unique opportunity to extend studies of  disk galaxy scaling laws to an entirely new regime, normally occupied by giant elliptical galaxies. 
In any case, the existence of super spirals demonstrates that the limit to spiral galaxy size and mass is much higher than previously thought, and that a high stellar mass can not be the primary cause of star-formation quenching. In fact, spiral galaxies with \mstar $\sim 10^{11}$ \msun\ may be most efficient at converting gas into stars, with mass fractions in stars approaching the cosmological baryon fraction \citep{posti19, diTeodoro22}.

In this paper, we present the first study of molecular gas in super spirals, derived from the CO(1-0) line intensity,  for a sample of 46 super spirals and  for a sample of 18 slightly less massive galaxies that are characterized by very broad  atomic hydrogen (HI) emission lines.  These data allow us to extend existing scaling relations to the mass regime of super spirals, find out how much molecular gas is available in these objects and whether the relation between molecular gas and SFR is comparable to   less massive galaxies. This will give us insight into how SF proceeds in the most massive galaxies  that have apparently escaped previous quenching mechanisms.

All rest-frame and derived quantities in this work assume a \citet{kroupa01} initial mass function and a cosmology with $\rm H_0$  = 70 km s$^{-1}$
 Mpc$^{-1}$, {  $\Omega_{\rm m} = 0.7$} , and $\Omega_{\rm \Lambda} = 0.3$. The distance are derived from redshifts in the CMB-frame.

\section{Sample and data}

\subsection{Samples}

\subsubsection{Sample of super spirals}

We selected the sample of super spirals primarily from the catalogs of \citet{ogle16} and \citet{ogle19b}. The galaxies in these catalogs were selected from the  Sloan Digital Sky Survey (SDSS) from the r-band with $L_{\rm r} > 8 L^*$ and $z< 0.3$. In addition, following \citet{ogle19a}, we selected additional objects from the 2 Micron All-sky Survey Extened Source Catalog \cite[2MASX;][]{Jarrrett+2000} which
allowed us to include more edge-on, dusty galaxies. These latter objects were selected for log(\mstar) $>$ 11.6 (estimated from the WISE band 1 luminosity and assuming a M/L ratio of 0.6), a slightly lower range in redshift of $z< 0.25$, and \diso $>$ 55 kpc. From both samples, we selected galaxies with SFR  $>$ 10 \msun yr$^{-1}$ \citep[calculated from the WISE band 3 and 4 luminosities, following][]{cluver14}  in order to increase the probability of detection with the IRAM 30m Telescope.
We selected in total 74 galaxies which were observed in CO(1-0) with the 30m telescope.

We then cleaned this sample by excluding 28 galaxies with a strong AGN, dominating the near-infrared and mid-infrared light and making the stellar mass and SFR determination uncertain (see Sect.~\ref{sec:AGN-from-WISE}). We present the molecular gas data for the AGNs, but we do not include the objects in the subsequent analysis. In this way, we end up with a sample of 46 star-forming super spiral galaxies. We call this sample the SS sample.

In addition, we included 18  slightly lower-mass (log(\mstar) $\gtrsim11$ \msun) galaxies that have very broad HI-lines and high peak rotation speeds ($>300$ km s$^{-1}$), indicating a large dynamical mass. These objects are more nearby than the super spiral sample (which are so rare that we do not find them in the local universe). We call this sample the "HI fast rotator" (HI-FR) sample. We include these objects because  of the possibility to  analyze also the HI content in a sample of galaxies with similar, albeit less extreme, properties as the super spirals, and because they fill  the stellar mass gap between the SS and the comparison sample.

\subsubsection{Comparison sample}

Several catalogs of noninteracting, nearby galaxies containing CO, HI, SFR, and \mstar\ exist in the literature; for example the AMIGA sample of isolated galaxies \citep{verdes-Montenegro05, Lisenfeld07, lisenfeld11}, the xCOLDGASS sample of mass- selected nearby galaxies \citep{saintonge11a, saintonge11b, saintonge17}, a catalogue of the ISM of normal galaxies \citep{bettoni03}, or an analysis of the scaling relations in DustPedia galaxies \citep{casasola20}. Here, we use the  xCOLDGASS sample for comparison because it is a representative sample of nearby galaxies, and the CO observations have been taken with the IRAM 30m telescope and have been processed in a similar way as for our sample which makes the comparison more reliable\footnote{The data for this sample has been retrieved from http://www.star.ucl.ac.uk/xCOLDGASS/data.html}. 
The  xCOLDGASS galaxy sample \citep{saintonge17}
is a mass-selected (\mstar $> 10^{9}$ \msun)  local sample of 532 nearby ($0.01 < z < 0.05$) galaxies. It was selected to be a representative sample for all galaxies  in the SDSS survey, based on the distribution  in the SFR-\mstar\ plane.  The HI fluxes 
were obtained from the xGASS survey \citep{catinella18}, a HI survey of 1179 observed with the Arecibo telescope.

The angular size of the xCOLDGASS is small enough to fit  almost completely inside the IRAM 30-m telescope beam width. {  A small aperture correction, \faper, with a mean value  of \faper $\sim$ 1.17, is applied in \citet{saintonge17} to the galaxies in xCOLDGASS in order to correct for the different fractions covered by the beam.}
For the aperture correction  the procedure defined in Lisenfeld et al. (2011) was followed, which is also adopted  in the present paper (Sect.~\ref{sec:molecular_gas_mass}) {  with a small difference in the choice of the assumed exponential scale length of the molecular gas distribution:}
For xCOLDGASS  an exponential \htwo\ distribution with a scale length  corresponding to the radius enclosing 50\% of the SFn as measured in the SDSS/GALEX photometry was adopted {  \citep{saintonge17}}. In the present paper, we also assume an  exponential distribution of the H$_2$, but with  exponential scale length $r_{\rm e} = 0.2  \times$ \riso\ {  (see Sect.~\ref{sec:molecular_gas_mass})}. We do not expect this relatively small difference to have any impact on our results, because the aperture corrections are small.

The molecular gas mass is calculated using a conversion factor \alphaco\ that varies as a function of metallicity and distance to the SFMS, following  \citet{accurso17}.  
Given the large variety of properties in the xCOLDGASS sample, this is the best choice. 
A factor of 1.36 for He and heavy metals is taken into account as for our sample.

The SFR of the xCOLDGASS galaxies follows the prescription of \citet{janowiecki17} and is based for most galaxies on a combination of WISE band 4 (or band 3) and GALEX NUV luminosities. The calculation of both the stellar mass and the SFR  are based on  a Chabrier IMF \citep{chabrier03}, which is very similar to the Kroupa IMF \citep{kroupa01} used in some of the prescriptions in the present paper. 

\subsection{Data}

\subsection{Molecular gas data}

\subsubsection{CO observations and data reduction with the IRAM 30m telescope}

Observations were carried out  between January and October 2020
with the Institut de Radioastronomie Milimetrique (IRAM) 30 m telescope on Pico Veleta within the projects 205-19 and 068-20.
In addition, we retrieved data for one object (UGC~06066) from the IRAM archive. 
It had been observed in project 070-12 (PI. M. Haynes).
We observed the redshifted $^{12}$CO(1-0)  in the central position of each galaxy.
We used  the dual polarization receiver EMIR in combination  with 
the autocorrelator FTS at  a   frequency resolution of 0.195 MHz  (corresponding to a  velocity resolution of $\sim$ 0.5 \kms\ at CO(1--0) at the frequency of our observations)
and with the autocorrelator  WILMA with a frequency resolution of  2MHz (corresponding to a  velocity resolution of  $\sim$ 5 \kms\  at CO(1--0)).
The observations were done in wobbler switching mode with a
wobbler throw of 80\arcsec\ in azimuthal direction. We confirmed for each galaxy 
that the off-position was well outside the galaxy.

The  broad bandwidth of the receiver (16 GHz) and backends (8 GHz for the FTS and 4 GHz for WILMA) allow the observations
of galaxies to be grouped into similar redshifts. 
The observed frequencies, taking into account
the redshift of the objects, range between 89.6~GHz and 110.5~GHz. 
Each object was observed until it was detected with a S/N  ratio of at least 5 or
until a root-mean-square noise (rms) of $\sim 1.5 $ mK (T$_{\rm mB}$) was achieved for a velocity resolution of 20  km s$^{-1}$ (only four
objects were undetected with a higher rms   between 1.6mK and 2.5 mK). 
The on-source integration times per object ranged between 20~minutes and 3 hours for most objects, and longer (6 hours) for UGC~06066.
Pointing was monitored on nearby quasars  every 60 -- 90  minutes.
During the observation period, the weather conditions were 
generally good, with a pointing accuracy better than 3-4~\arcsec.
Data taken in poorer conditions was rejected.
The mean system temperature for the observations
was 130~K for CO(1-0) 
on the $T_{\rm A}^*$ scale. At 100 GHz 
the  IRAM forward efficiency, $F_{\rm eff}$, is  0.95 
and the  beam efficiency, $B_{\rm eff}$, is 0.79.
The half-power beam size for CO(1-0) ranges between  22.5$^{\prime\prime}$ (for 110.5 GHz)  and 27.6$^{\prime\prime}$ (for 89.6 GHz).
All CO spectra and luminosities are
presented on the main beam temperature scale ($T_{\rm mb}$) which is
defined as $T_{\rm mb} = (F_{\rm eff}/B_{\rm eff})\times T_{\rm A}^*$.

The data were reduced in the standard way via the CLASS software
in the GILDAS package\footnote{http://www.iram.fr/IRAMFR/GILDAS}.
We first discarded poor scans and data taken in poor weather conditions (e.g.,  with large pointing uncertainties)
and then subtracted a constant or linear baseline.
Some observations taken with  the FTS backend  were affected by platforming,
that is the baseline level changed abruptly at one or two positions along the band. This effect could be
reliably corrected because the baselines in between these (clearly visible) jumps were linear and could
be subtracted from the different parts individually, using the {\it FtsPlatformingCorrection5.class}
 procedure provided by IRAM.
We then averaged the spectra and 
smoothed them  to resolutions of  10, 20 and 40~\kms.

We present the detected spectra  in Appendix A.
For each spectrum, we  visually determined the zero-level line widths, if detected. 
The velocity-integrated spectra were calculated by summing the individual channels in between
these limits. 
 For nondetections we set an upper limit as

\begin{equation}
I_{\rm CO} < 3 \times {\rm rms} \times \sqrt{\delta \rm{V} \ \Delta V},
\end{equation}

\noindent where $\delta \rm{V}$ is the channel width (in kilometers per second), $\Delta$V the zero-level line width (in kilometer
per second), and rms the root mean square noise (in Kelvin).
For the nondetections, we assumed a line width of $\Delta$V = 700  \kms\ which is close to the mean velocity width found
for CO(1-0) in the sample (mean $\Delta \rm{V}$ = 708 \kms\ with a standard deviation of 227 \kms).
We considered spectra with a S/N ratio  of the velocity integrated intensity $> 5$ as firm detections and those with a S/N rato in the range
of 3-5 as tentative detections.
The results of our CO(1-0)  observations are listed in Table~\ref{tab:ico}. We have 77 detections 42 SS, 15 HI-FR and 20 AGNs), 7  tentative detections (2 SS, 1 HI-FR and 4 AGNs) and 8 nondetections  (2 SS, 2 HI-FR and 4 AGNs).
In addition to the statistical error of the velocity-integrated line intensities, a calibration error of 
15 \% for CO(1-0) 
has to be taken into account \citep[see][]{lisenfeld19}. 

In addition to the central pointing, we mapped four objects (NGC~2713,  NGC~5790, UGC~08902, and UGC~12591) at various positions
along the major axis. The spacing between the pointings is 11\arcsec (about half the FWHM of the beam at 110~GHz) and the total number of pointings per galaxies ranged between 3 and 6. We show the individual spectra of the mapped galaxies in Appendix~\ref{app:maps}.

\begin{table}
\caption{\label{tab:ico} Velocity-integrated CO intensities (central pointings)}
\begin{tabular}{llllll}
\noalign{\smallskip} \hline \noalign{\medskip}
Galaxy name &  rms\tablefootmark{a} &  $I_{\rm CO(1-0)}$\tablefootmark{b} & det\tablefootmark{c}  &   $\rm \Delta \rm{V}_{CO(1-0)}$\tablefootmark{d}  \\
   & [mK]  & [K \kms]& & [\kms] \\
\noalign{\smallskip} \hline \noalign{\medskip}
2MFGC12344 &   0.88 &   1.57 $\pm$   0.13 & 0 &   1018 \\ 
OGC~139 &   1.67 & $<$   0.60 & 1 &    700 \\ 
OGC~217 &   1.67 &   1.39 $\pm$   0.14 & 0 &    347 \\ 
OGC~290 &   1.46 &   1.02 $\pm$   0.13 & 0 &    403 \\ 
... & ...& ...& ... & ... \\
\noalign{\smallskip} \hline \noalign{\medskip}
\end{tabular}
\tablefoot{
\tablefoottext{a}{Root-mean-square noise at a velocity resolution of 40 \kms.}
\tablefoottext{b}{Velocity integrated intensity and statistical error of the CO(1-0)  line.}
\tablefoottext{c}{Detection code: 0 = detection (S/N $\gtrsim$ 5), 2 = tentative detection (S/N $\approx$ 3-5), 1 = nondetection.}
\tablefoottext{d}{Zero-level line width. The uncertainty is roughly given by the velocity resolution ($\sim 20$  \kms).}
The full table is available online at the CDS.
}
\end{table}

\subsubsection{Aperture correction}
\label{sec:molecular_gas_mass}

In most of our observations with the IRAM 30m telescope we only observed the galaxies in their central pointing.
Since the galaxies in our sample are in general small, the central pointing covers a  large fraction of the galaxy.
However, this fraction is different for each galaxy depending on its size.  We therefore need to apply a correction for emission outside the beam.
We carried out this aperture correction in the same way as described in \citet{lisenfeld11}, 
assuming an exponential distribution of the CO flux:

\begin{equation}
S_{\rm CO}(r) = S_{\rm CO,center}\propto \exp(-r/r_{\rm e}) ,
\label{eq:Ico_r}
\end{equation}
where $S_{\rm CO,center}$ is the CO(1-0) flux in the central position derived from
the measured \ico\ applying the $T_{\rm mB}$-to-flux  conversion factor of the IRAM 30m telescope (5 Jy/K).
\citet{lisenfeld11} adopted  an exponential scale length of $r_{\rm e} = 0.2 \times r_{\rm 25}$, where \riso\ is 
the major optical isophotal radius at 25 mag arcsec$^{-2}$,  from different  studies  of  local spiral galaxies
 \citep{2001PASJ...53..757N, 2001ApJ...561..218R, leroy08}
 and from their own CO data.  
 Very similar values for $r_{\rm e}$/\riso\ were found by \citet{boselli14a} ($r_{\rm e}$/\riso $\sim
  0.2$)
 and \citet{casasola17} ($r_{\rm e}$/\riso = 0.17$\pm$ 0.03) from an analysis of nearby mapped galaxies.

Thus, we adopt $r_{\rm e}$ = 0.2$\times$\riso\ in 
 eq.~\ref{eq:Ico_r} and use this
 distribution to calculate the expected CO flux  from the entire disk, $S_{\rm CO,tot}$, taking the galaxy inclination into account,
by 2D integration over the exponential galaxy disk \citep[see][for more details]{lisenfeld11}.
\citet{boselli14a} generalized this method to three dimensions by taking the finite thickness of galaxy disks  
into account.
Except for edge-on galaxies ($i>80^\circ$) the 3D method gives basically the same result as the 2D approximation,
and also for edge-on galaxies the difference is $<5\%$ for $z_{\rm CO}/\Theta < 0.1$ ($z_{\rm CO}$ being the scale height of the CO perpendicular to
the disk and $\Theta$ the beam size). We therefore  consider the 2D aperture correction to be  sufficient.

The resulting aperture correction factors, \faper, defined as the ratio between $S_{\rm CO,center}$ and the total
aperture-corrected flux $S_{\rm CO,tot}$,
lie between 1.03 and 6.26 with a mean (median) value of 1.46 (1.13). 
There are 5 objects in the sample for which neither values for the inclination nor \riso\ were found. We adopted the median value of the sample, \faper = 1.13, for them.
The values of  \faper\ are
listed in Table~\ref{tab:mhtwo_extra}.

\subsubsection{Molecular gas mass and \alphaco} 

\begin{table}
\caption{\label{tab:mhtwo_extra} Extrapolated molecular gas mass}
\begin{tabular}{lllll}
\noalign{\smallskip} \hline \noalign{\medskip}
Galaxy name & $z$\tablefootmark{a}  &  $D_L$\tablefootmark{b} & log(\mmol)\tablefootmark{c}  &  \faper  \\
   &  & [Mpc]  & [ \msun ] & \\
\noalign{\smallskip} \hline \noalign{\medskip} 
2MFGC12344 &  0.141 &    665 &  10.58 $\pm$   0.15 &   1.17 \\ 
OGC~139 &  0.247 &   1244 & $<$  10.66  &   1.15 \\ 
OGC~217 &  0.249 &   1254 &  10.99 $\pm$   0.16 &   1.05 \\ 
OGC~290 &  0.296 &   1528 &  11.01 $\pm$   0.17 &   1.05 \\ 
... & ...& ...& ... & ... \\
\noalign{\smallskip} \hline \noalign{\medskip}
\end{tabular}
\tablefoot{
\tablefoottext{a}{Redshift, $z$, from SDSS DR9 or DR13 \citep[see][]{ogle16,ogle19a}.}
\tablefoottext{b}{Luminosity distance, calculated adopting  $\rm H_0$  = 70 km s$^{-1}$
 Mpc$^{-1}$, $\Omega_{\rm m} = 0.3$ , {  $\Omega_{\rm \Lambda} = 0.7$.}} 
\tablefoottext{c} {Extrapolated molecular gas mass, except for UGC~12591 where the total mapped molecular gas mass is listed.}
The full table is available online at the CDS.
}
\end{table}

We calculated the molecular gas mass 
from the CO(1-0) luminosity, \lco , following \citet{solomon97} as:
\begin{equation}
L^\prime_{\rm CO} [{\rm K \, km\, s^{-1} pc^{-2}}]= 3.25 \times 10^7\, S_{\rm CO,tot} \nu_{\rm rest}^{-2} D_{\rm L }^{2} (1+z)^{-1},
\label{eq:lco}
\end{equation}
where  $S_{\rm CO, tot}$ is the aperture-corrected CO line  flux (in Jy \kms), 
$D_{\rm L }$ is the luminosity distance in Mpc, $z$ the redshift,
and $\nu_{\rm rest}$ is the rest frequency of the line in gigahertz.
We then calculated the molecular gas mass, \mmol\ (including a mass fraction of helium and heavy metals of a factor 1.36) as:
\begin{equation}
M_{\rm mol} [M_\odot]= \alpha_{\rm CO} L^\prime_{\rm CO} .
\label{eq:mmol}
\end{equation}

 The conversion factor \alphaco\ is known to vary as  a function of metallicity. The most drastic variations occur in  low-metallicty galaxies (12+log(O/H) $\lesssim 8.4$), where \alphaco\  increases steeply as a function of decreasing metallicity \citep[see][]{bolatto13}.  A considerably lower value of \alphaco\ should be applied in   starbursting galaxies lying well above ($\sim$ 1 dex) the SFMS. They are  characterized by  high surface densities which change the conditions of the ISM. In addition, \citet{accurso17} has shown that \alphaco\ varies as a function of the distance to the MS  for nonstarbursting galaxies, with higher values above the MS due to the stronger radiation field, and lower values below it. The effect produces a small correction of up to 12~\% and should only be applied to nonstarbursting galaxies.
 
 Based on the mass-metallicity relation, our SS+HI-FR sample is expected to have slightly super-solar metallicities. The difference is not expected to be very large because the metallicity approaches constant values for stellar masses above $\sim 10^{10.5}$ \mstar, independent of the exact method of measuring the metallicity \citep[see][]{kewley08, mannucci10}. Adopting the prescription of \citet{mannucci10} (their eq. 2), we derive, based on the stellar mass and SFRs of the SS+HI-FR sample, a metallicity of  12+log(O/H) $\sim$ 9.0. With a solar metallicity of  of  12+log(O/H) = 8.69 \citep{asplund09} this gives a metallicities of  a factor 2 higher than in the Solar neighborhood.  
 
 We use the  metallicity dependence of \alphaco\ from the prescription of  \citet{accurso17} (their eq. 25) and of \citet{bolatto13} (their eq. 31) to predict the expected \alphaco\ in SS+HI-FR galaxies. We ignore the  dependence on the surface density included in the prescription of \citet{bolatto13} because SS+HI-FR galaxies are not in the starburst regime.  We neither consider a possible dependence on the distance from the SFMS included in the prescription of \citet{accurso17} in order to keep the method simple and because the effect is small. We discuss the validity of our choice in Sect.~\ref{sec:discussion-uncertainties}. We predict  \alphaco = 2.95 from \citet{accurso17} (adopting 12+log(O/H) = 8.8 which is the maximum value for which their prescription is valid and which they recommend for higher metallicites) and \alphaco = 3.5 from \citet{bolatto13}. In addition, we take into account of the results of \citet{wolfire10} who calculated the fraction of dark gas, that is the fraction of molecular gas in a molecular cloud that does not contain CO, as a function of different parameters. They find  (their Fig. 10) that  this fraction decreases from roughly 30\% (40\%) for Solar metallicity and a mean surface density of $1.5\times 10^{22}$ cm$^{-2}$ ($0.75\times 10^{22}$cm$^{-2}$) to value of 17\% (25\%) for a factor 1.9 higher metallicity. The decrease in dark mass fraction, $f_{\rm DG}$, is thus a factor of 1.6 -1.8. This translates, adopting a simple picture in which \alphaco $\propto (1-f_{\rm DG})^{-1}$, to a value of  \alphaco\ between 3.5 - 3.6, in agreement with the relation of \citet{bolatto13}.
 
 Based on these predictions we adopt  \alphaco = 3~\msun/(K \, km s$^{-1}$ pc$^{-2})$ as a reasonable  estimate for our galaxies, which is a factor 1.4 lower than the Galactic value \citep[\alphacogal = 4.3 ~\msun/(K \, km s$^{-1}$ pc$^{-2}$, ][]{bolatto13}. We note that this value is on the lower end of the range of \alphaco\ predicted from the method considered above, which makes our derivation of \mmol\ conservative in the sense that we do not expect to overestimate \mmol\ with this choice of \alphaco.
 {  Our adopted value of \alphaco\ closely corresponds to what the \citet{accurso17} prescription would predict for galaxies of this mass and metallicity and is therefore consistent with the \alphaco\ adopted for the comparison sample xCOLDGASS.}
 The extrapolated molecular gas masses calculated with this conversion factor are  listed in Tab.~\ref{tab:mhtwo_extra}.

\subsubsection{Mapped molecular gas mass}
\label{sec:compare_maps}

Four galaxies were mapped with 3-6 pointings along the major axis. For these, we derived the total flux from the average \ico\ by applying an adjusted  $T_{\rm mB}$-to-flux 
conversion factor of 5 Jy/K $\times$ (mapped area/area of the CO(1-0) beam). Then, we calculated the total molecular mass from eq.~\ref{eq:lco} and eq.~ \ref{eq:mmol}.
In Table~\ref{tab:mmol_map} the mapped molecular gas masses, $M_{\rm mol, map}$,  are listed and compared to the extrapolated values.

For all objects except UGC~12591, $M_{\rm mol, map}$ is smaller than the extrapolated molecular gas mass which is not surprising because the 
mapping only covers part of the major axis (see column 2 in  table~\ref{tab:mmol_map}). UGC~12591 was mapped furthest, 
out to \riso. Here, the mapped molecular gas mass is only slightly (10\%) higher than the extrapolated value, showing that the extrapolation works 
well even for this relatively large object (\riso = 45\arcsec , \faper = 2.3).
For UGC~12591, we use the mapped molecular gas mass, $M_{\rm mol, map}$, instead of the extrapolated value in the analysis of this paper.

\begin{table}
\caption{\label{tab:mmol} Molecular gas mass from mapped objects}
\begin{tabular}{lccc}
\noalign{\smallskip} \hline \noalign{\medskip}
Galaxy name &    $\left(\frac{r_{\rm map}}{r_{\rm 25}}\right)$\tablefootmark{a}  &   log(\mmol$_{\rm, map}$)\tablefootmark{b} & $\left(\frac{ M_{\rm mol, map}}{M_{\rm mol}}\right)$\tablefootmark{c}\\
   &  &  [ \msun ] & \\
\noalign{\smallskip} \hline \noalign{\medskip} 
NGC~2713 &  0.3 & 9.26  &  0.4\\
NGC~5790 & 0.6 & 9.53  & 0.6\\
UGC~08902 & 0.6  & 9.96  &0.8 \\
UGC~12591 & 1.0  & 9.55  & 1.1 \\
\noalign{\smallskip} \hline \noalign{\medskip}
\end{tabular}
\tablefoot{
\tablefoottext{a}{Ratio between maximum radial distance of the CO pointings to the radius of the galaxy at a surface brightness of 25 mag arcsec$^2$.}
\tablefoottext{b}{Decimal logarithm of the mapped molecular gas mass. }
\tablefoottext{c}{Ratio between mapped and extrapolated molecular gas mass.}
}
\label{tab:mmol_map}
\end{table}

\subsection{Atomic gas mass}

For the fast HI rotators, we obtained the velocity integrated HI fluxes, $S_{\rm HI}$,  from the Alfalfa survey \citep{haynes18}
 and calculated the atomic gas mass as:
 
 \begin{equation}
 M_{\rm HI} = \frac{2.36\cdot 10^5}{(1+z)^2} \cdot \left( \frac{S_{\rm HI}}{\rm Jy\, km s^{-1}} \right) \left(  \frac{D_{\rm L}}{\rm Mpc} \right)^2,
 \end{equation}
\citep[see][]{meyer17, saintonge22}.
No correction for Helium and metals is included. For UGC~12521 the value from di Teodoro et al. (2022) (their Tab. 1), adapted to our distance and no Helium, 
is used.

\subsection{WISE data}
\label{sec:Wise-data}
\subsubsection{WISE photometry}

WISE galaxy measurements come from the WISE Extended Source Catalog 
\cite[WXSC;][]{jarrett13,jarrett19}.
It utilizes 
 custom image mosaic construction 
 of the four WISE bands: 3.4, 4.6, 12, and 23\,$\mu$m
\citep{jarrett+2012} which preserves native resolution.
It catalogues 
 complete resolved source characterization that includes careful contaminant removal, local background estimation, size and orientation, a suite of photometric, surface brightness, and radial profile measurements
\citep[see][]{jarrett13,jarrett19}.

Based on these maps we estimated
total fluxes  by modeling the emission profile in each band,
constructing
axi-symmetric radial profiles,  which were fitted with a double-Sersic function to represent
the spheroidal and disk population distributions, extrapolated to several disk scale lengths
to determine the total emission.  

We then derived rest-frame fluxes  using SED modeling of the observed-frame fluxes.  As described in \cite{jarrett19, jarrett23}, a suite of composite templates (ranging across all morphological types) are (1+z) scaled to the redshift of the object and fit  
to the measurements.  The best match is then used to provide observed-to-rest flux corrections.
Errors in the corrections are driven by the photometric quality, number of available measurements to define the SED, and the finite set of templates.
Based on the analysis in \citet[]{yao+2022}, the k-correction imparts less than 5-10\% uncertainty for most sources that have redshifts $<$ 0.3
 \citep[see the Appendix in][]{yao+2022}.

In  Tab.~\ref{tab:wise} we list the total measured, and  the k-corrected fluxes in the four WISE bands.

\begin{table*}
\caption{\label{tab:wise} WISE fluxes and classification}
\resizebox{\textwidth}{!}{%
\begin{tabular}{llllllllll}
\noalign{\smallskip} \hline \noalign{\medskip}
Galaxy name & $F_{\rm W1,obs}$\tablefootmark{a} & $F_{\rm W1,kcorr}$\tablefootmark{b} &
$F_{\rm W2,obs}$\tablefootmark{a} & $F_{\rm W2, kcorr}$\tablefootmark{b} & $F_{\rm W3, obs}$\tablefootmark{a}  & $F_{\rm W3, kcorr}$\tablefootmark{b} & $F_{\rm W4,obs }$\tablefootmark{a} & $F_{\rm W4, kcorr}$\tablefootmark{b} & Type\tablefootmark{c} \\
   &  [mJy] &[mJy] &[mJy] &[mJy] &  [mJy] &[mJy] &[mJy] &[mJy] & \\
\noalign{\smallskip} \hline \noalign{\medskip} 
2MFGC12344 &   3.14 $\pm$   0.10 &   4.56 $\pm$   0.05 &   1.81 $\pm$   0.09 &   2.75 $\pm$   0.03 &   3.24 $\pm$   0.60 &   4.17 $\pm$   0.19&   5.52 $\pm$   0.99 &   5.54 $\pm$   0.89 & SS\\ 
OGC~139 &   0.90 $\pm$   0.04 &   1.68 $\pm$   0.02 &   0.57 $\pm$   0.04 &   1.27 $\pm$   0.02 &   1.47 $\pm$   0.12 &   2.65 $\pm$   0.10&   2.20 $\pm$   0.80 &   2.65 $\pm$   0.59 & SS\\ 
OGC~217 &   0.72 $\pm$   0.03 &   1.23 $\pm$   0.02 &   0.51 $\pm$   0.04 &   1.10 $\pm$   0.02 &   4.23 $\pm$   0.23 &   5.26 $\pm$   0.16&  14.19 $\pm$   1.20 &   6.67 $\pm$   0.69 & SS\\ 
OGC~290 &   0.59 $\pm$   0.03 &   1.20 $\pm$   0.02 &   0.41 $\pm$   0.03 &   1.07 $\pm$   0.01 &   2.28 $\pm$   0.20 &   3.04 $\pm$   0.15&   6.77 $\pm$   1.13 &   3.43 $\pm$   0.57 & SS\\ 
... & ... &... &... &... &... &... &... &... &... \\
\noalign{\smallskip} \hline \noalign{\medskip}
\end{tabular}
}
\tablefoot{
\tablefoottext{a} {Photometrically measured fluxes and photometrical error.}
\tablefoottext{b}{Fluxes with applied k-corrections (as described in Sec.~\ref{sec:Wise-data}).}
\tablefoottext{c}{Galaxy type
(SS= super spiral, HI = HI fast rotator, AGN = AGN dominated galaxy). The distinction between AGN and SF galaxies (i.e, SS+HI-FR) was done based on the WISE colours as described in Sect.~\ref{sec:AGN-from-WISE}.}
Fluxes with a signal-to-noise ratio $<3$ are considered upper limits in the analysis. 
The full table is available online at the CDS.
}
\end{table*}

\subsubsection{Determination of AGN activity from WISE colors}
\label{sec:AGN-from-WISE}

\begin{figure}
   \centering
\includegraphics[width=8.cm,trim=0.cm 0.cm 0cm 0cm,clip]{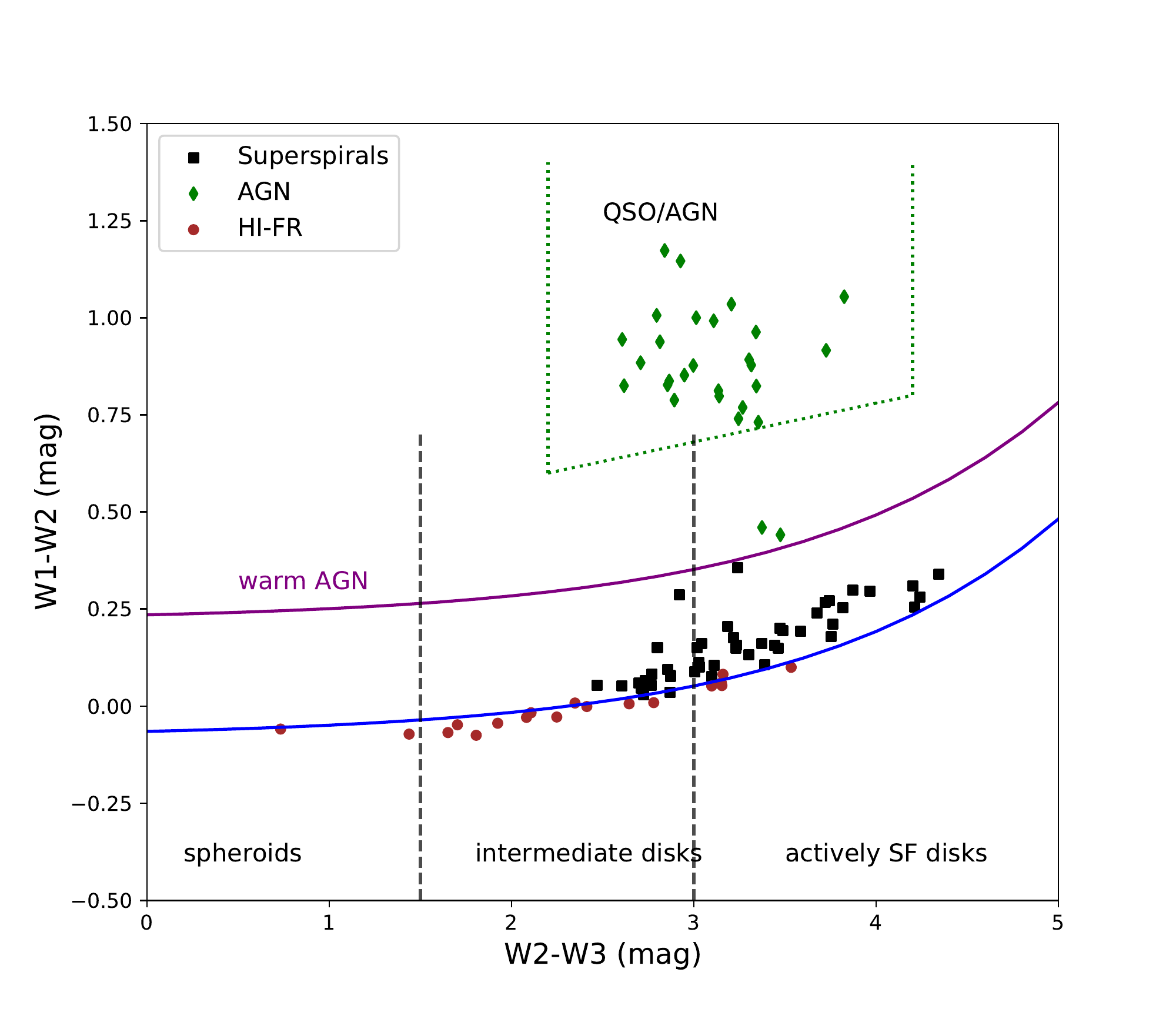}
      \caption{WISE color magnitude plot for the SS, HI fast rotator and AGN galaxies, following the classification scheme of Jarrett et al. 2019 (their Fig. 10).  The green dotted lines indicate the zone
      populated by QSO/AGN, following \citet{jarrett11}. The blue line gives the sequence of SF galaxies  \citep[eq. 1 from][]{jarrett19}, from quiescent objects (low [W2-W3]) to actively 
      star-forming objects (high [W2-W3]).   The purple line, labeled "warm AGN", indicates the region where low-level Seyferts and Liners reside \citep[see][]{jarrett11}. We adopt this as 
      the dividing line between  star-forming  and AGN dominated galaxies and flag galaxies above this line as AGN-dominated.
 }
                \label{fig:wise-colors}
   \end{figure}

WXSC mid-IR  colors can be used to separate  quiescent,  actively SF or AGN dominated galaxies. We use the W1-W2 and W2-W3 colours and  the classification of \citet{jarrett17}, as presented in  \citet[][their Fig. 10]{jarrett19},  to separate galaxies with dominant AGN 
emission in the  mid-IR (Fig.~\ref{fig:wise-colors}).
We use the prescription from \citet[][their eq. 1 ]{jarrett19}  to define the mid-IR star-forming sequence:

\begin{equation}
[W1-W2] = 0.015  \times \exp([W2-W3]/1.38) - 0.08
\end{equation}
and define galaxies as AGN dominated if they lie above the "warm AGN" line  in Fig.~\ref{fig:wise-colors} which is offset by +0.3 mag from the mid-IR star-forming sequence.
We exclude AGN dominated objects from our analysis because we cannot derive reliable values for the SFR and the stellar mass since the mid-IR luminosities are
to a large extent due to AGN and not stellar emission. Based on this criterion, 28 galaxies are AGN dominated. In  Tab.~\ref{tab:wise} the resulting classification codes are listed.

\subsection{GALEX data}

Near-ultraviolet (NUV) images from the Galaxy Evolution Explorer (GALEX) satellite were extracted from the Mikulski Archive for Space Telescopes (MAST) GALEX GR6/7 archive\footnote{see http://galex.stsci.edu/GR6/} for the 55 SS+HI-FR sample.  
Nine galaxies did not match any GALEX observation because some regions of the sky were not observed due to either bright UV source avoidance or because of very high stellar density close to the plane of the Milky Way. In some cases, 
several different images contained the same target object. In such cases we used the image with the longest integration time for our analysis.

NUV emission was extracted from the GALEX science images in counts s$^{-1}$ over a circular aperture corresponding to the isophotal diameter D$_{25}$ obtained from the NASA Extragalactic 
Database (NED; D$_{25}$ is the B-band isophotal diameter at a surface brightness of 25 mag arcsec$^2$). Background subtraction was achieved by subtracting the counts s$^{-1}$ in the background image evaluated over the same area. Surface brightness 
profiles were also extracted to ensure that D$_{25}$ was a good representation of the main body of UV emission for each object. In almost all cases, a large fraction of the UV light was captured inside this diameter. 
Each image was visually inspected to ensure that there were no bright contaminating stars within the aperture. Only in two cases were bright stars found near the edge of the aperture, and these were masked to preserve 
the quality of the photometry. Uncertainties in the measured fluxes were evaluated by adding both shot noise and background uncertainty together in quadrature. Background uncertainties were difficult to measure directly from the images, often leading 
to unrealistically small values,  and so we assumed a conservative average background uncertainty of 2$\%$ (and 5$\%$ for FUV) based on documentation provided by GALEX home page. In addition,  a calibration error of 14.8\% \citep{gildepaz07} has to be added in quadrature.

Conversion from count s$^{-1}$, c,  to AB magnitde, M(AB)$_{NUV}$  followed the standard relation given in the GALEX User Manual M(AB)$_{nuv} = -2.5 \times \log_{10}(c) + 20.08$. These photometric  magnitudes were corrected for Galactic extinction assuming a value for E(B-V) determined from \citet{schlegel98} with the additional recalibration corrections of \citet{schlafly11}.
Following \citet{bianchi11} we assumed a Galactic extinction curve and A$_{NUV}$/E(B-V) = (R$_{NUV}$ = 7.95).
The measured fluxes, together with their photometrical errors, are listed in Tab.~\ref{tab:galex}.

We also applied a k-correction following \citet{chilingarian10, Chilingarian12}\footnote{We used the online-calculator at http://kcor.sai.msu.ru/}. The k-correcting was small, less than 10\% for 47 objects, and between 10\% and 25\% for the remaining eight objects.

\begin{table}
\caption{Measured GALEX NUV flux and photometric error, SFR and \mstar.}
\label{tab:galex}
\begin{tabular}{llll}
\noalign{\smallskip} \hline \noalign{\medskip}
Galaxy name & $F_{\rm NUV}$
& log(SFR) & log(\mstar) \\
   &  [$\mu$Jy] & [\msun\ yr$^{-1}$] & \msun  \\
\noalign{\smallskip} \hline \noalign{\medskip} 
2MFGC12344 &   58.0 $\pm$    1.4 &   0.99  &  11.65 \\ 
OGC~139 &   38.2 $\pm$    0.9 &   1.14  &  11.65 \\ 
OGC~217 &   66.7 $\pm$    4.6 &   1.89  &  11.56 \\ 
OGC~290 &   64.2 $\pm$    1.8 &   1.78  &  11.65 \\ 
... & ...& ...& ... \\
\noalign{\smallskip} \hline \noalign{\medskip}
\end{tabular}
\tablefoot{
SFR and \mstar\ calculated as described in Sect.~\ref{subsec:sfr-mstar}.
The full table is available online at the CDS.}
\end{table}
 
\subsection{Star formation rate and stellar mass}
\label{subsec:sfr-mstar}

For both the calculation of the SFR and the stellar mass different prescriptions exist in the literature. Normally, the stellar mass  is derived from the  near-infrared  emission and the SFR can be derived from the ultraviolet (UV), combined with the  mid-infrared (to probe dust-enshrouded SF). 

In the present work, we therefore tested and compared different methods (see Appendix C and D) to ensure that the used prescription gives consistent results for the  SS+HI-FR and the comparison sample which cover different stellar mass ranges.  None of the existing SFR or \mstar\ prescriptions has been tested so far in the high stellar mass range of super spirals. As shown by \citet{leroy19}, the coefficients of the prescriptions have a dependence on stellar mass, and therefore we need to test as well as possible that the existing methods hold for higher masses.  
Apart from comparing different prescriptions, we also compare them to SED fitting with CIGALE \citep{boquien19} in order to derive both the SFR and the stellar mass in an independent way (Appendix E).

\subsubsection{Star formation rate}
\label{subsubsec:sfr}

In the present paper, we calculate the SFR from GALEX and WISE data, in order to probe both dust-free and dust-enshrouded SF. 
It is important to use  the same method  for  all samples of our study. We decided to follow the method used in xCOLDGASS  to calculate their  \sfrbest\  parameter \citep{saintonge17}. \sfrbest\  was  calculated following a "SFR ladder"  \citep[see][]{janowiecki17}.  A combination of GALEX NUV and WISE luminosities was used (preferentially W4,  and,  if not detected, W3) for all galaxies with good WISE and GALEX data, and for the remaining cases (30\% of the galaxies) the SFR was derived from SED fitting. 

We use  a very similar prescription for the SS+HI-FR galaxies. We calculate the SFR from W4+NUV \citep[eq. 3][]{janowiecki17} for those galaxies with good (S/N > 3) data for both the NUV and W4  bands (42 galaxies). For galaxies with good NUV data but poor W4 data, we use eq. 4 of \citet{janowiecki17} and calculate the SFR from W3+NUV (13 galaxies). For the remaining 8 galaxies with neither good W4 data nor good or existing NUV data we calculate the SFR only from W3 data alone.  Here, we use the prescription by \citet{cluver17} (their eq. 4), lowered by a 0.2 dex in order to guarantee a consistent normalization (see Appendix D). Thus, we use the following formulae for the SFR  (in order of decreasing preference):

\begin{equation}
SFR_{W4+NUV, J17} [\rm M_\odot yr^{-1}] = L_{\rm NUV} 10^{-43.29} + L_{\rm W4,dust} 10^{-42.70}   \\
\end{equation}
\begin{equation}
SFR_{W3+NUV, J17} [\rm M_\odot yr^{-1}] = L_{\rm NUV} 10^{-43.29} + L_{\rm W3,dust} 10^{-42.89}
\end{equation}
\begin{equation}
SFR_{W3, C17} [M_\odot yr^{-1}] =  0.889 L_{\rm W3,dust} 10^{-42.89} 10^{-41.54}    
\label{eq:SFR-W3_C17}
\end{equation}
where $ L_{\rm NUV}$  is the luminosity  of the GALEX NUV band and $L_{\rm W3,dust}, L_{\rm W4,dust}$  are the luminosities from the dust contribution to the WISE W3 and W4 bands. The latter  are obtained from the total luminosities in these bands after subtracting the stellar continuum based on the W1 luminosity, $L_{\rm W1} $, calculated following Jarrett et al. (2011) as in \citet{cluver17} as $L_{\rm W3,dust} = 0.158 \times L_{\rm W1}$ and $L_{\rm W4,dust} = 0.059\times L_{\rm W1} $ (very similar to the coefficients of Janowiecki ($L_{\rm W3,dust, J17} = 0.201 \times L_{\rm W1} $, and $L_{\rm W4,dust, J17} = 0.044 \times L_{\rm W1}$).
All luminosities are defined as $\nu L_\nu$ and  are in units of erg s$^{-1}$. 

As shown  in Appendix E, this definition of the SFR agrees well with the results from CIGALE for the SS+HI-FR sample. In Appendix D, we compare our  prescriptions for both xCOLDGASS and SS+HI-FR with the prescriptions of \citet{leroy19} and \citet{cluver17} and find in general good correlations,  \citep[albeit with a constant offset  in the case of][]{cluver17}). From this comparison we conclude that the systematic uncertainty in the SFR is about 0.2 dex.

\subsubsection{Stellar mass}

The stellar mass can be well traced by the mid-infrared emission and it is frequently derived  from the WISE 3.4~\mi\ (W1) luminosity. For this,  a  stellar mass-to-light ratio,  $\Upsilon_{\ast}^{3.4} $ (in units \msun/\lwonesun)\footnote{We use, as \citet{leroy19} and \citet{cluver14}, a value of  \lwonesun = $\rm 1.6 \times 10^{32} erg^{-1}$}
has to be adopted, which depends, 
however, considerably on the properties of a galaxy, in particular the age of the stellar population. Typical values range between   $\Upsilon_{\ast}^{3.4} \approx 0.1 - 0.7 $ \msun/\lwonesun\  \citep[e.g.,][]{leroy19}. 
There are different prescriptions to calculate the stellar mass from the mid-infrared luminosities. Some use simply a constant  mass-to-light ratio $\Upsilon_{\ast}^{3.4}$ \citep[e.g., ][]{eskew12}, whereas other use values of $\Upsilon_{\ast}^{3.4}$ that depend on  mid-IR color 
\citep[e.g., ][]{jarrett13, cluver14, jarrett23}, or sSFR \citep{leroy19}. 

All these prescriptions have not been tested in the mass range of super spiral galaxies.
Therefore, in Appendix D, we compare different prescription  for the xCOLDGASS and the super spiral sample, and in Appendix E we compared the  prescriptions to CIGALE.
For the SS+HI-FR sample, we find a good correlation of the stellar mass derived from CIGALE and those derived with $\Upsilon_{\ast}^{3.4} = 0.5$.
There is also a good correlation of the CIGALE results with the stellar mass  of \citet{leroy19}, albeit with a small offset of  0.1  dex. Considering the uncertainties and in order to keep the derivation of the stellar mass simple, we use a constant $\Upsilon_{\ast}^{3.4} = 0.5$ for our SS+HI-FR sample.
For the xCOLDGASS sample, mostly  for consistency with other studies, we use the stellar mass provided in \citet{saintonge17} which was taken from the SDSS DR7 MPIA-JHU catalog.  Good correlations with the prescription of \citet{leroy19}   and with \citet{cluver14} exist (for the latter with a constant offset of 0.3 dex).

\section{Results}

The goal of this study is to compare the molecular gas mass, stellar mass and SFR of very massive, star-forming galaxies to  those of galaxies with lower stellar masses. In order to properly compare our SS+HI-FR sample to the comparison sample, we need to (i) take into account that the SS galaxies are further away than the HI-FR and xCOLDGASS galaxies. Both the molecular gas fraction (\fmol = \mmol/\mstar) and the sSFR have a strong dependence on redshift $z$ \citep[e.g., ][]{genzel15, tacconi18} and we need to correct for this trend in order to carry out a meaningful comparison. (ii) Many properties of a galaxy depend very sensitively on the distance to the SFMS. We subsequently analyse our results with respect to this parameter. There are different prescriptions for the SFMS in the literature,  mostly due to differences in the way how to calculate the  SFR, and also due to details of the sample selection. We adopt the prescription of \citet{janowiecki20} which was derived from the xCOLDGASS sample.

\begin{table*}
\caption{\label{tab:mean-values} Mean value, its error, and median value for the SS and HI-FR sample. Upper limits are  treated as detections.}
\begin{tabular}{lcccccccc}
\noalign{\smallskip} \hline \noalign{\medskip}
 &    \multicolumn{2}{c}{SS}  &   \multicolumn{2}{c}{FR-HI (SFMS)}  & \multicolumn{2}{c}{FR-HI (below SFMS)} & \multicolumn{2}{c}{xCOLDGASS (SFMS)\tablefootmark{b}} \\
 & \multicolumn{2}{c}{(n = 46)} & \multicolumn{2}{c}{(n= 6)} & \multicolumn{2}{c}{(n= 12)} & \\
 & mean (err)  & median & mean  (err) & median & mean  (err) & median & mean  (err) & median \\
\noalign{\smallskip} \hline \noalign{\medskip} 
log(sSFR)$_{\rm zcorr}$  (yr$^{-1}$) &  -10.73 (0.04) & -10.81 & -10.57 (0.04)  &  -10.59 & -11.20 (0.06)  & -11.15 & -- & --  \\
log(\fmolzcorr )& -1.39 (0.02) & -1.38 & -1.27 (0.04)  & -1.28 & -1.87 (0.09) & -1.83 & -- & -- \\
 \taudep (yr) & 9.29 (0.03)  & 9.33  & 9.28 (0.04)  & 9.27 & 9.32 (0.06) & 9.30 & 8.98 (0.02) & 9.00 \\
\taudep (SFR$_{\rm C17}$)\tablefootmark{a}  (yr) & 9.32 (0.02)  & 9.33 & 9.27 (0.02) & 9.28 & 9.26 (0.15) & 9.26 & 8.90 (0.02) & 8.87 \\
$\mu_*$ \msun\ kpc$^{-2}$ &8.69 (0.04) & 8.68 & 8.93(0.15) & 8.85 & 9.06(0.06) & 9.09 & 8.42 (0.05) & 8.45 \\ 
log(\fhi)& --  & -- & -0.78 (0.09)  & -0.86 & -1.00 (0.11) & -0.99 & -- & -- \\
log(\mmol/\mhi)  & --  & -- & -0.40 (0.11)  & -0.27 & -0.82 (0.11) & -0.79 & -0.48 (0.05) & -0.48 \\
\noalign{\smallskip} \hline \noalign{\medskip}
\end{tabular}
\tablefoot{
\tablefoottext{a} {Gas depletion time calculated with the prescription of \citet{cluver17}, shifted by +0.2 dex.}
\tablefoottext{b}{Values are only given for quantities that have a weak dependence on \mstar.}
}
\end{table*}

\subsection{Redshift-dependence of sSFR and \mmol/\mstar}
\label{sec:z-dependence}

\begin{figure}
   \centering
\includegraphics[width=8.cm,trim=0.cm 0.cm 0cm 0cm,clip]{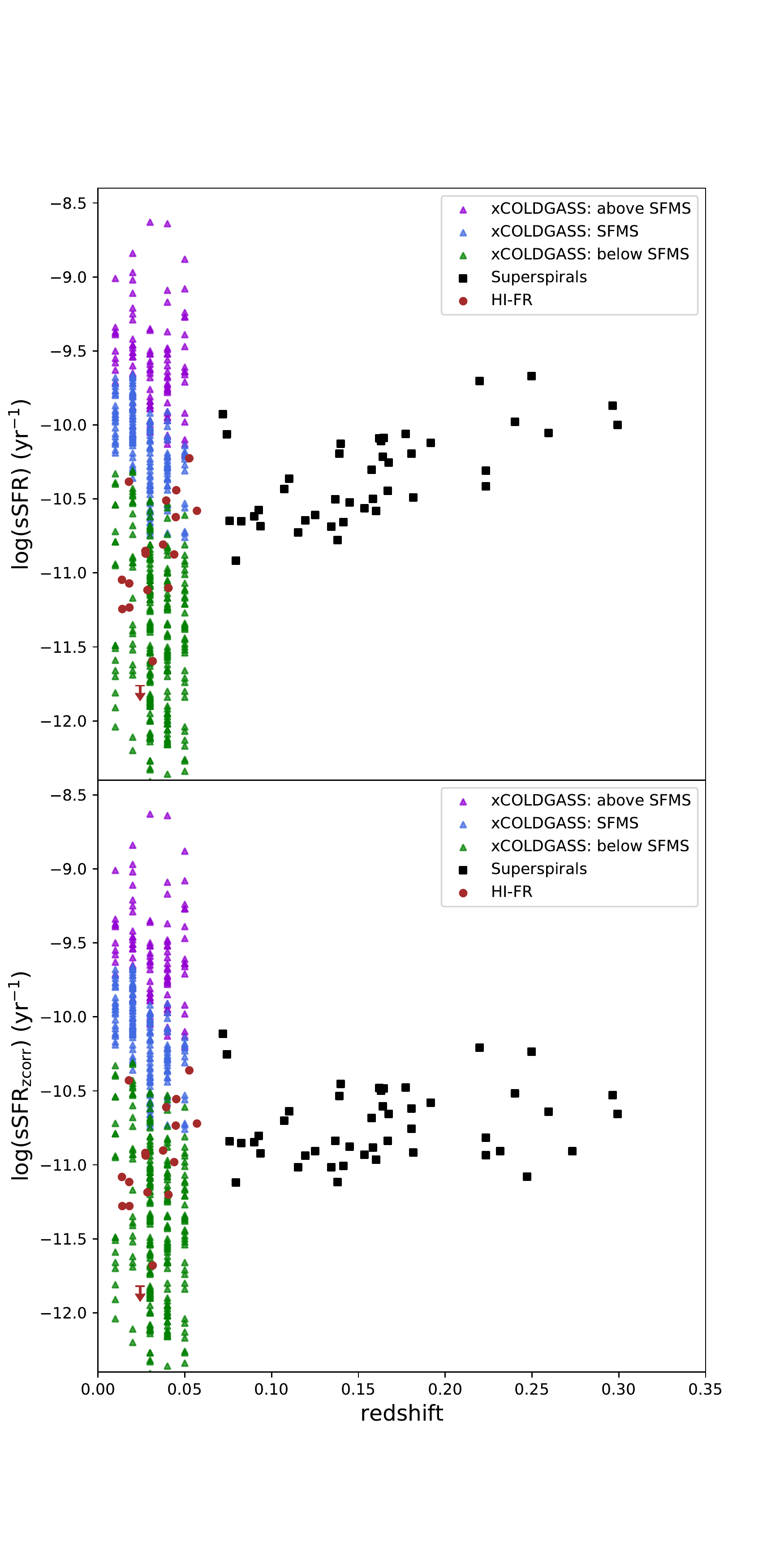}
      \caption{Redshift dependence of the sSFR. {\it Upper panel:} Specific SFR  as a function of redshift for the SS, HI-FR and the xCOLDGASS sample.
      {\it Lower panel:} The specific SFR for the SS has been adjusted to $z=0$ following the $z$-dependence of the SFMS by \citet{speagle14}.  sSFR for xCOLDGASS and HI-FR are the same as in the upper panel. 
 }
                \label{fig:ssfr_vs_redshift}
   \end{figure}

\begin{figure}
   \centering
\includegraphics[width=8.cm,trim=0.cm 0.cm 0cm 0cm,clip]{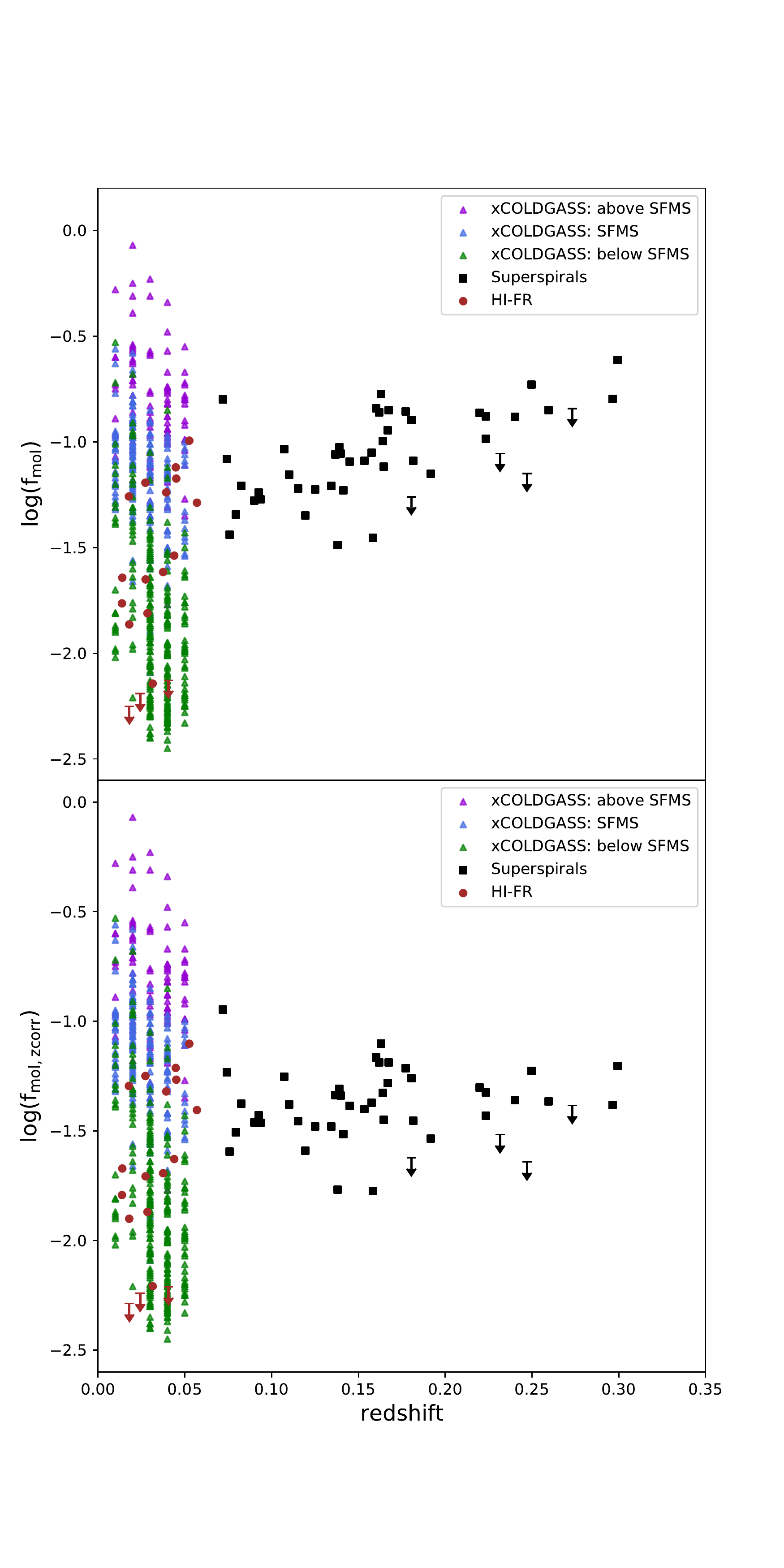}
      \caption{Redshift dependence of the molecular gas fraction. {\it Upper panel:} The molecular gas fraction \fmol\ (= \mmol/\mstar) as a function of redshift for the SS, HI-FR galaxies and the xCOLDGASS sample (only detections in CO for xCOLDGASS in order not to overload the figure).
      {\it Lower panel:} The molecular gas fraction \fmol\  for the SSs has been adjusted to $z=0$ following the  nonlinear $z$-dependence found by \citet[][their Tab. 3]{tacconi20}. \fmol\ for xCOLDGASS and HI-FR are the same as in the upper panel. }
                \label{fig:smmol_vs_redshift}
   \end{figure}

Both the sSFR and the molecular gas mass fraction (\mmol/\mstar) are known to have a strong dependence on the redshift. This can  be clearly seen  for galaxies in the SS sample   (Figs.~\ref{fig:ssfr_vs_redshift} and Fig.~\ref{fig:smmol_vs_redshift}, upper panels). We need to correct for this redshift dependence in order to compare the SS sample to the $z\approx 0$ (HI-FR and xCOLDGASS).

\citet{speagle14} studied the SFMS  of galaxies at different redshifts and derived a prescription for the SFMS as a function of $z$ (sSFR$_{\rm MS,S14}(M_*, z)$). We adopt this prescription \citep[as  cited in][their eq. 1]{tacconi18}) to derive a sSFR reprojected to $z=0$ for the SS, by applying
sSFR$_{\rm zcorr}$ = sSFR $\cdot$   (sSFR$_{\rm SFMS,S14}(M_*, z=0)$/sSFR$_{\rm SFMS,S14}(M_*, z)$).
In a similar way, we reproject the molecular gas fraction of the SS galaxies to  $z=0$, by applying the  nonlinear relation of \citet[][from their Tab. 3, see also their Fig. 5]{tacconi20} which is, due to the curved shape of the $z$-dependence, more appropriate for low $z$ galaxies than the general linear relation \fmol  $ \propto (1+z)^{-2.5}$ \citep{tacconi18}. We thus correct the molecular gas fraction as 
\fmolzcorr = \fmol + $3.62 \cdot (0.66^2 -( \log(1+z)+0.66)^2)$.
 In the following analysis, we always use the redshift-corrected values of the sSFR and the  molecular gas mass fraction, except for the calculation of the depletion time which is based on observed values of SFR and \mmol.
 
 In Figs.~\ref{fig:ssfr_vs_redshift} and Fig.~\ref{fig:smmol_vs_redshift} (lower panels) we show the corresponding relations for the z-corrected quantities. The applied correction eliminate the trends of both sSFR and \fmol\ with $z$ to a large extent, although a weak relation with redshift is still visible (a linear least-square fit yields sSFR$_{\rm zcorr} \propto (1+z)^{1.4}$ and \fmolzcorr $\propto (1+z)^{0.84}$).

\begin{figure}
   \centering
\includegraphics[width=8.cm,trim=0.cm 0.cm 0cm 0cm,clip]{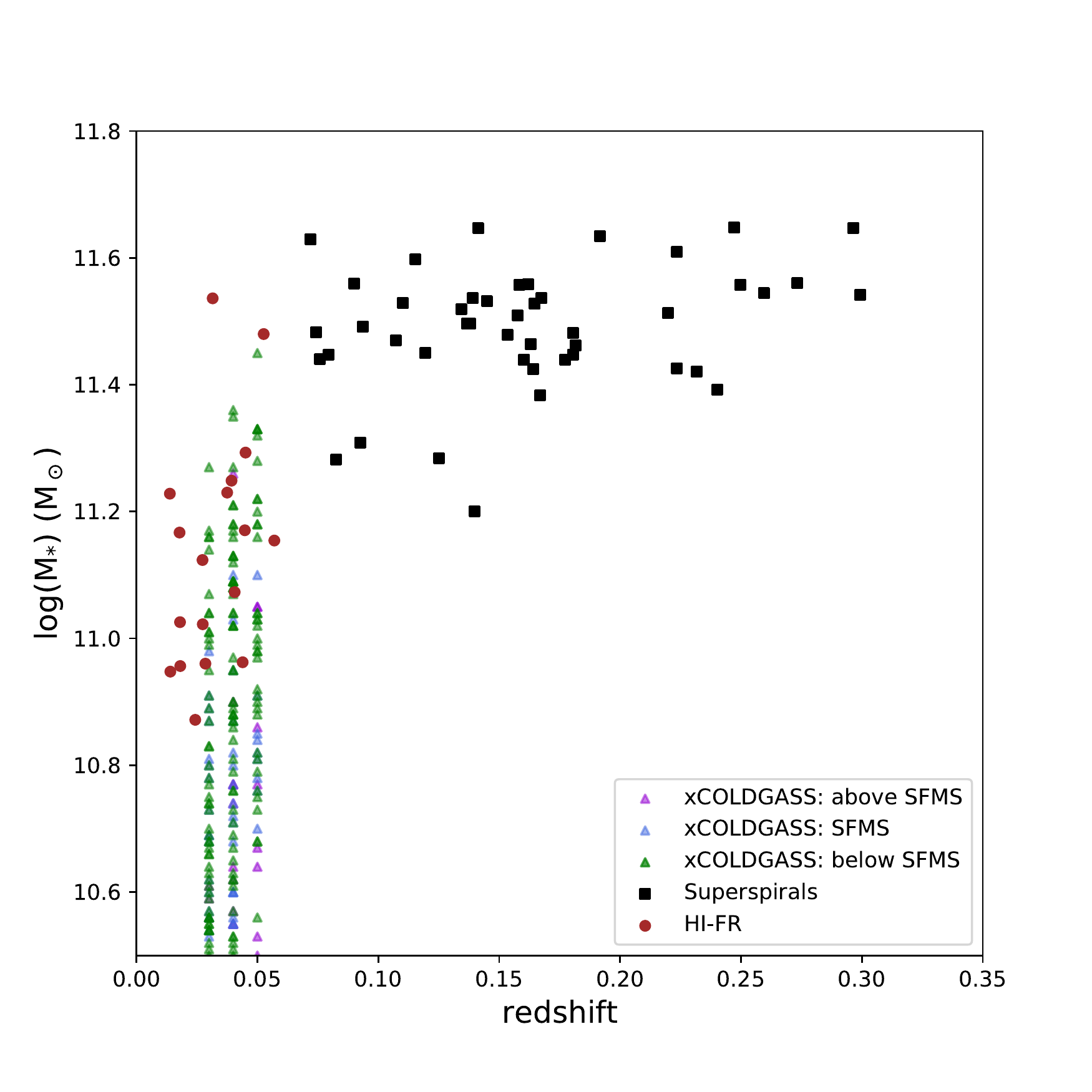}
      \caption{\mstar\  as a function of redshift for the super spirals, HI-FR galaxies and the xCOLDGASS sample.
      Only a weak trend of \mstar\ with redshift is visible (\mstar $\propto (1+z)^{1.16}$). 
 }
                \label{fig:mstar_vs_redshift}
   \end{figure}

Fig.~\ref{fig:mstar_vs_redshift} shows the relation of the stellar mass with redshift. There is only a weak  trend with redshfit (\mstar $\propto (1+z)^{1.16}$), showing that in our sample there is a weak tendency for the more massive galaxies to be more distant. 

\subsection{Star-forming main sequence}

\begin{figure}
   \centering
\includegraphics[width=8.cm,trim=0.cm 0.cm 0cm 0cm,clip]{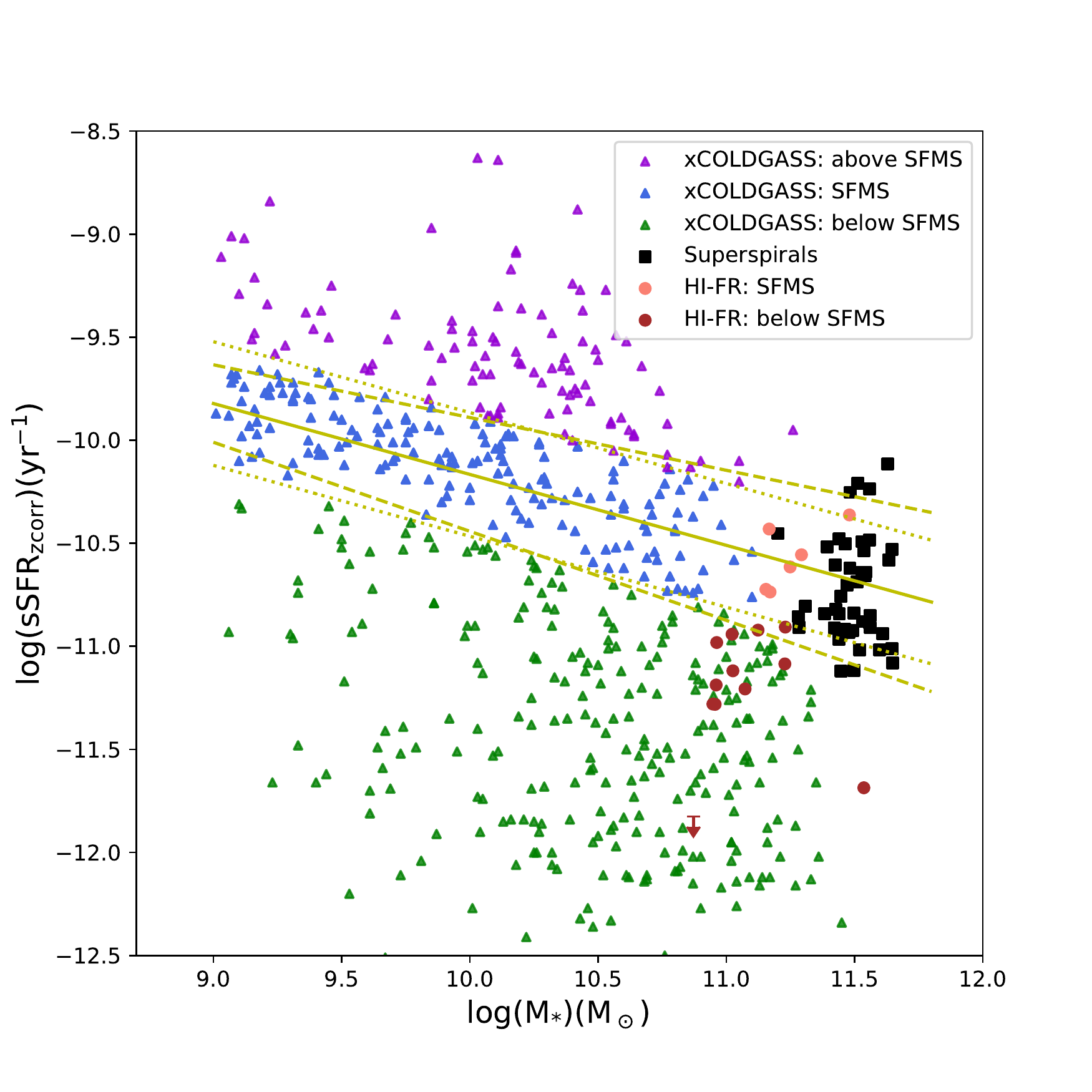}
      \caption{sSFR  as a function of stellar mass  for the SS, HI-FR and the xCOLDGASS sample.
      The sSFR of the SS galaxies is adjusted  to $z = 0$ according 
      \citet{speagle14}  as explained in Sect. ~\ref{sec:z-dependence}. The full yellow line denotes the SFMS from \citet{janowiecki20}, derived for the xCOLDGASS sample, and the dashed yellow line shows its 1$\sigma$ scatter. The dotted yellow line show a distance of 0.3~dex from the SFMS which is adopted, following  \citet{janowiecki20}, to define SFMS galaxies.}
                \label{fig:ssfr-vs-mstar-zcorr}
   \end{figure}

Figure~\ref{fig:ssfr-vs-mstar-zcorr} shows the relation between the SFR and stellar mass. The properties of a galaxy are determined to a large extent from its position in this plane, and in particular whether the galaxy lies on, above or below the SFMS. 
We include the SFMS together with its width, defined as the 1$\sigma$ scatter, derived for the xCOLDGASS sample by \citet[their equations 1 and 2]{janowiecki20}.  This relation is practically identical to that derived by \citet{leroy19} for a sample of ~15000 nearby galaxies. Following \citet{janowiecki20}, we  split the sample into ``starburst"  objects ($> $ 0.3 dex  above the SFMS), SFMS objects (within $\pm$ 0.3 dex  of the MS) and quiescent objects (more than  0.3 dex below the SFMS). \citet{janowiecki20} distinguished within this quiescent subsample furthermore between transitioning objects (between 0.3 dex and  1.55 dex below the MS) and red-sequences objects (more than  1.55 dex below the MS). We do not include the latter distinction, because none of our SS+HI-FR objects lies in the quiescent regime.

Figure~\ref{fig:ssfr-vs-mstar-zcorr} shows that the SSs follow very well the extrapolation of the SFMS derived by Janowieski, with practically all objects lying within the 1~$\sigma$ width. This means that in spite of their large mass, SSs are forming stars at a rate which puts them on the same SF relation as lower-mass spirals.
On the other hand, the sample of fast HI rotators contains galaxies which lie on the SFMS and galaxies which are well below, in the range of transitioning galaxies. 

In the following, when appropriate,  we distinguish between star-forming and transitioning HI-FR galaxies as those that are on the SFMS (within$\pm$ 0.3 dex) or more than 0.3 dex  below the SFMS. With respect to the SS galaxies, we consider them all as belonging to the MS. In addition, we define the distance to the SFMS as $\rm \triangle(SFMS) = log(sSFR) (yr^{-1}) - \log(sSFR_{\rm MS, Jan20}) (yr^{-1})$, where sSFR$_{\rm MS, Jan20}$  is the SFMS from \citet{janowiecki20}. In Tab.~\ref{tab:mean-values} the mean and median values, as well as the standard deviation for the sSFR of the SS and HI-FR samples are given.

\subsection{Molecular gas mass fraction}
\label{sec:mol_gas_fraction}

\begin{figure}
   \centering
\includegraphics[width=8.cm,trim=0.cm 0.cm 0cm 0cm,clip]{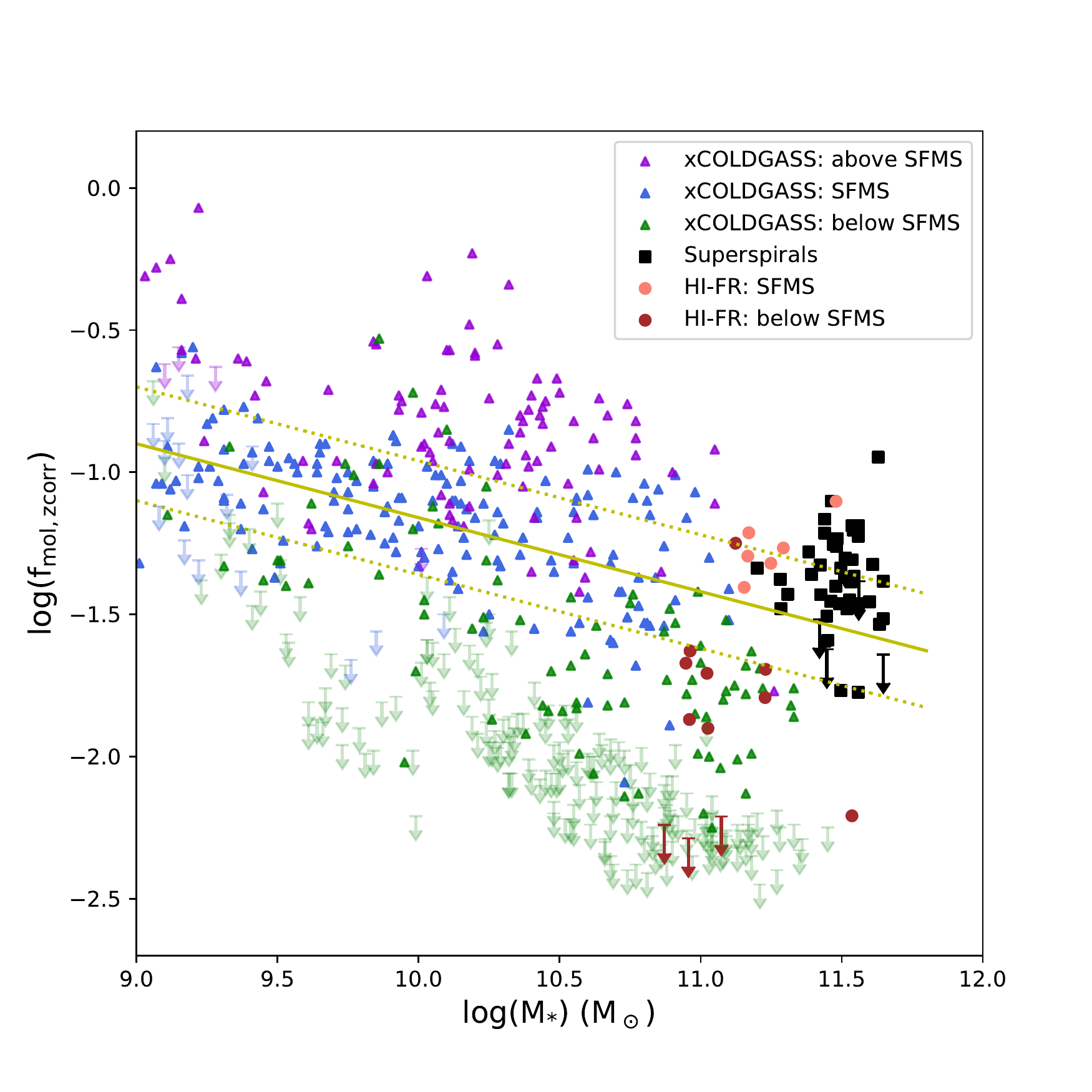}
      \caption{Molecular gas mass fraction (\fmolzcorr),  as a function of stellar mass  for the super spirals, HI-FR and the xCOLDGASS sample. The molecular gas mass  of the SS galaxies is reprojected to $z=0$ following \citet{tacconi18} as explained in Sect. ~\ref{sec:z-dependence}.
       }
\label{fig:mmol_zcorr_over_mstar_vs_mstar.pdf}
   \end{figure}

Figure~\ref{fig:mmol_zcorr_over_mstar_vs_mstar.pdf} shows the scaling relation  between the molecular gas fraction \mmol/\mstar\ and the stellar mass. Included is, as a yellow line, the scaling relation found by \citet{janowiecki20} for the xCOLDGASS sample, (which they called the H$_2$ main sequence, H$_2$MS) and  its 0.2~dex widths which was derived as the standard deviation of the SFMS galaxies in this relation.

The molecular gas fractions of SS galaxies lie  mostly above the scaling relation found for lower-mass SFMS galaxies (the mean value of \fmolzcorr\ of SS, see Tab.~\ref{tab:mean-values} is roughly 0.2 dex above the value of the H$_2$MS at the stellar mass of SS). This means that SS galaxies have a large reservoir of molecular gas, higher than what is expected for  SFMS galaxies of their mass, if one extrapolated from lower masses.
FR-HI galaxies that lie on the SFMS, also have a relatively high molecular gas mass fractions, lying in the upper half of the H$_2$MS, whereas FR-HI galaxies below the SFMS also have molecular gas fractions below the   H$_2$MS . 

\begin{figure}
   \centering
\includegraphics[width=8.cm,trim=0.cm 0.cm 0cm 0cm,clip]{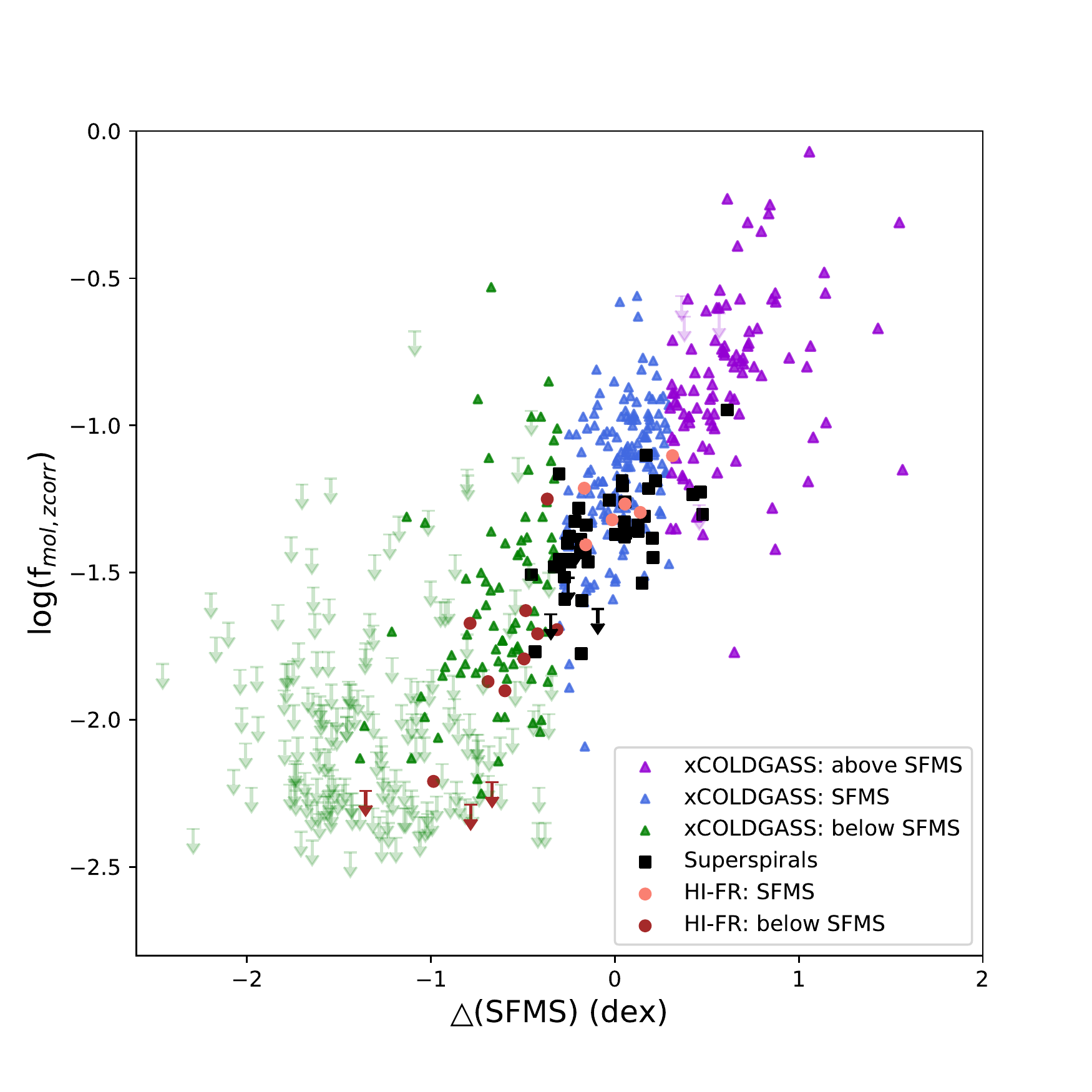}
      \caption{Molecular gas mass fraction (\fmolzcorr),  as a function of distance to the SFMS  for the  SS,  HI-FR and the xCOLDGASS sample. The molecular gas mass  of the SS galaxies is reprojected to $z=0$ following \citet{tacconi18} as explained in Sect. ~\ref{sec:z-dependence}. 
       }                     \label{fig:mmol_zcorr_over_mstar_vs_distSFMV.pdf}
   \end{figure}

Fig.~\ref{fig:mmol_zcorr_over_mstar_vs_distSFMV.pdf} displays the molecular gas mass fraction as a function of the distance to the SFMS. Here, SS and  HI-FR galaxies follow the same trend as galaxies from the comparison sample. This means that SS+HI-FR galaxies have the molecular gas fraction that corresponds to their SF activity. Taken together, these two relations suggest that the decrease of \fmolzcorr\ with stellar mass for star-forming disk galaxies is less than what is suggested from the extrapolation of the  H$_2$MS relation  from lower-mass galaxies. In other words, \fmolzcorr\ for the highest stellar masses seems to be biased low when only considering the xCOLDGASS data. If we include the SS+HI-FR galaxies together with the xCOLDGASS sample  and again fit the relation (considering only galaxies on the  SFMS) we derive \fmolzcorr = (-0.18 $\pm$ 0.02) $\times$  (log(\mstar)-9) - (0.95$\pm$ 0.4), slightly flatter than the relation in \citet{janowiecki20} (\fmolzcorr = (-0.26 $\pm$ 0.03) $\times$  (log(\mstar)-9) - (0.90$\pm$ 0.18)).

\subsection{Molecular gas depletion time}
\label{sec:tau_mol_gas}

\begin{figure}
   \centering
\includegraphics[width=8.cm,trim=0.cm 0.cm 0cm 0cm,clip]{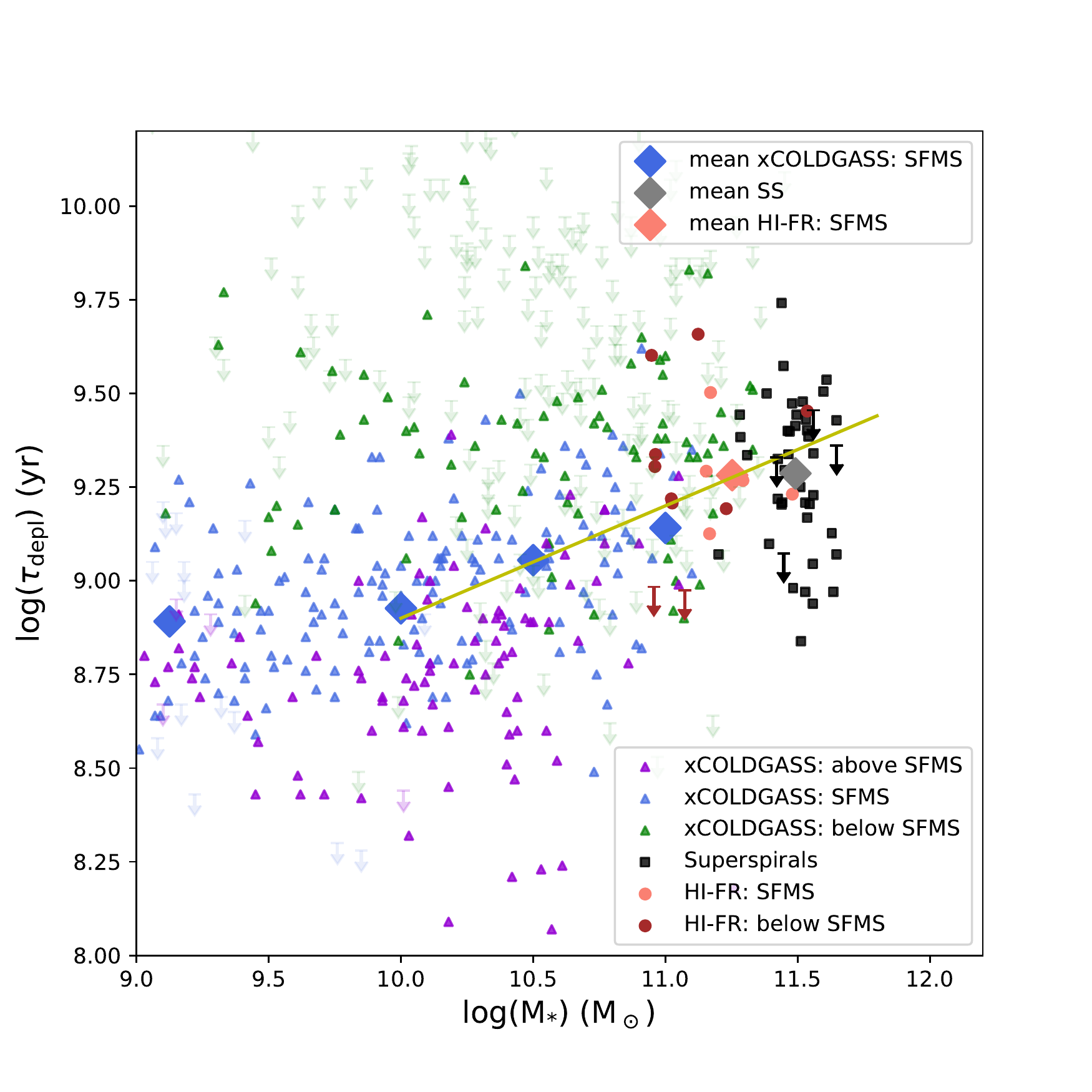}
      \caption{Gas depletion time \taudep = \mmol/SFR  as a function of stellar mass  for the SS, HI-FR and the xCOLDGASS sample.
       The yellow line gives, for illustration,  the relation from \citet[][their Fig. 8, log(\taudep) = 0.203 log(\mstar) +6.79]{saintonge22} .
}
                \label{fig:taudep-vs-mstar}
   \end{figure}

Figure~\ref{fig:taudep-vs-mstar} shows the depletion time, \taudep =  \mmol/SFR, as a function of stellar mass. Here, neither the SFR nor the molecular gas mass are corrected for redshift.

\begin{figure}
   \centering
\includegraphics[width=8.cm,trim=0.cm 0.cm 0cm 0cm,clip]{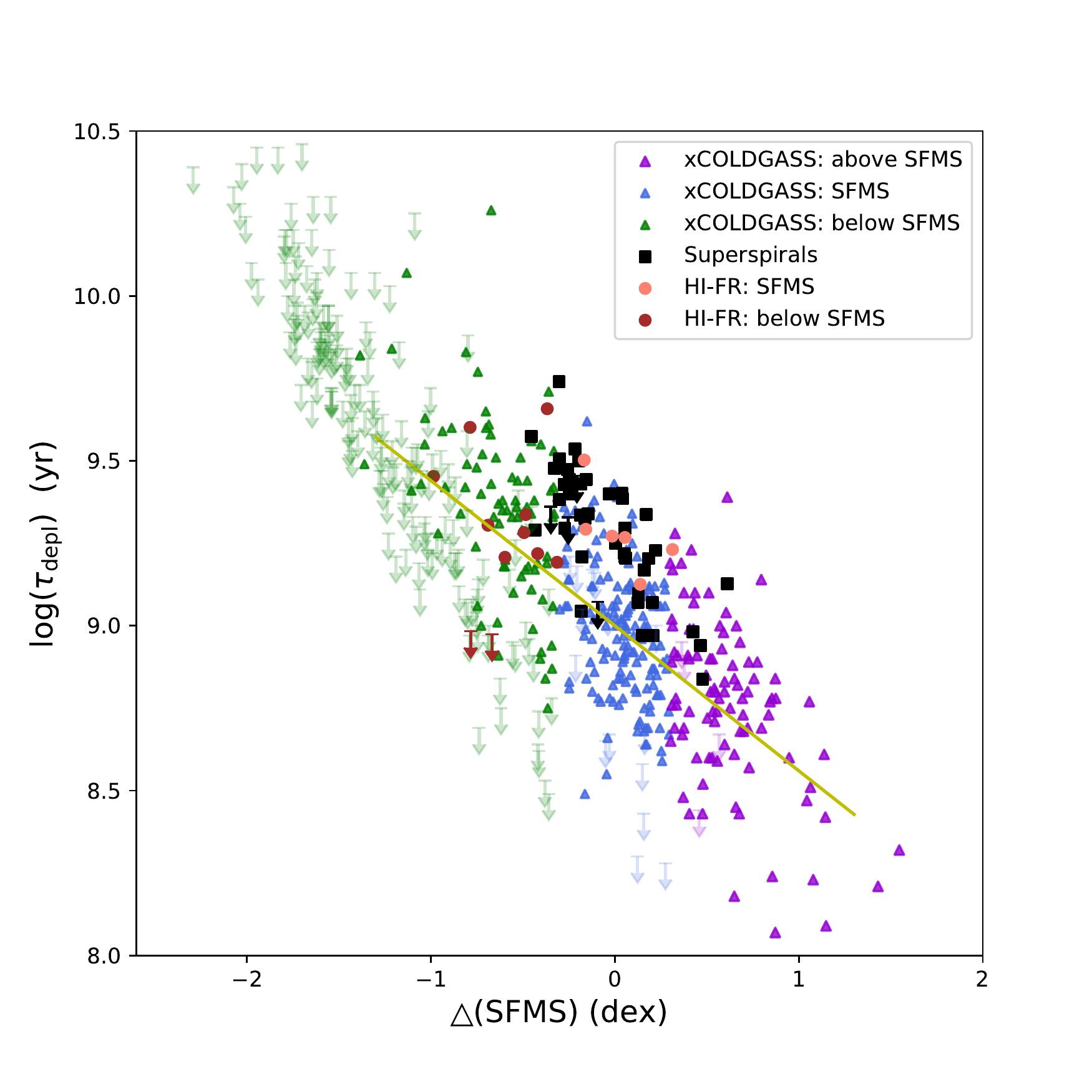}
      \caption{The gas depletion time \taudep = \mmol/SFR  as a function of distance to the SFMS  for the SS, HI-FR  and the COLDGASS sample.
              The yellow line gives, for illustration, the relation of \taudep  $\propto \triangle$(SFMS)$^{-0.44}$, \citep{tacconi18} . }
                \label{fig:taudep-vs-distMS}
   \end{figure}

The SS and HI-FR galaxies  have longer  gas depletion times (mean value  log(\taudep) $ \sim 9.3$ yr , see Tab.~\ref{tab:mean-values}) than the comparison sample (mean log(\taudep) of xCOLDGASS for galaxies on the SFMS is  9.0 yr ).
In general, only a weak trend of \taudep\ with \mstar\ has been found in the literature, both for nearby galaxies (Saintonge et al. 2017) and at higher redshifts, up to $z = 4$  \citep{genzel15, tacconi18}. We include in Fig.~\ref{fig:taudep-vs-mstar} as a yellow line a relation of log(\taudep) $\propto$ \mstar$^{0.203}$ from \citep[][their Fig. 8]{saintonge22},
which  fits the trend in all the  samples reasonably well. The long depletion time of the SS+HI-FR sample might thus be  the continuation of a trend with stellar mass that is also seen in the xCOLDGASS galaxies above a mass  of about log(\mstar) $\sim$  10.5. 
In Fig. ~\ref{fig:taudep-vs-distMS} we show \taudep\  as a function of the distance to the SFMS. In studies of close-by galaxies \citep{saintonge17, janowiecki20} and high-z galaxies \citep{tacconi18} the depletion time has been found to depend on the distance to the SFMS, with quiescent galaxies having considerably higher values for \taudep. We include in the figure the relation found by Tacconi et al. 2018 (\taudep  $\propto \triangle$(SFMS)$^{-0.44}$). The relation fits very well the xCOLDGASS galaxies, but the SS+HI-FR galaxies lie considerably above it, showing that they have a rather long depletion time for their position on the SFMS.

\begin{figure}
   \centering
\includegraphics[width=8.cm,trim=0.cm 0.cm 0cm 0cm,clip]{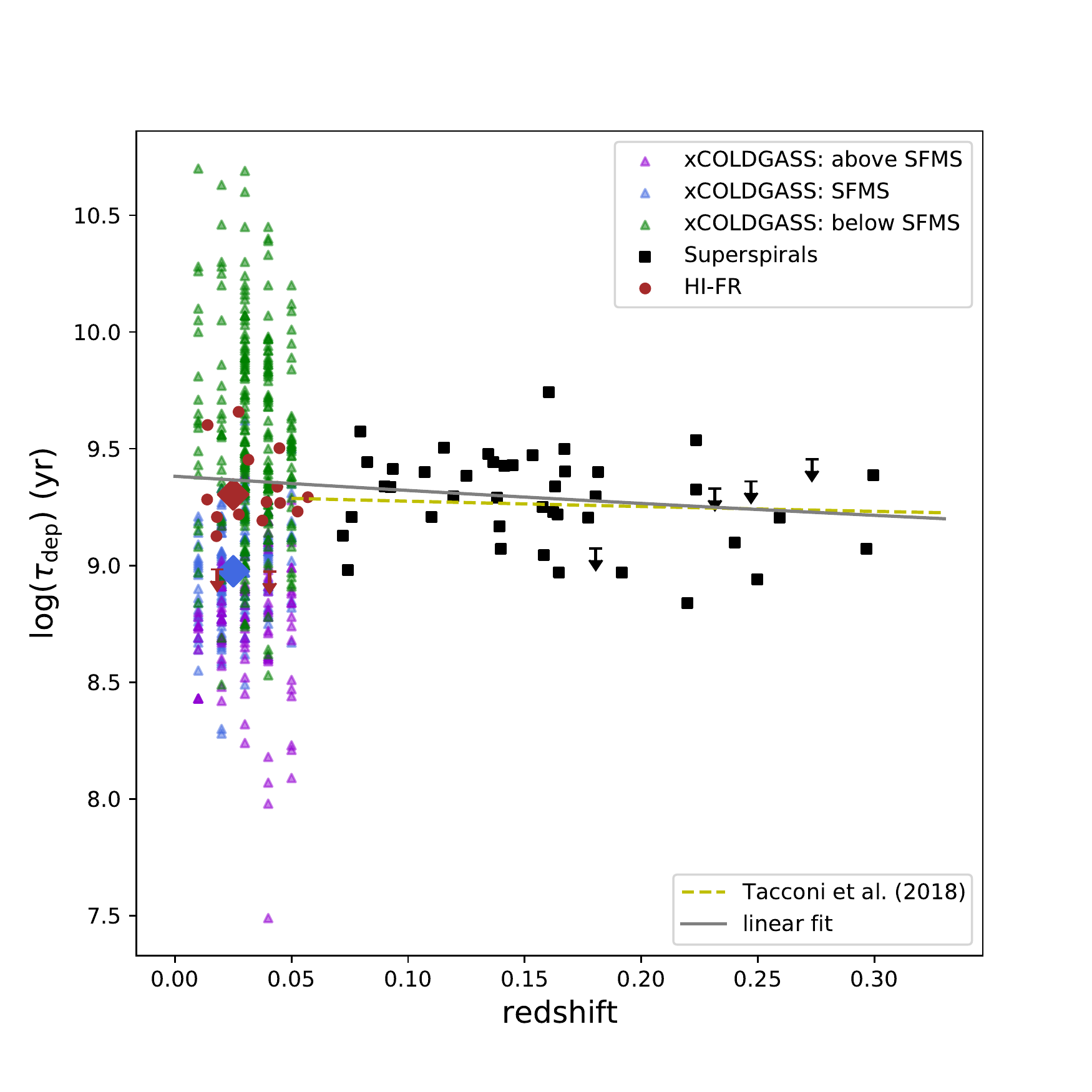}
      \caption{The gas depletion time \taudep = \mmol/SFR  as a function of redshift  for the SS, HI-FR  and the COLDGASS sample. The yellow line gives, for illustration, the relation of \taudep  $\propto (1-z)^{-0.6}$ from  \citet{tacconi18}, and the gray line is the fit to our data (\taudep  $\propto (1-z)^{-0.02}$).
                    The blue diamond is the mean value of the xCOLDGASS galaxies that lie on the SFMS and the brown diamond is the mean value of the HI-FR sample. The errors of both means are smaller than the symbol size.}
                \label{fig:taudep-vs-redshfit}
   \end{figure}

Finally, Fig.~\ref{fig:taudep-vs-redshfit} shows \taudep\ as a function of redshift. As expected, only a weak trend is visible.
\citet{genzel15} and \citet{tacconi18} found that  \taudep\ decreases with redshift \citep[\taudep $\propto (1+z)^{-0.6}$,][]{tacconi18} which we include for illustration in the plot (yellow line), together with the best-fit  relation to the data (\taudep $\propto (1+z)^{-0.02}$).  Within the redshift range covered by the SS sample, the effect of this weak relation is small with a difference in  \taudep\ of $< 0.1$ dex.
There is a clear offset  between the SS+HI-FR sample and the xCOLDGASS sample even for low redshifts.  We conclude, therefore, that the difference in \taudep\ between xCOLDGASS and SS+HI-FR is not due to redshift, but rather due to intrinsic properties of the galaxies. We further discuss this in Sect~\ref{sec:discussion}.

\subsection{Atomic and molecular gas in HI fast rotators}

\begin{figure}
   \centering
\includegraphics[width=8.cm,trim=0.cm 0.cm 0cm 0cm,clip]{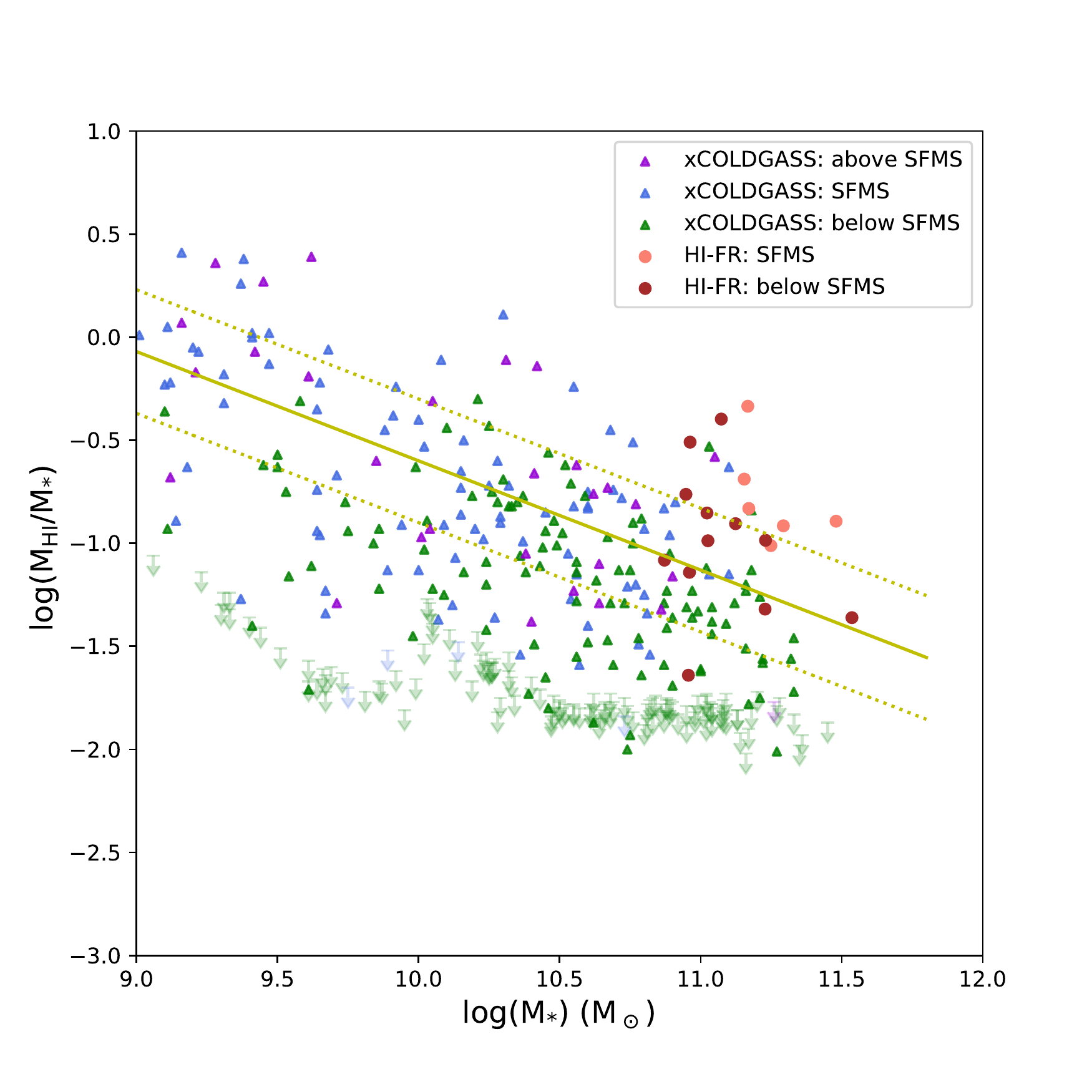}
\includegraphics[width=8.cm,trim=0.cm 0.cm 0cm 0cm,clip]{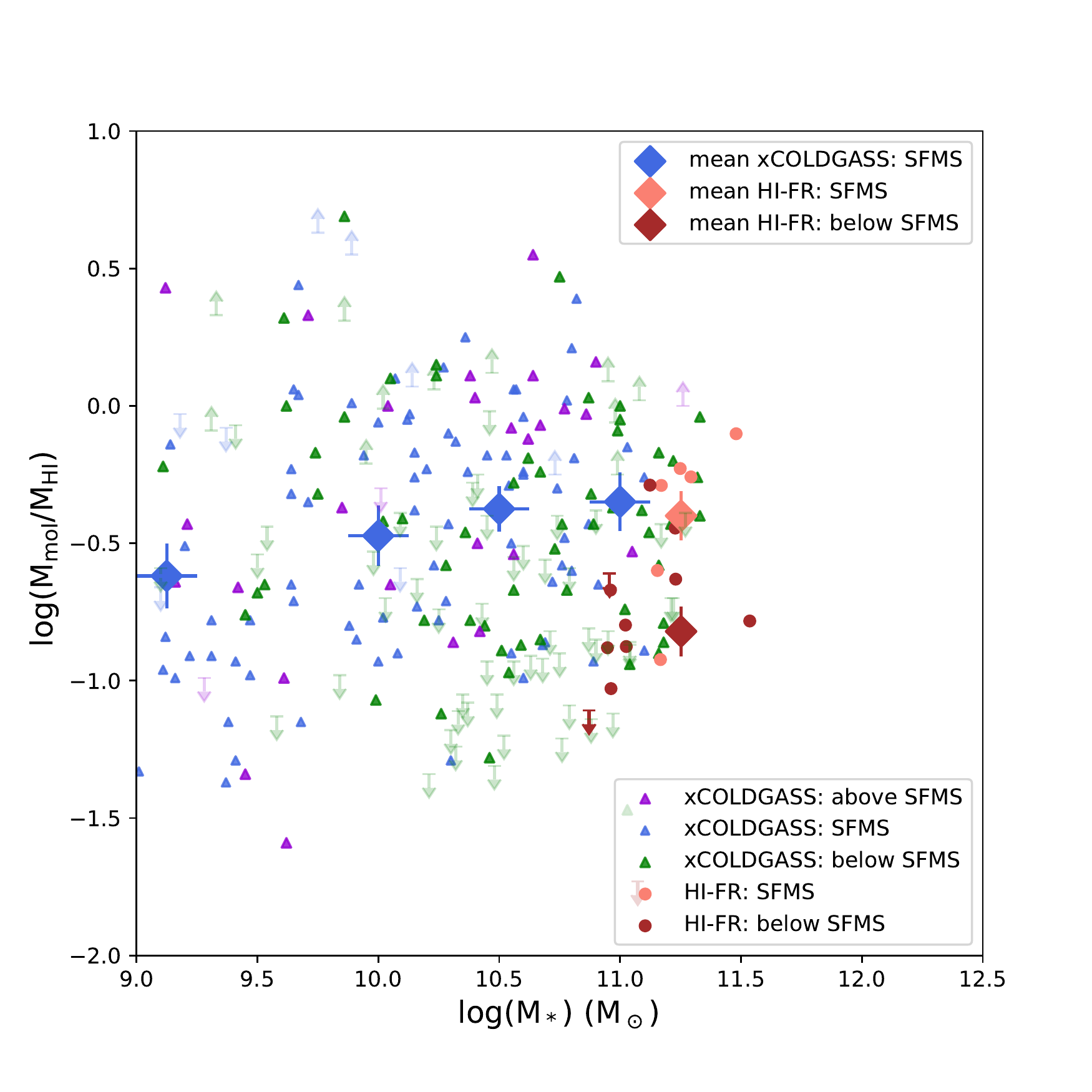}
      \caption{The atomic gas mass fraction ({\it upper panel})  and 
      the molecular-to-atomic mass ratio ({\it lower panel})  as a function of stellar mass  for the  HI-FR  and the xCOLDGASS sample.
       The mean values of the HI-FR and of the different xCOLDGASS samples  are indicated as diamonds, with the length of the vertical bar indicating the error of the mean, and the length of the horizontal bar the width of the chosen mass interval.  
Only galaxies with detections in HI are considered for the mean value of \mhi/\mmol. 
The yellow line in the upper panel indicates the scaling relation found by Janowieski et al. (2020) together with the 0.3 dex width adopted by these authors to characterize the HI main sequence.
}
                \label{fig:mhi-vs-mstar}
   \end{figure}

Figure~\ref{fig:mhi-vs-mstar} shows the properties of the atomic gas in the HI-FR galaxies. Due to their higher redshift, no HI data exists for SS galaxies. In the upper panel we show the atomic gas fraction, \fhi = \mhi/\mstar, together with the scaling relation found by \citet{janowiecki20} for xCOLDGASS galaxies. The atomic gas fraction of HI-FR objects lies clearly above the relation found for lower-mass galaxies, both for SFMS and for below-SFMS objects (see Tab.~\ref{tab:mean-values}), which can be explained by the selection of the sample. Detected galaxies from the ALFALFA survey tend to be HI bright, and in addition we selected galaxies with very broad HI spectra.

The molecular-to-atomic gas fraction for galaxies on the SFMS, on the other hand, is similar to xCOLDGASS. Together with the relations found for the molecular gas in Sect.~\ref{sec:mol_gas_fraction}, this shows that actively star-forming HI-FR objects are in general gas-rich, both for atomic and  for molecular gas. The transformation from atomic to molecular gas seems to be similar as in lower-mass galaxies.

\section{Discussion}
\label{sec:discussion}

The major  findings of our analysis are: (i) In agreement with the earlier results of \citet{ogle19a},  super spiral galaxies have a SFR that puts them on the SFMS established from lower-mass galaxies. 
(ii) They contain large amounts of  molecular gas. Adopting a lower than Galactic  conversion factor \alphaco = 3~\msun/(K \, km s$^{-1}$ pc$^{-2}$  (expected for these high-mass, high-metallicity objects), we find that the molecular gas mass is slightly above  what is expected from their mass (extrapolated from lower-mass scaling relations) and roughly what is expected from their distance to the SFMS. (iii) Their molecular gas depletion time is following the mass trend found from lower-mass galaxies, but  lies above what is expected from their distance to the SFMS. In the following we are going to discuss how robust these results are and what we can learn from super spirals with respect to the processes driving galaxy evolution.

\subsection{Uncertainties in our analysis}
\label{sec:discussion-uncertainties}

\subsubsection{Uncertainties in the molecular gas mass} 
\label{sec:discussion-uncertainties-alphaco}

The molecular gas mass has two uncertainties: The value of the chosen \alphaco\  and the redshift correction that we applied. We note that the redshift correction does not affect the molecular gas depletion time because we use the uncorrected molecular gas mass and SFR to calculate it.

We determined \alphaco\ based on its expected metallicity dependence. Both the dependence of \alphaco\ on metallicity as well as the derivation of metallicty based on the mass-metallicity relation are somewhat uncertain. We based our choice of \alphaco\ on the comparison of different prescriptions, derived from different methods. \citet{accurso17} derived their prescription from a combined analysis of [\ion{C}{II}] and CO(1-0)  data together with radiative transfer modeling. The results from \citet{wolfire10}, on which the prescription of \citet{bolatto13} is based, are from theoretical models of molecular clouds. The difference in \alphaco\ between the different studies is only $\sim 20\%$. The influence of the exact value of the metallicity in the high-mass and high-metallicity range is also relatively small because the relation between \alphaco\ and metallicity converges to a fixed value \citep[2.9~\msun/(K \, km s$^{-1}$ pc$^{-2}$) in the prescription of][]{bolatto13} for high metallicities. Therefore, the uncertainties in \alphaco\ due to uncertainties in the metallicity of our objects are not very large, most likely less than 20-30\%.

Our choice of \alphaco\ depends on the assumption that molecular clouds are on average the same in super spirals as in lower-mass galaxies. More detailed studies are necessary to confirm this assumption, in particular higher resolution observations. There are two parameters that are  indicators for possible differences: (i) The distance to the SFMS \citep{accurso17} which determines the  intensity of the ultraviolet radiation field that can photo-dissociate CO in the outer layer of molecular clouds. However, this effect is relatively small close to the SFMS. We have tested the prescription of \citet{accurso17} by taking into account the distance to the SFMS and found only negligible  differences for \mmol\ in our sample. (ii) The total surface density (stars + gas) which is much higher in starburst galaxies, leading to an inter-cloud medium as dense as molecular clouds. This is not the case in our SS+HI-FR sample which are not in a starburst phase and have surface densities in the same range as xCOLDGASS galaxies. 

Finally, the molecular gas mass fraction is known to increase with redshift. We have corrected the SS galaxies  for this effect, following relations found for other samples, spanning a larger redshift range than SS \citep{genzel15, tacconi18}. The resulting  $z$-corrected values for \fmolzcorr\ (Fig.~\ref{fig:smmol_vs_redshift})  are left with only a very small dependence on redshift. We therefore consider the redshift correction for the molecular gas mass fraction (as well as for the sSFR) adequate.

\subsubsection{Uncertainties in SFR and  \mstar}

We derive the SFR with the same prescriptions for both SS and comparison sample in order to avoid biases. In Appendix C we contrast different prescriptions for the calculation of the SFR and conclude that typical uncertainties can be up to  0.1 - 0.2  dex. This small difference does not alter our conclusion that SS galaxies follow the SFMS.
In addtion, we can test whether the difference in \taudep\ between SS and xCOLDGASS galaxies would have  been smaller if we had used a different prescription for the SFR.
In order to do this, we calculated the SFR for all samples (SS, HI-FR and xCOLDGASS) following the method of \citet{cluver17}, offset by 0.2 dex (eq.~\ref{eq:SFR-W3_C17}), which is the method of those tested in Appendix C with the largest difference. Fig.~\ref{fig:taudep_vs_mstar-SFRcluver17} shows the result. We still find that \taudep\ of the SS sample is considerably higher than that of  xCOLDGASS (see Tab.~\ref{tab:mean-values}) and that there is a trend of increasing \taudep\ with stellar mass. Thus, we consider this a robust result.

\begin{figure}
\centering
\includegraphics[width=8.cm,trim=0.cm 0.cm 0cm 0cm,clip]{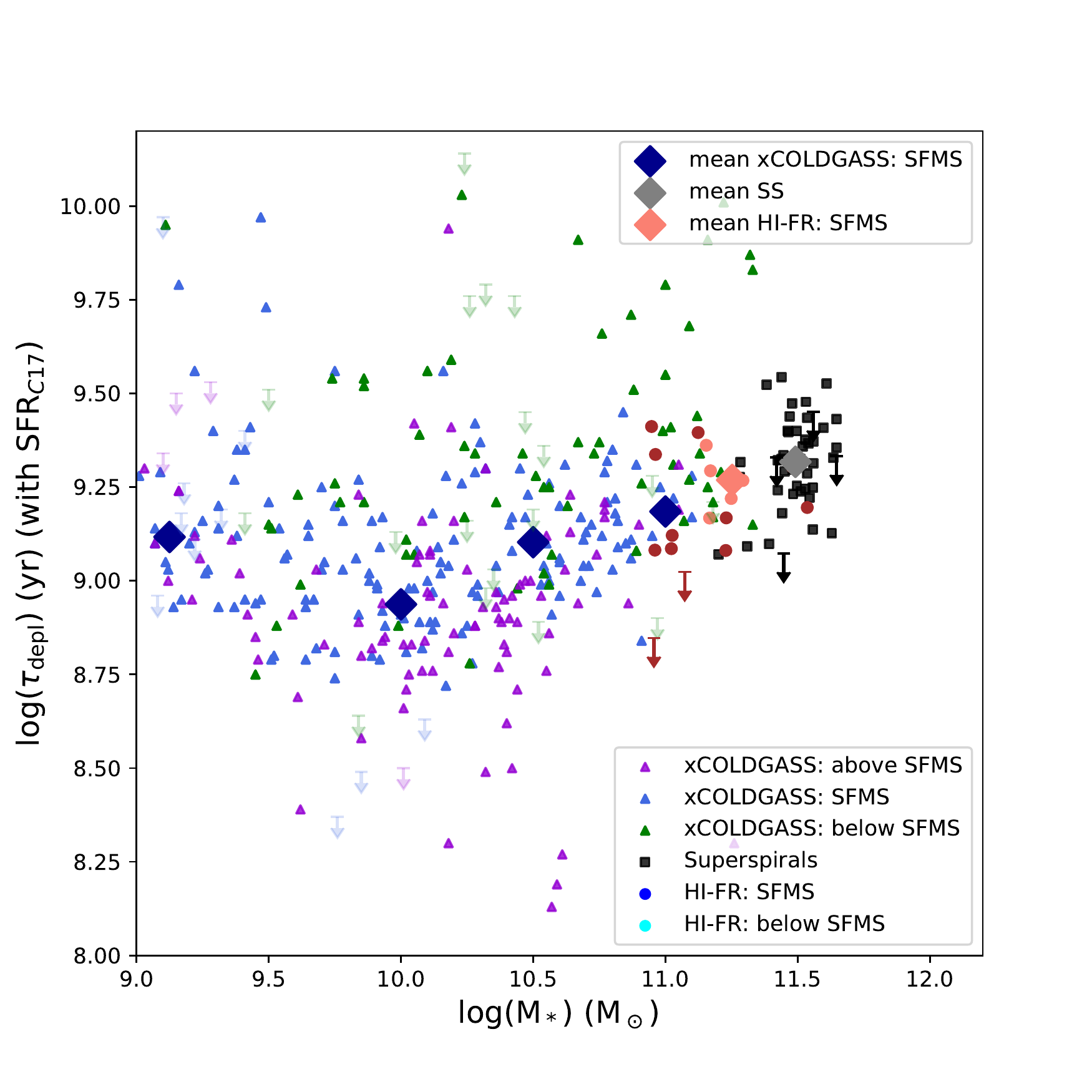}
\caption{The gas depletion time \taudep = \mmol/SFR  as a function of stellar mass  for the super spirals, fast HI rotators and the COLDGASS sample. The SFR has been calculated with the prescription of  Cluver et al. (2017), offset by 0.2 dex  (eq.~\ref{eq:SFR-W3_C17}).
}
\label{fig:taudep_vs_mstar-SFRcluver17}
\end{figure}

{In Appendix D we compare several methods to calculate \mstar\ and conclude that a constant  mass-to-light  ratio of $\Upsilon_{\ast}^{3.4} = 0.5$  is the best choice for the SS+HI-FR sample, whereas for xCOLDGASS with its wide range of properties, we used, also for consistency with previous studies, the MPA/JHU value provided in \citet{saintonge17}. In appendix D we concluded that the expected uncertainty is $\sim 0.2$ dex.} This is not expected to affect the conclusion of this paper in a considerable way.

\subsection{Super spirals and galaxy evolution}

\begin{figure}
\centering
\includegraphics[width=8.cm,trim=0.cm 0.cm 0cm 0cm,clip]{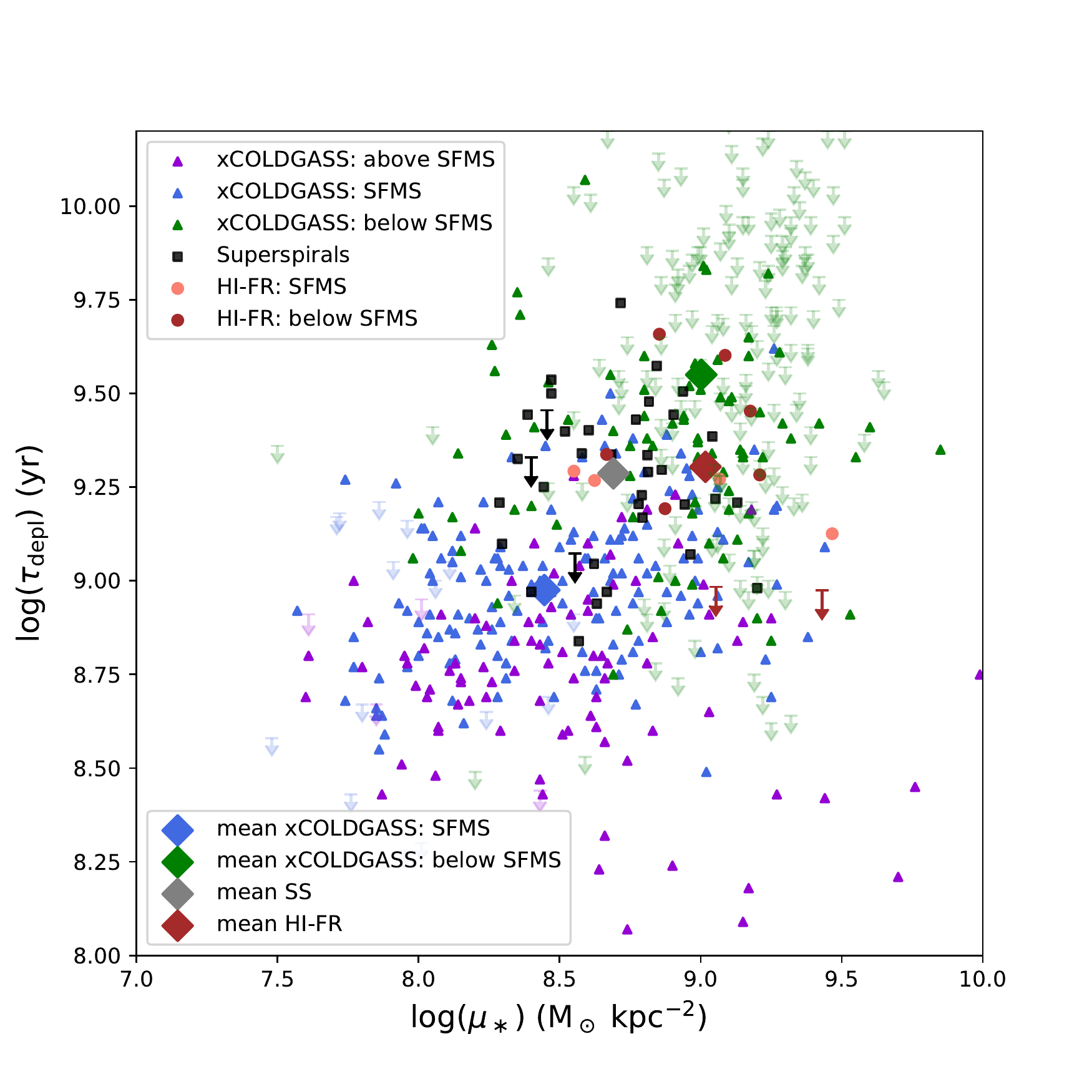}
\caption{The gas depletion time \taudep = \mmol/SFR  as a function of stellar mass  surface density ($\mu_* = M_*/(2\pi r_{p, 50,z}^2$) for the SS, HI-FR and the xCOLDGASS sample.
}
\label{fig:taudep_vs_must}
\end{figure}

Scaling relations between the gas mass (both atomic and molecular), stellar mass and SFR give insight into  how galaxies evolve. The relation between SFR and \mstar\ for star-forming galaxies (the SFMS) has a slope of less than 1, indicating that massive galaxies are not just scaled-up  versions of low mass galaxies, instead they presently form less stars per existing stellar mass.   
There could be various reasons for this trend: (i) a decreasing amount of  gas (atomic + molecular) with \mstar, (ii) a decreasing amount of molecular gas with \mstar\ or (iii) a decreasing efficiency of converting the molecular gas into stars (equivalent to an increasing molecular gas depletion time, \taudep).
Both the atomic and the molecular gas fraction decrease with stellar mass, showing that indeed the gas supply is decreasing with stellar mass. In addition, both the atomic and the molecular gas mass decrease below the SFMS \citep{saintonge17}.
This strongly indicates that the gas supply and in particuar the molecular gas fraction is a driving factor that determines the SFR and the growth of galaxies.  Studies of spatially resolved observations confirm this by showing  that the surface density of the molecular gas mass correlates stronger with \mstar\ than the SFR, suggesting that the molecular gas  is the driving parameter for the SFMS \citep{lin19}.

However, not only \mmol\ is a relevant parameter, but also \taudep\  depends on the position of a galaxy on the SFR-\mstar\ plane. Whereas the dependence on stellar mass is weak, \taudep\ strongly varies  inversely with distance from the SFMS (i.e, it increases below the SFMS).  A long depletion time  can have different reasons. To start with,  CO is a molecule with a low critical density  ($\sim $3000 cm$^{-3}$). This means that it can form in relatively low-density gas, so that  molecular gas probed by CO(1-0) might  be diffuse, unbound gas of relatively low density, and not necessarily in the form of  Giant Molecular Clouds. 
Another possible  reason for the long  gas depletion time in massive or below-SFMS galaxies could be a high  bulge fraction which can quench SF due to a steep gravitational gradient ("morphological quenching," \citeauthor{martig09} \citeyear{martig09} or "gravitational quenching," \citeauthor{genzel14} \citeyear{genzel14}) or dynamical effects ("dynamical quenching," \citeauthor{gensior20} \citeyear{gensior20}).
  Since the bulge fraction generally increases  with stellar mass, this effect could explain the increase of \taudep\ with \mstar. Finally, environmental effects can perturb the gas and affect its ability to form stars as seen in Tidal Dwarf Galaxies \citep{lisenfeld16, querejeta21} or galaxy interactions \citep{lisenfeld17, braine03, appleton22}.
  
How do super spiral galaxies fit into this picture? They are actively star-forming objects, lying on the SFMS which is unusual for their stellar mass. Our observations showed that they are very gas rich, with molecular gas masses that even put them above the scaling relation extrapolated from lower stellar mass galaxies. This confirms a picture in which the molecular gas mass is a driving factor for maintaining the SF in a galaxy. A second results of our study is that \taudep\ is longer than for lower-mass galaxies, with SS galaxies following the  trend of \taudep\ with \mstar\ established from lower-mass galaxies, but lying above from what would be expected from their position on the SFMS.  This suggests that the efficiency of the molecular gas to form stars is indeed decreasing with stellar mass, even for galaxies on the SFMS.

What could be the reason for the relatively long \taudep\ in super spiral galaxies? For super spirals a high bulge fraction does not seem to be the reason for the long depletion time because they have on average rather small bulges with  bulge-to-total mass ratios mostly  between 0.1-0.2 \citep{ogle19b}.  In Fig.~\ref{fig:taudep_vs_must} we show the depletion time as a function of stellar mass surface density, $\mu_* = M_*/(2\pi r_{p, 50,z}^2)$ with $r_{p, 50,z}$ being the radius encompassing 50\% of  $z$-band flux band, which is a proxy for the dominance of bulges. Super spirals indeed have high stellar mass surface densities, spanning a wide range between  $\log(\mu_* ) \approx 8.3 - 9.2$ \msun kpc$^{-2}$.
This range is, however, entirely in overlap with the surface densities of xCOLDGASS star-forming galaxies which have  shorter values of \taudep\ and also the mean values of $\mu_*$  are close (see Tab.~\ref{tab:mean-values}).  On the other hand xCOLDGASS galaxies below the  SFMS and FR-HI have considerably higher value of $\log(\mu_*)$. Thus, the difference in \taudep\ between SFMS xCOLDGASS and SSs does not seem to have it origin in the respective dominance of the bulge.
There is no evidence that environmental effects play a role, or that galaxy interactions \citep[although present, as a large fraction of super spirals have multiple nuclei,][]{ogle19a} have had a major impact on their ISM. Possibly the properties of GMCs change in disks of these high stellar mass, but observations with a higher spatial resolution would be necessary to test this.  

\section{Conclusions}

We present and analyze CO(1-0) observations of a sample of 46  super spiral (SS) galaxies, that is to say very massive (log(\mstar) $\gtrsim$ 11.5 \msun), actively star-forming disk  galaxies.  
 In addition, we include a sample of somewhat less massive disk galaxies with data for the atomic hydrogen and very broad HI spectra from ALFALA (HI fast rotator HI galaxies, HI-FR). These samples are not  representative for their mass range, but instead they are rare objects. (Super spirals make up 6 \% of the galaxies in their mass range). Their interest consists primarily in their existence and in the fact that the  analysis of their properties can provide insights into possible galaxy evolution pathways.

We analyze the relation between SFR, \mstar\ and \mmol\ (and \mhi\ for the HI-FR sample), and compare the properties, after correcting for the  expected increase of the SFR and molecular gas fraction. with redshift, to the local sample xCOLDGASS \citep{saintonge17}. 
Our main results  are:

\begin{itemize}

\item We confirm earlier results \citep{ogle19b} that super spiral galaxies form stars following the SFMS, albeit with values ranging over  $\sim$ one  order of magnitude in sSFR. 

\item Our observations show that super SSs contain large amounts of molecular gas. Adopting a conversion factor of \alphaco = 3~ \msun/(K\,km s$^{-1}$ pc$^{-2}$) (which is a factor of 1.4 lower than the Galactic conversion factor, including helium), appropriate for the expected higher metallicity of super spirals, we find a mean (redshift-corrected) molecular gas fraction, log(\fmolzcorr)  $= -1.36 \pm 0.02$. This molecular gas is higher than expected from the scaling relation with \mstar\ found for xCOLDGASS SFMS galaxies  \citep{janowiecki20}, but lies well within the scaling relation with the distance from the SFMS. 

\item The mean value of the gas depletion time, log(\taudep) = log(\mmol/SFR) is $9.30 \pm 0.03$ yr,   higher than the value of SFMS xCOLDGASS galaxies in the highest mass bins and following a weak  trend with \mstar\ found by other studies \citep[e.g.,][]{saintonge17}. 
The depletion times of SS+HI-FR galaxies lies slightly above the relation with the distance from the SFMS for xCOLDGASS galaxies.

\item The atomic gas mass fraction (\fhi = \mhi/\mstar) of the HI-FR galaxies lies above the scaling relation derived from lower-mass galaxies, which can be explained by a selection effect since we chose galaxies detected by ALFALFA and with broad HI lines. The molecular-to-atomic gas mass ratio of the HI-FR galaxies belonging to the SFMS is in the same  range as found for somewhat lower-mass galaxies  (10$^{10}$ \msun $<$ \mstar $< 10^{10}$ \msun, suggesting the conversion from atomic to molecular gas proceed in the same way as for lower-mass galaxies.

\end{itemize}

Our results taken together allow the following conclusions about SF in SSs and  galaxy evolution in general:

\begin{itemize}

\item A high stellar mass by itself is not a reason for the quenching of SF. If sufficient gas is present - as we found in SS galaxies - and mergers are rare during the lifetime of a galaxies, disk galaxies can grow to large sizes and masses. 

\item The relation between SFR, \mmol\   and \mstar\ are slightly different than for lower-mass galaxies. Super spiral galaxies are more molecular gas rich than what is expected from their stellar mass, showing that the scaling relations derived from lower-mass galaxies are too steep at this high-mass end. In addition, super spirals have longer molecular gas  depletion times than what is expected from their position on the SFMS, suggesting that stellar mass is an additional relevant parameter. The latter results  indicates that the properties of the molecular clouds, or the galactic environment in which they are embedded, might be changing as a function of stellar mass. 
\end{itemize}

In conclusion, we find that SSs with their extreme properties allow us to derived more precise scaling relations that can help to better understand SF and galaxy evolution.

\begin{acknowledgements}
We thank the referee for the careful revision of the manuscript and constructive comments. UL acknowledge support by the research projects
 AYA2017-84897-P and PID2020-114414GB-I00 from the Spanish Ministerio de Econom\'\i a y Competitividad,
from the European Regional Development Funds (FEDER)
and the Junta de Andaluc\'ia (Spain) grants FQM108.
This work is based on observations carried out under project numbers  205-19 and 068-20 with the IRAM 30m telescope. 
IRAM is supported by INSU/CNRS (France), MPG (Germany) and IGN (Spain). This research made use of the ``K-corrections calculator'' service available at http://kcor.sai.msu.ru/. This research made use of Astropy, a community- developed core Python (http://www.python.org) package for Astronomy \citep{astropy13, astropy18}; ipython \citep{perez07}; matplotlib \citep{hunter07}; SciPy, a collection of open source software for scientific computing in Python \citep{virtanen20}; 
and NumPy, a structure for efficient numerical computation \citep{vanderwalt11}. This publication makes use of data products from the Wide-field Infrared Survey Explorer, which is a joint project of the University of California, Los Angeles, and the Jet Propulsion Laboratory/California Institute of Technology, funded by the National Aeronautics and Space Administration. This work was made possible by the NASA/IPAC Extragalactic Database and the NASA/ IPAC Infrared Science Archive, which are both operated by the Jet Propulsion Laboratory, California Institute of Technology, under contract with the National Aeronautics and Space Administration. We acknowledge the usage of the HyperLeda database (http://leda.univ-lyon1.fr).
\end{acknowledgements}


\bibliographystyle{aa}
\bibliography{biblio_ss}

\begin{appendix}

\section{CO(1-0) spectra of central pointings}
\label{app:spectra}

\FloatBarrier


\begin{figure*}[htb!]
\centerline{
\includegraphics[width=4.5cm]{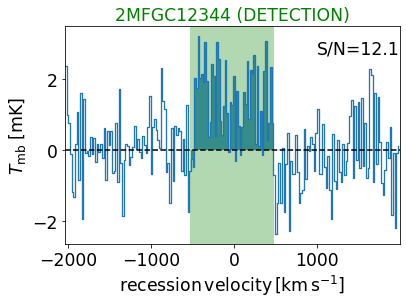} 
\includegraphics[width=4.5cm]{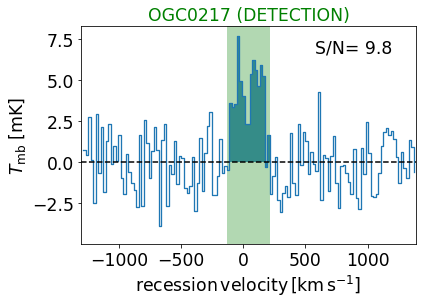} 
\includegraphics[width=4.5cm]{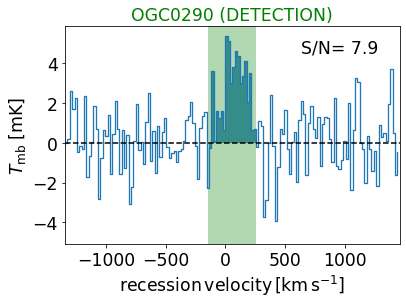} 
\includegraphics[width=4.5cm]{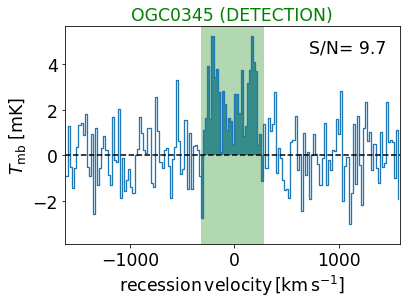} 
}
\quad
\centerline{
\includegraphics[width=4.5cm]{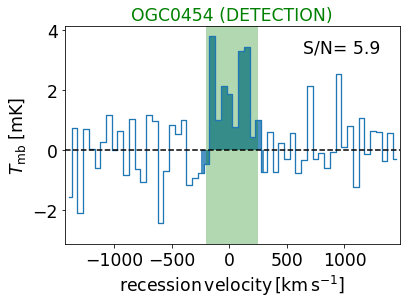} 
\includegraphics[width=4.5cm]{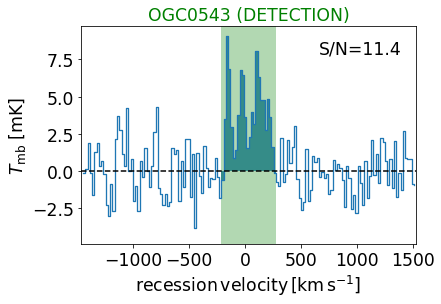} 
\includegraphics[width=4.5cm]{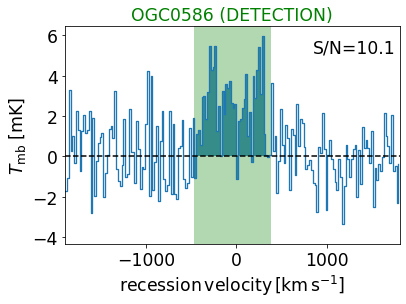} 
\includegraphics[width=4.5cm]{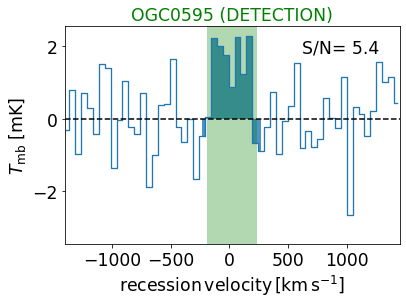} 
}
\quad
\centerline{
\includegraphics[width=4.5cm]{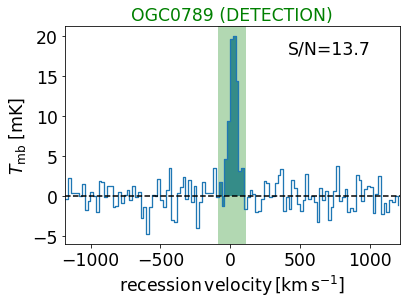} 
\includegraphics[width=4.5cm]{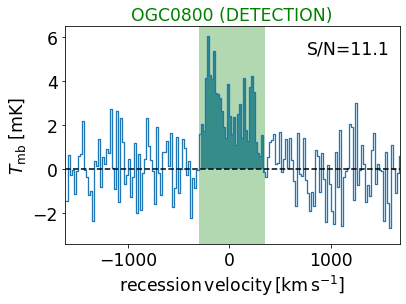} 
\includegraphics[width=4.5cm]{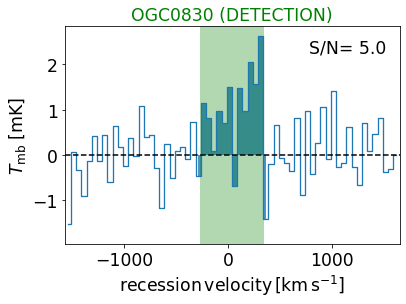} 
\includegraphics[width=4.5cm]{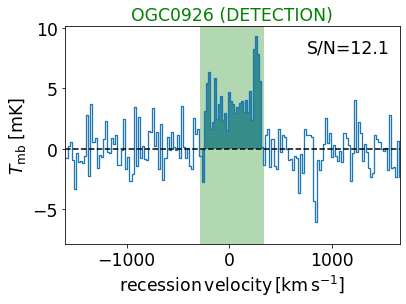} 
}
\quad
\centerline{
\includegraphics[width=4.5cm]{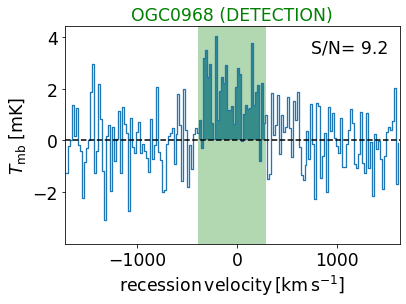} 
\includegraphics[width=4.5cm]{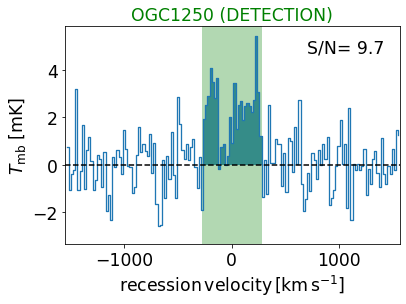} 
\includegraphics[width=4.5cm]{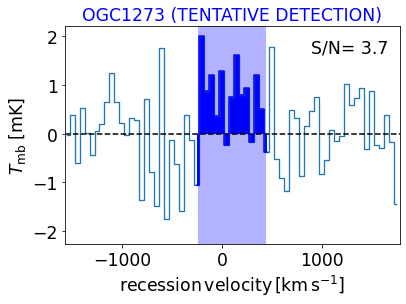} 
\includegraphics[width=4.5cm]{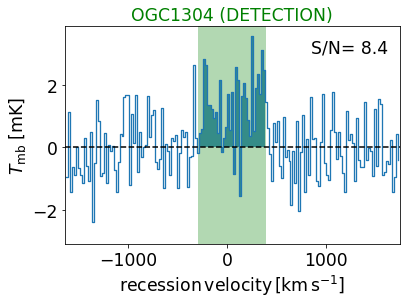} 
}
\quad
\centerline{
\includegraphics[width=4.5cm]{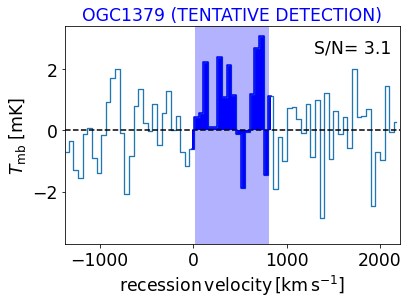} 
\includegraphics[width=4.5cm]{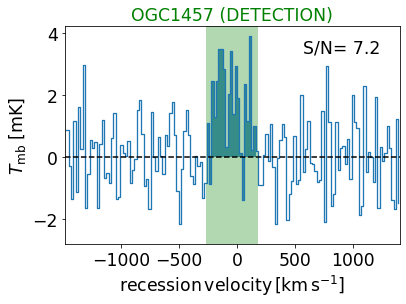} 
\includegraphics[width=4.5cm]{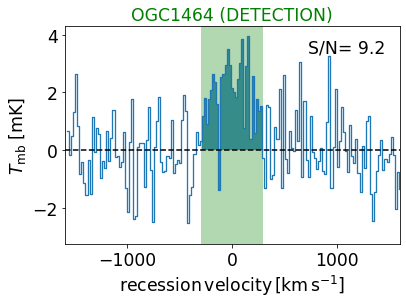} 
\includegraphics[width=4.5cm]{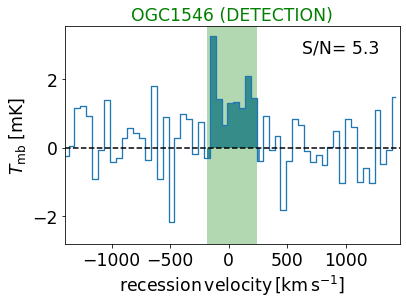} 
}
\quad
\centerline{
\includegraphics[width=4.5cm]{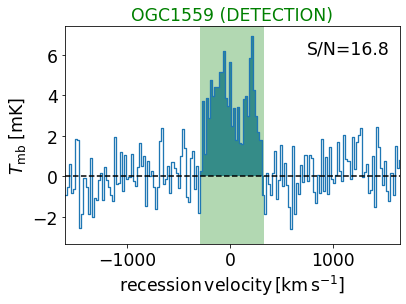} 
\includegraphics[width=4.5cm]{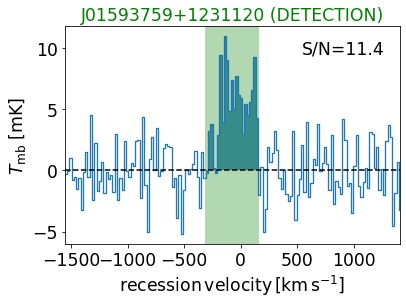} 
\includegraphics[width=4.5cm]{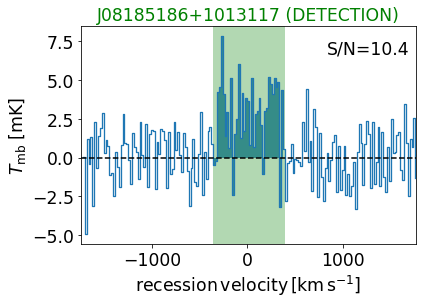} 
\includegraphics[width=4.5cm]{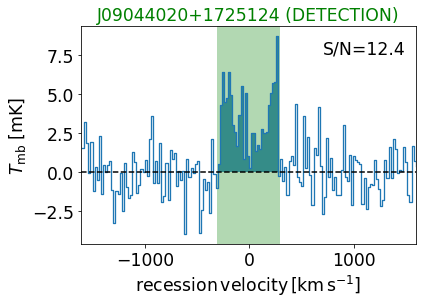} 
}
\caption{Observed spectra of the SS galaxies. {The signal-to-noise ratio S/N = \ico/error(\ico) is indicated in the upper left corner.} The velocity resolution is 20~\kms\ {  for objects with S/N $\gtrsim$ 7 and 50~\kms\  for objects with S/N $\lesssim $ 7 }. The x-axis gives the velocity  
{  relative to the (optical) recession velocity, $v_{\rm rec} = cz$}, where $z$ is the SLOAN redshift. The coloured shaded area represents the region over which the line is integrated to determine the total flux.}
\end{figure*}

\begin{figure*}
\centerline{
\includegraphics[width=4.5cm]{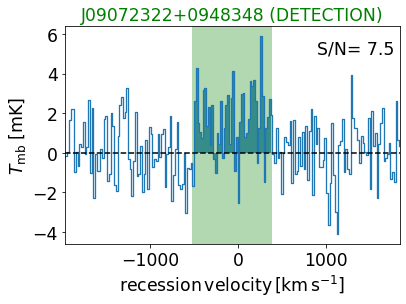} 
\includegraphics[width=4.5cm]{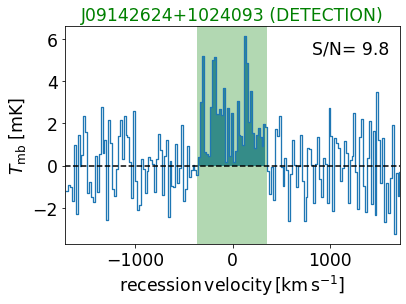} 
\includegraphics[width=4.5cm]{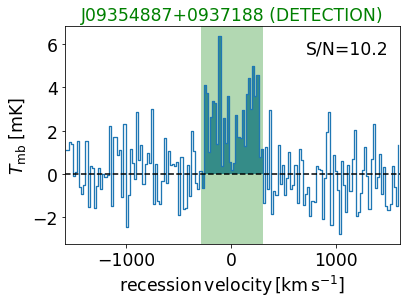} 
\includegraphics[width=4.5cm]{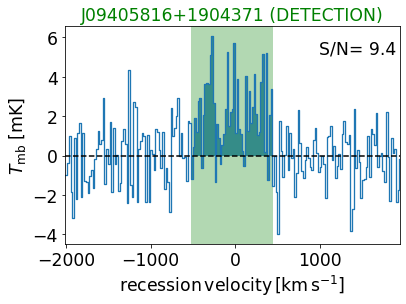} 
}
\quad
\centerline{
\includegraphics[width=4.5cm]{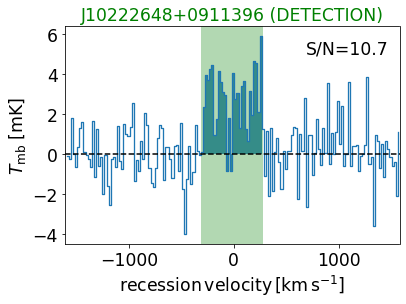} 
\includegraphics[width=4.5cm]{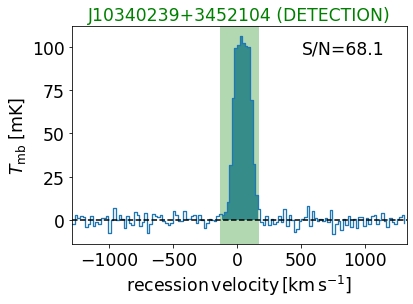} 
\includegraphics[width=4.5cm]{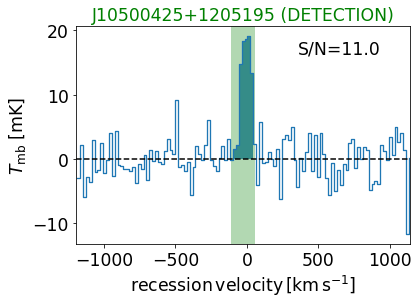} 
\includegraphics[width=4.5cm]{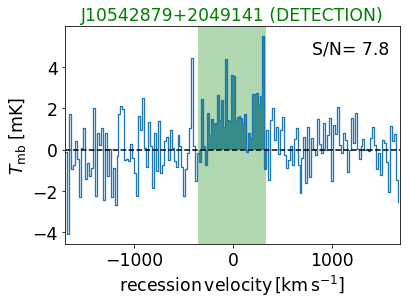} 
}
\quad
\centerline{
\includegraphics[width=4.5cm]{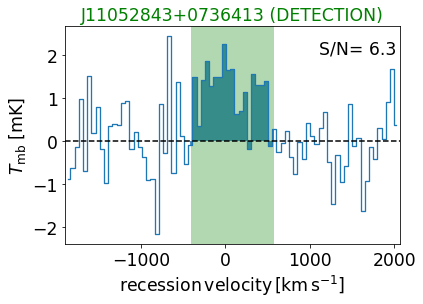} 
\includegraphics[width=4.5cm]{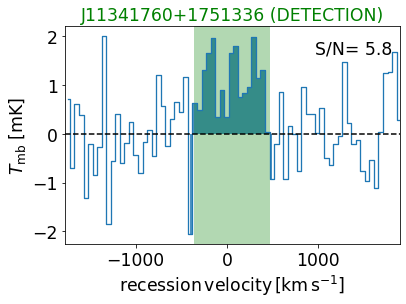} 
\includegraphics[width=4.5cm]{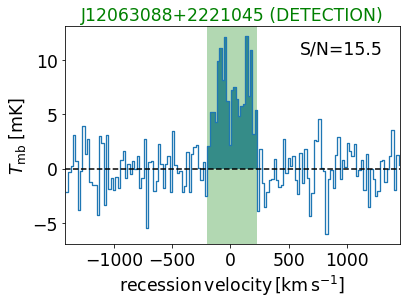} 
\includegraphics[width=4.5cm]{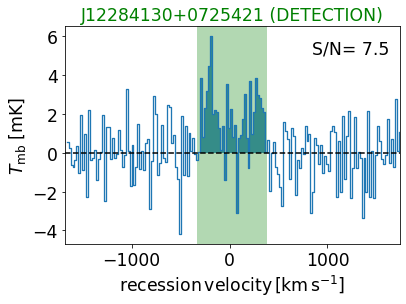} 
}
\quad
\centerline{
\includegraphics[width=4.5cm]{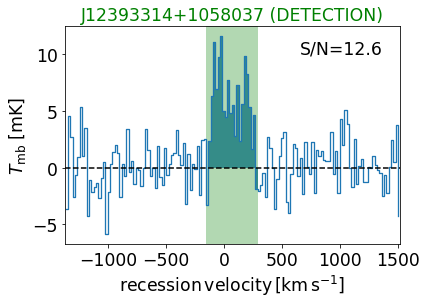} 
\includegraphics[width=4.5cm]{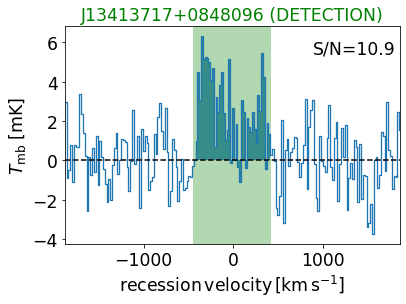} 
\includegraphics[width=4.5cm]{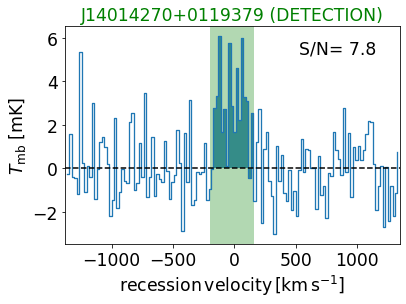} 
\includegraphics[width=4.5cm]{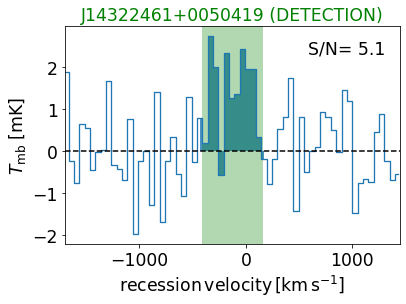} 
}
\quad
\centerline{
\includegraphics[width=4.5cm]{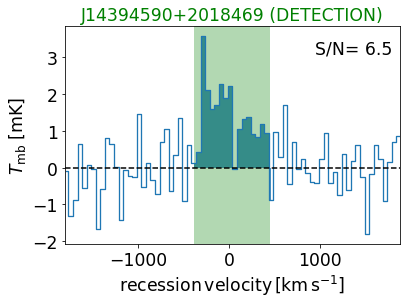} 
\includegraphics[width=4.5cm]{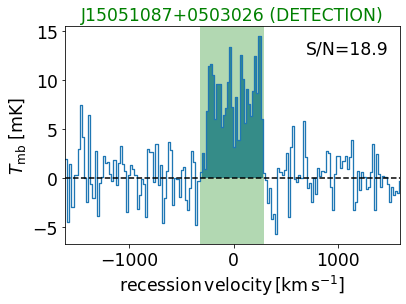} 
\includegraphics[width=4.5cm]{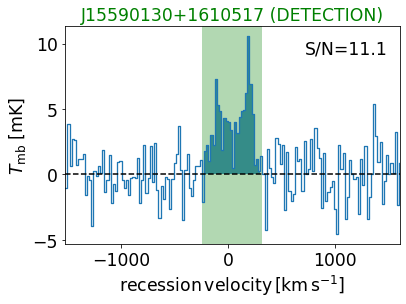} 
\includegraphics[width=4.5cm]{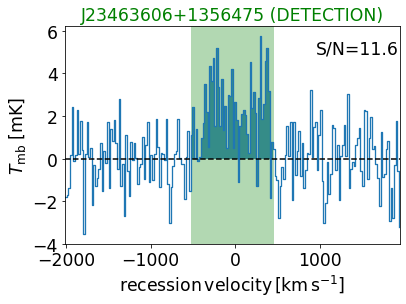} 
}
\addtocounter{figure}{-1} 
\caption{continued}
\end{figure*}

\begin{figure*}
\centerline{
\includegraphics[width=4.5cm]{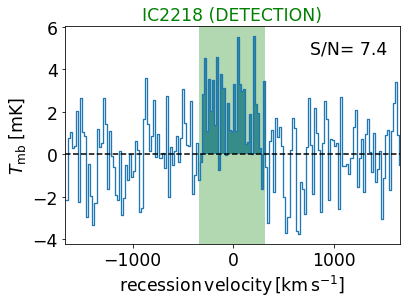} 
\includegraphics[width=4.5cm]{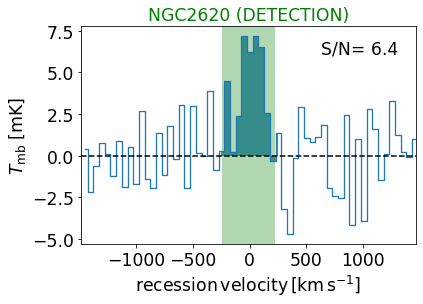} 
\includegraphics[width=4.5cm]{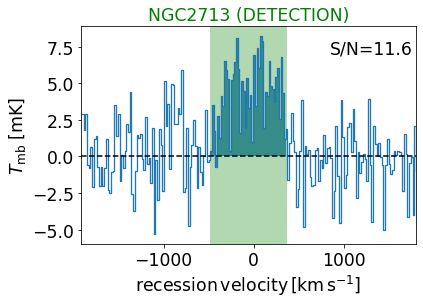} 
\includegraphics[width=4.5cm]{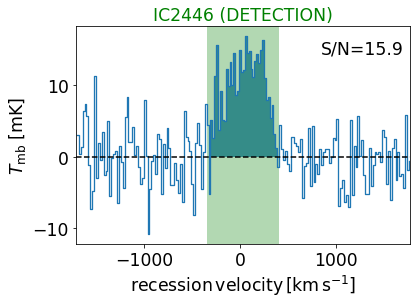} 
}
\quad
\centerline{
\includegraphics[width=4.5cm]{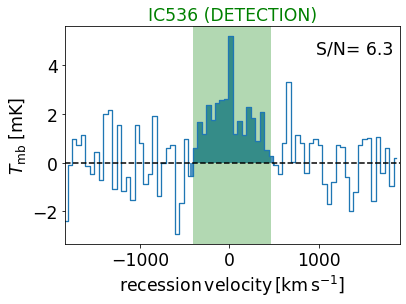} 
\includegraphics[width=4.5cm]{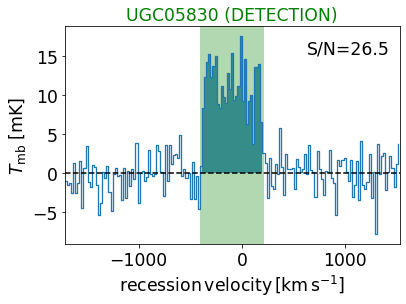} 
\includegraphics[width=4.5cm]{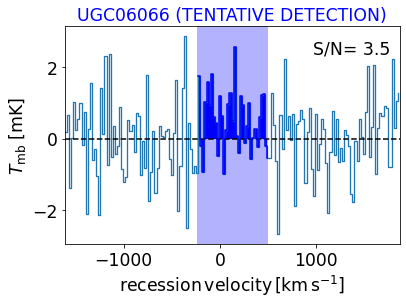} 
\includegraphics[width=4.5cm]{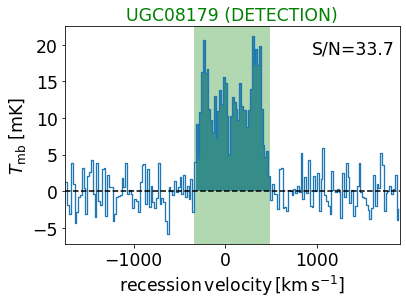} 
}
\quad
\centerline{
\includegraphics[width=4.5cm]{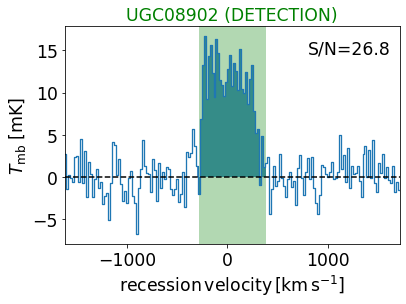} 
\includegraphics[width=4.5cm]{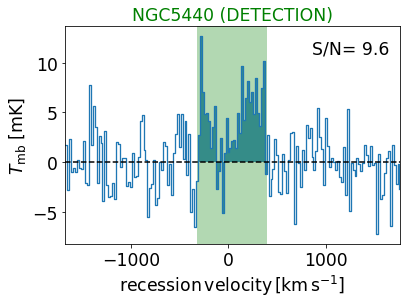} 
\includegraphics[width=4.5cm]{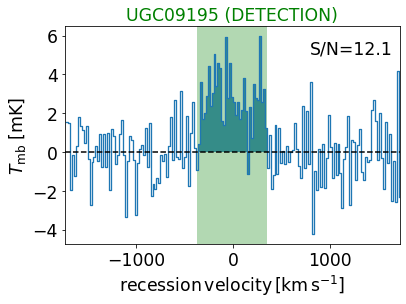} 
\includegraphics[width=4.5cm]{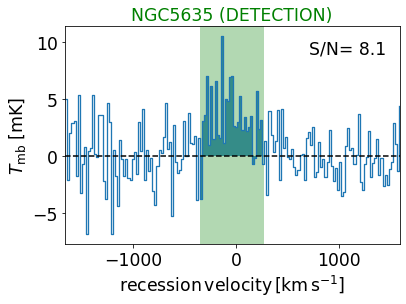} 
}
\quad
\centerline{
\includegraphics[width=4.5cm]{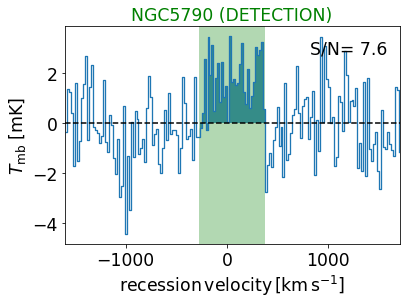} 
\includegraphics[width=4.5cm]{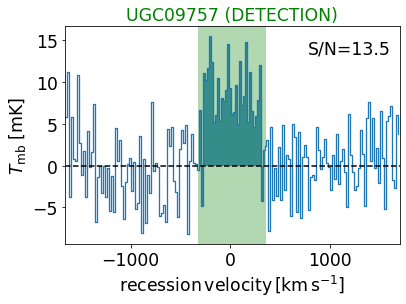} 
\includegraphics[width=4.5cm]{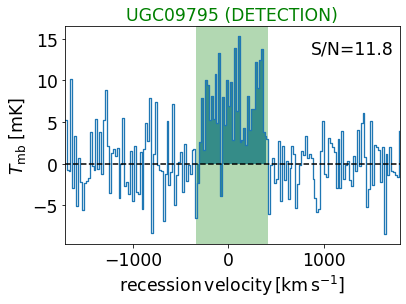} 
\includegraphics[width=4.5cm]{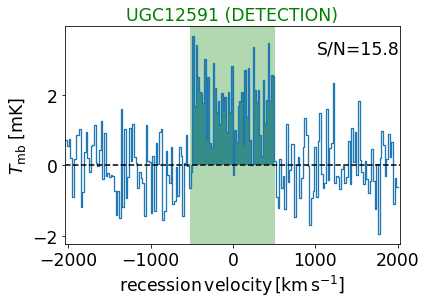} 
}
\caption{Observed spectra of the HI-FR galaxies. {  The signal-to-noise ratio S/N = \ico/error(\ico) is indicated in the upper left corner.} The velocity resolution is 20~\kms\ {  for objects with S/N $\gtrsim$ 7 and 50~\kms\  for objects with S/N $\lesssim $ 7 }. The x-axis gives the velocity  
{  relative to the (optical) recession velocity, $v_{\rm rec} = cz$}, where $z$ is the SLOAN redshift. The coloured shaded area represents the region over which the line is integrated to determine the total flux.
 The spectra are for the central emission, except for NGC~2713, NGC~5790, UGC~08902  and UGC~12591 for which the spectrum  averaged over the positions along the major axis are shown.}
\end{figure*}

\FloatBarrier

\newpage

\begin{figure*}
\centerline{
\includegraphics[width=4.5cm]{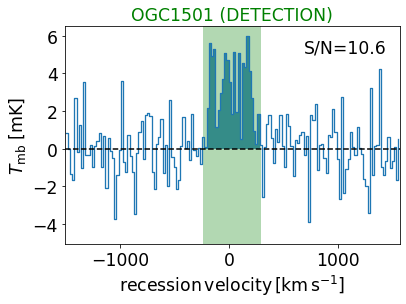} 
\includegraphics[width=4.5cm]{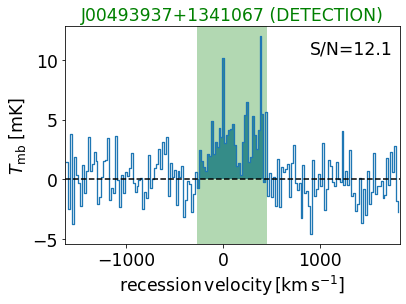} 
\includegraphics[width=4.5cm]{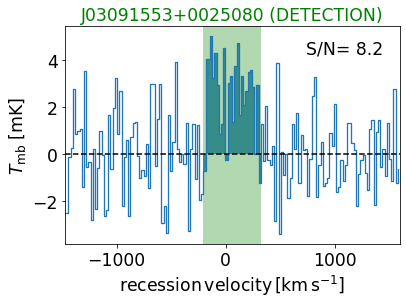} 
\includegraphics[width=4.5cm]{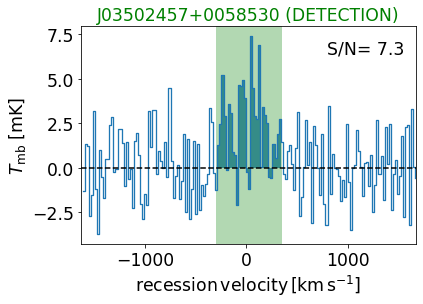} 
}
\quad
\centerline{
\includegraphics[width=4.5cm]{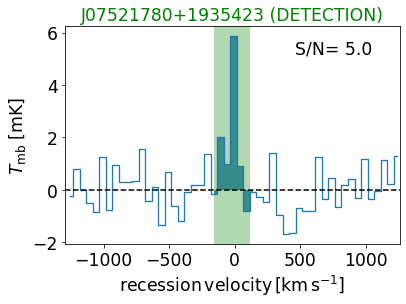} 
\includegraphics[width=4.5cm]{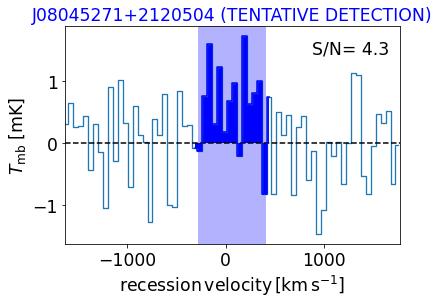} 
\includegraphics[width=4.5cm]{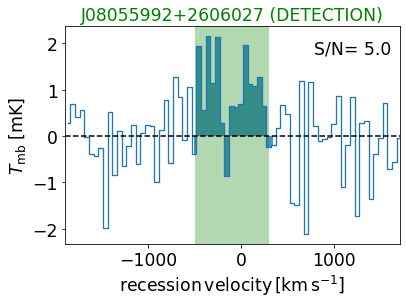} 
\includegraphics[width=4.5cm]{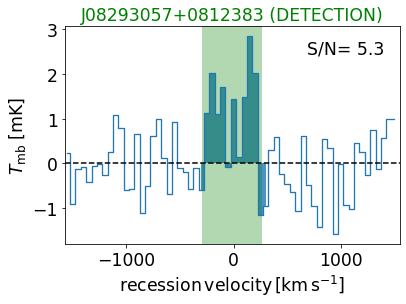} 
}
\quad
\centerline{
\includegraphics[width=4.5cm]{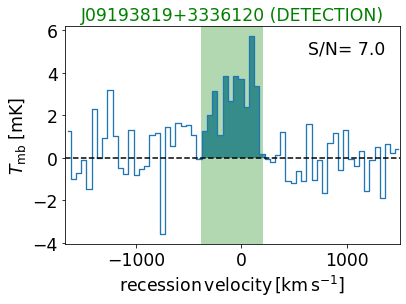} 
\includegraphics[width=4.5cm]{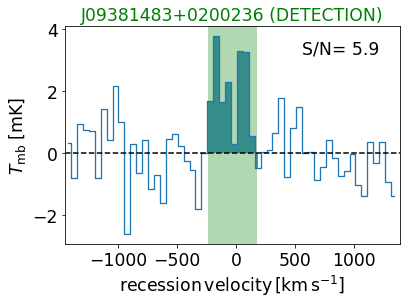} 
\includegraphics[width=4.5cm]{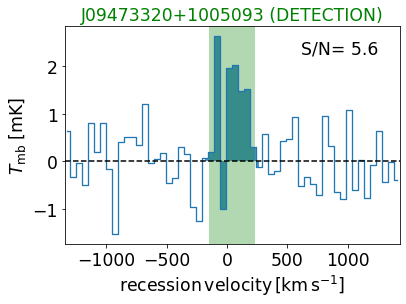} 
\includegraphics[width=4.5cm]{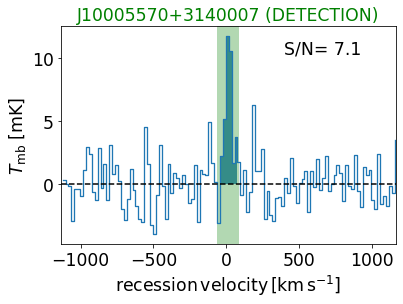} 
}
\quad
\centerline{
\includegraphics[width=4.5cm]{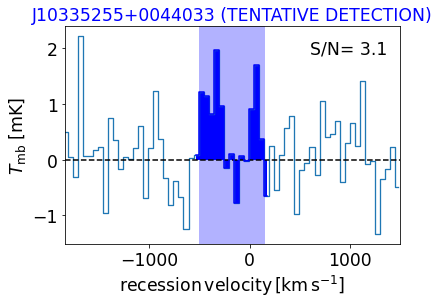} 
\includegraphics[width=4.5cm]{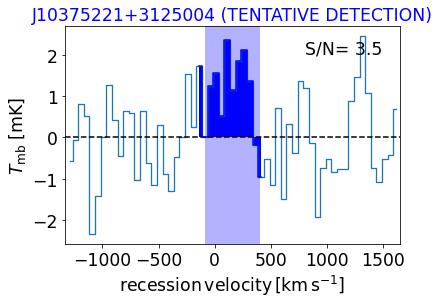} 
\includegraphics[width=4.5cm]{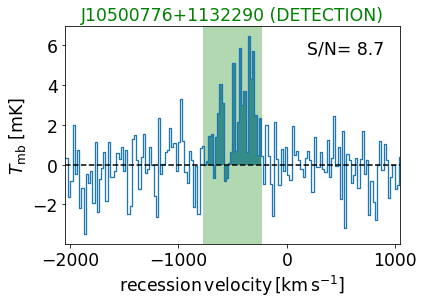} 
\includegraphics[width=4.5cm]{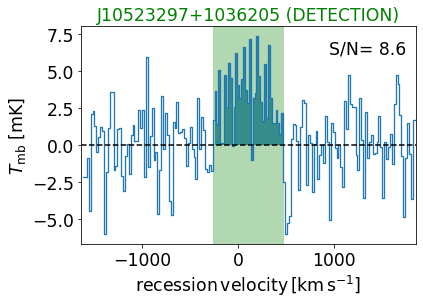} 
}
\quad
\centerline{
\includegraphics[width=4.5cm]{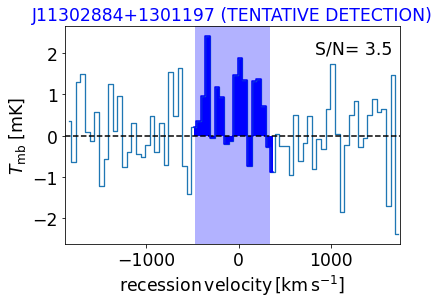} 
\includegraphics[width=4.5cm]{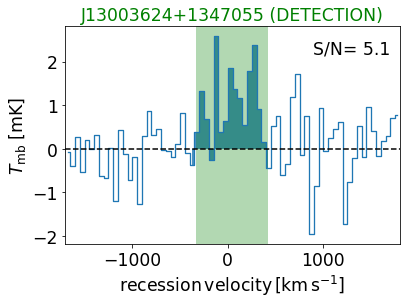} 
\includegraphics[width=4.5cm]{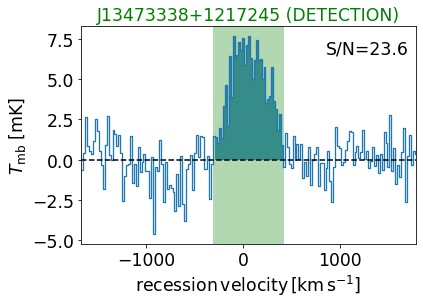} 
\includegraphics[width=4.5cm]{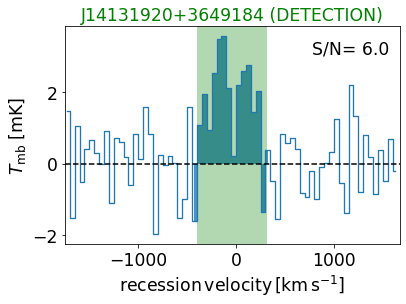} 
}
\quad
\centerline{
\includegraphics[width=4.5cm]{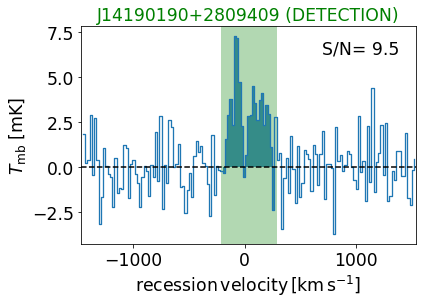} 
\includegraphics[width=4.5cm]{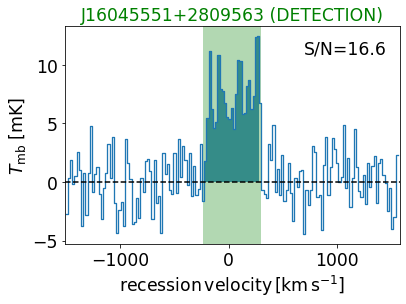} 
\includegraphics[width=4.5cm]{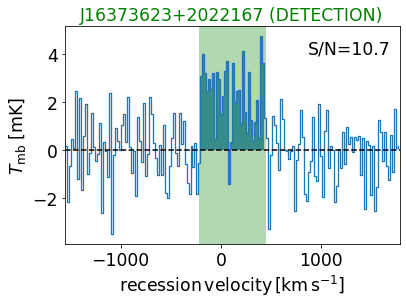} 
\includegraphics[width=4.5cm]{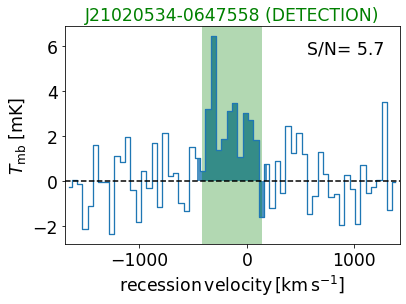} 
}
\caption{Observed spectra of AGNs. {  The signal-to-noise ratio S/N = \ico/error(\ico) is indicated in the upper left corner.} The velocity resolution is 20~\kms\ {  for objects with S/N $\gtrsim$ 7 and 50~\kms\  for objects with S/N $\lesssim $ 7 }. The x-axis gives the velocity  
{  relative to the (optical) recession velocity, $v_{\rm rec} = cz$}, where $z$ is the SLOAN redshift. The coloured shaded area represents the region over which the line is integrated to determine the total flux.}

\end{figure*}

\FloatBarrier

\section{Mapped galaxies}
\label{app:maps}

\FloatBarrier

\begin{figure*}[htb!]
\centerline{
\includegraphics[width=4.5cm]{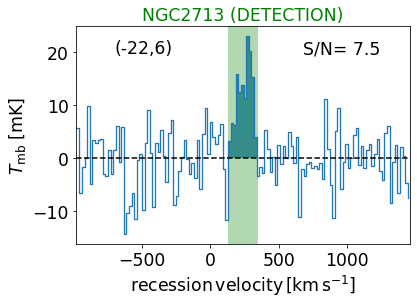} 
\includegraphics[width=4.5cm]{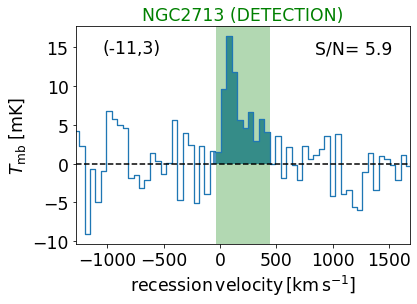} 
\includegraphics[width=4.5cm]{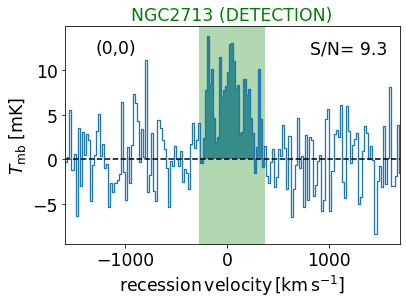} 
}
\centerline{
\includegraphics[width=4.5cm]{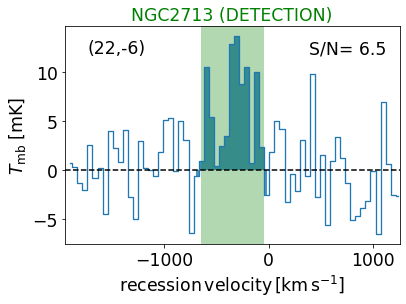} 
\includegraphics[width=4.5cm]{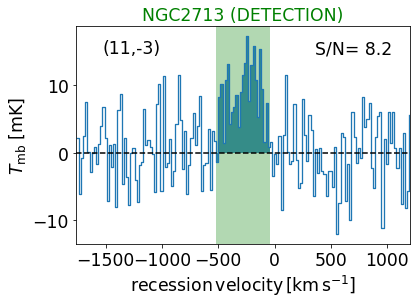} 
}
\caption{Observed spectra along major axis of NGC~2713. The velocity resolution is 20~\kms\ {  for objects with S/N $\gtrsim$ 7 and 50~\kms\  for objects with S/N $\lesssim $ 7}. The offset in arcsec is given in the upper left corner.}
\end{figure*}

\begin{figure*}
\centerline{
\includegraphics[width=4.5cm]{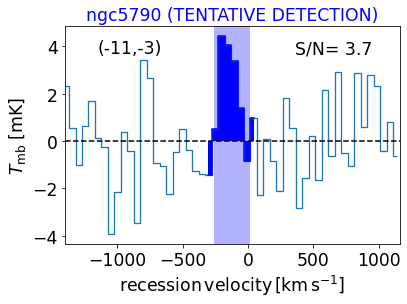} 
\includegraphics[width=4.5cm]{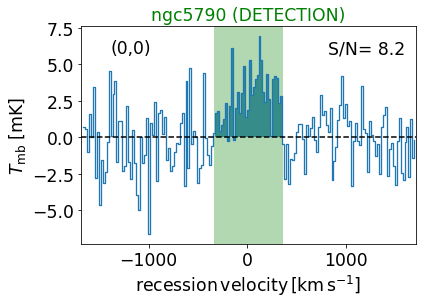} 
\includegraphics[width=4.5cm]{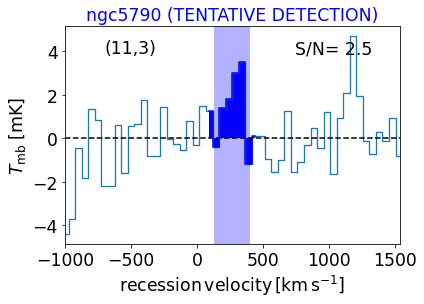} 
}
\caption{Observed spectra along major axis of NGC 5790. The velocity resolution is 20~\kms\ {  for objects with S/N $\gtrsim$ 7 and 50~\kms\  for objects with S/N $\lesssim $ 7}. The offset in arcsec is given in the upper left corner.}
\end{figure*}

\begin{figure*}
\centerline{
\includegraphics[width=4.5cm]{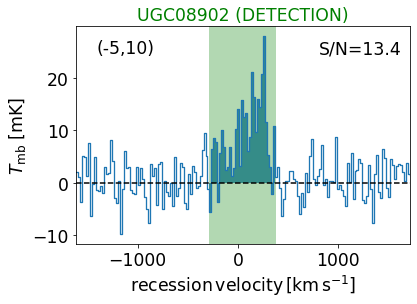} 
\includegraphics[width=4.5cm]{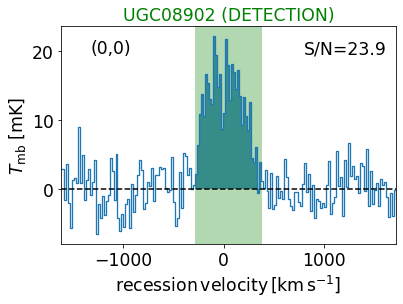} 
\includegraphics[width=4.5cm]{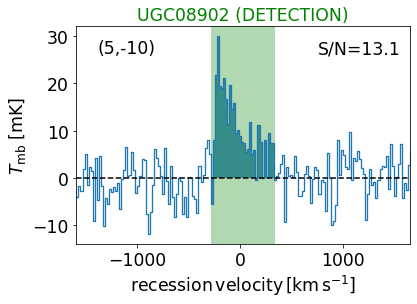} 
}
\caption{Observed spectra along major axis of UGC08902. The velocity resolution is 20~\kms. The offset in arcsec is given in the upper left corner.}
\end{figure*}

\begin{figure*}
\centerline{
\includegraphics[width=4.5cm]{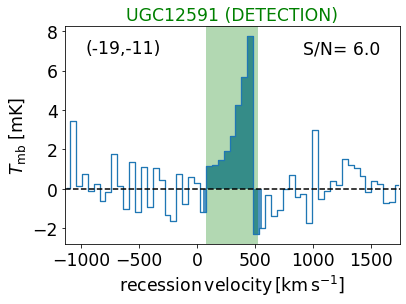} 
\includegraphics[width=4.5cm]{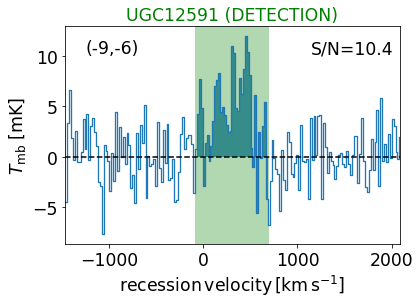} 
\includegraphics[width=4.5cm]{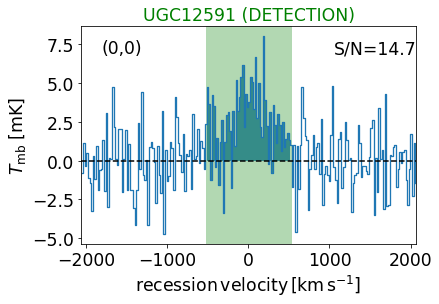} 
}
\centerline{
\includegraphics[width=4.5cm]{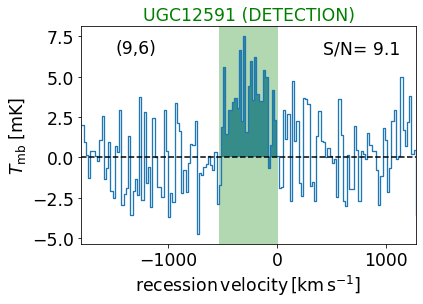} 
\includegraphics[width=4.5cm]{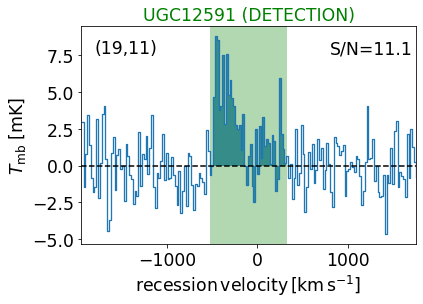} 
\includegraphics[width=4.5cm]{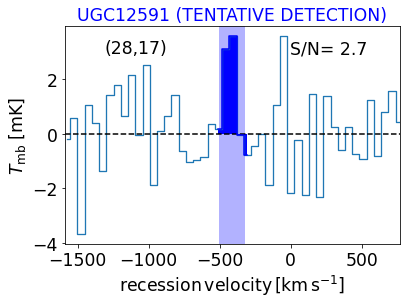} 
}
\caption{Observed spectra along major axis of UGC 12591. The velocity resolution is 20~\kms\ {  for objects with S/N $\gtrsim$ 7 and 50~\kms\  for objects with S/N $\lesssim $ 7}. The offset in arcsec is given in the upper left corner.}
\end{figure*}

\FloatBarrier

\section{Comparison of methods to calculate the  SFR}
\label{app:sfr}

In order to measure the SFR, the most reliable methods combine direct emission from massive stars (as UV or \halpha) and emission from dust to probe dust-enshrouded SF. For massive galaxies, the second part is usually dominant so that methods that solely rely on the dust emission give very reliable results as well. In this section, we are going to compare the hybrid SFR tracer \sfrbest\ (see Sect.~\ref{subsubsec:sfr}) to the hybrid SFR tracer from \citet{leroy19} and the 
monocromatic SFR tracer from \citet{cluver17}.

Both the WISE W3 and the W4 bands can be used as sensitive SF tracers. \citet{cluver17} derived monocromatic SFR prescription for both the W3 and W4 band for the combined SINGS and KINGFISH sample. They showed  that W3 is an excellent tracer for the SFR. In contrast to the {\it Spitzer} 8~\mi\ band, which is dominated by PAH emission \citep{calzetti07, engelbracht08}, the WISE W3 band at 11~\mi\ only has a contribution of $\sim$ 30\% PAH emission, the rest being hot dust and stellar emission. Therefore, after correction for the stellar emission, \citet{cluver17} found a lower scatter for the SFR derived from W3  compared to the SFR derived from the stellar-continuum corrected W4 band. Here, we use the prescription based on the stellar-continuum subtracted W3 emission following eq.~\ref{eq:SFR-W3_C17}.

\citet{leroy19} derived the coefficients for the SFR prescription based on GALEX and WISE data for a sample of $\sim 100\, 000$ galaxies with masses up to $\sim 10^{11}$ \msun\ by comparing the luminosities  to SFRs derived from CIGALE by \citet{salim18}. Due to the large number of galaxies in their sample they could study trends of these coefficients with respect to  other parameters as the stellar mass, WISE colours or the sSFR. In contrast to \citet{cluver17}, they found  that the W4 band has a  higher stability as a SFR tracer, i.e. that the  W4  coefficients  depend less on other parameters than for W3. This difference between \citet{cluver17}  and \citet{leroy19}  might be due to the fact that the Leroy prescriptions are based on the total WISE luminosities, i.e. without subtracting the stellar continuum. The stellar continuum has a larger contribution in the W3 than in the W4 band. Thus, the higher  dependence of W3 on other parameters  found by \citet{leroy19}   might in reality be the effect of a  varying stellar contribution in the W3 band. Taking both studies into account, we conclude that both the W3 and W4 band are reliable tracers for the SFR, especially when a correction for the stellar continuum is done.
We test the prescription of Leroy et al. (2019), based on W3, W4 and NUV (their  Table. 7): 

\begin{equation}
SFR_{W4+NUV, L19} [M_\odot {\rm yr}^{-1}] = L_{\rm NUV} 10^{-43.24} + L_{\rm W4,dust} 10^{-42.79}   \\
\label{eq:SFR-L19W4}
\end{equation}
\begin{equation}
SFR_{W3+NUV, L19} [M_\odot {\rm yr}^{-1}] = L_{\rm NUV} 10^{-43.24} + L_{\rm W3,dust} 10^{-42.86}
\label{eq:SFR-L19W3}
\end{equation}

\begin{figure}
\centering
\includegraphics[width=8.cm,trim=0.cm 0.cm 0cm 0cm,clip]{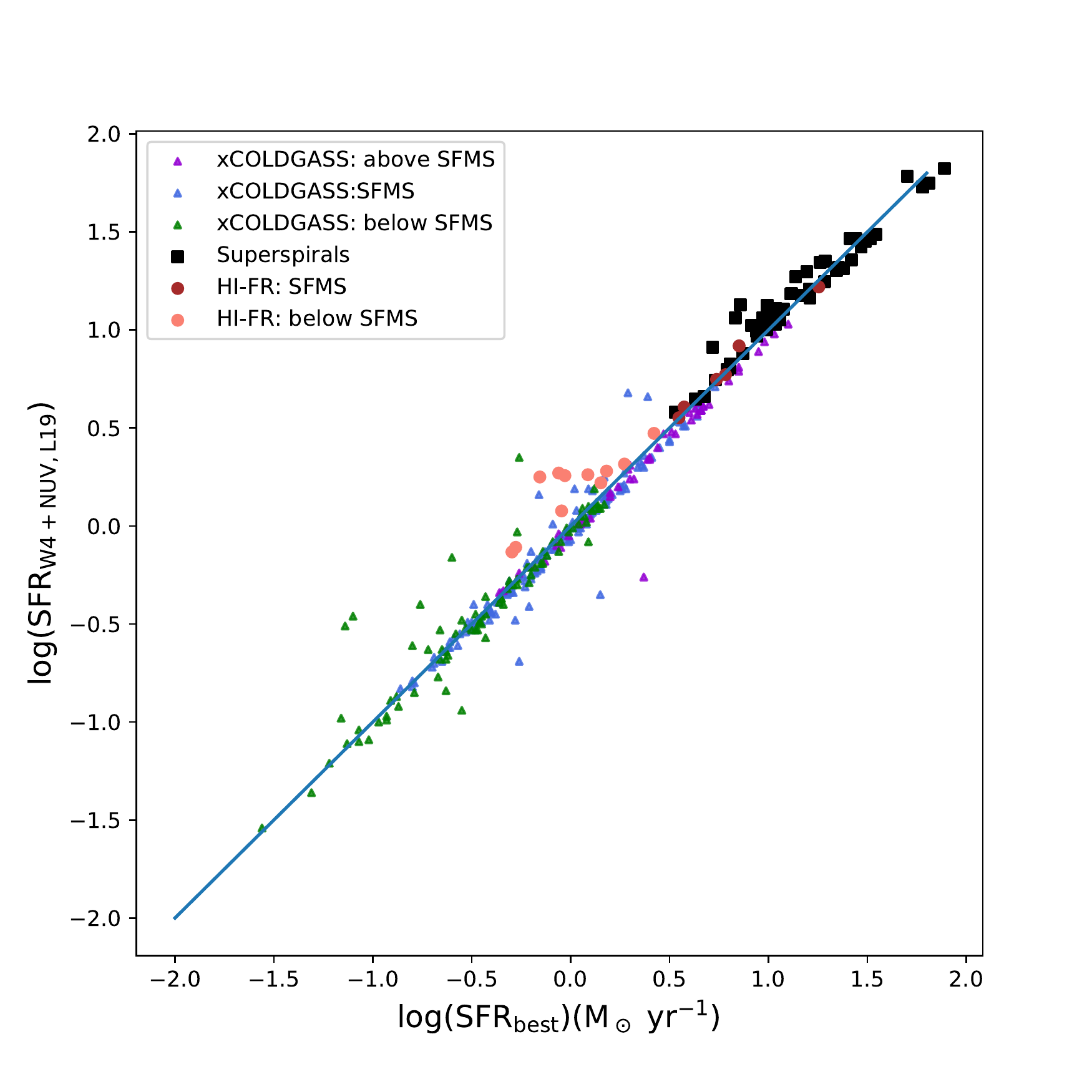}
\caption{Comparison of  \sfrbest\  to the prescription of \citet{leroy19} (see eq.~\ref{eq:SFR-L19W4} ). The blue line is the unity line to guide the eye.}
\label{fig:compare_SFR_W4+NUV-L19-vs-SFRBEST}
\end{figure}

\begin{figure}
\centering
\includegraphics[width=8.cm,trim=0.cm 0.cm 0cm 0cm,clip]{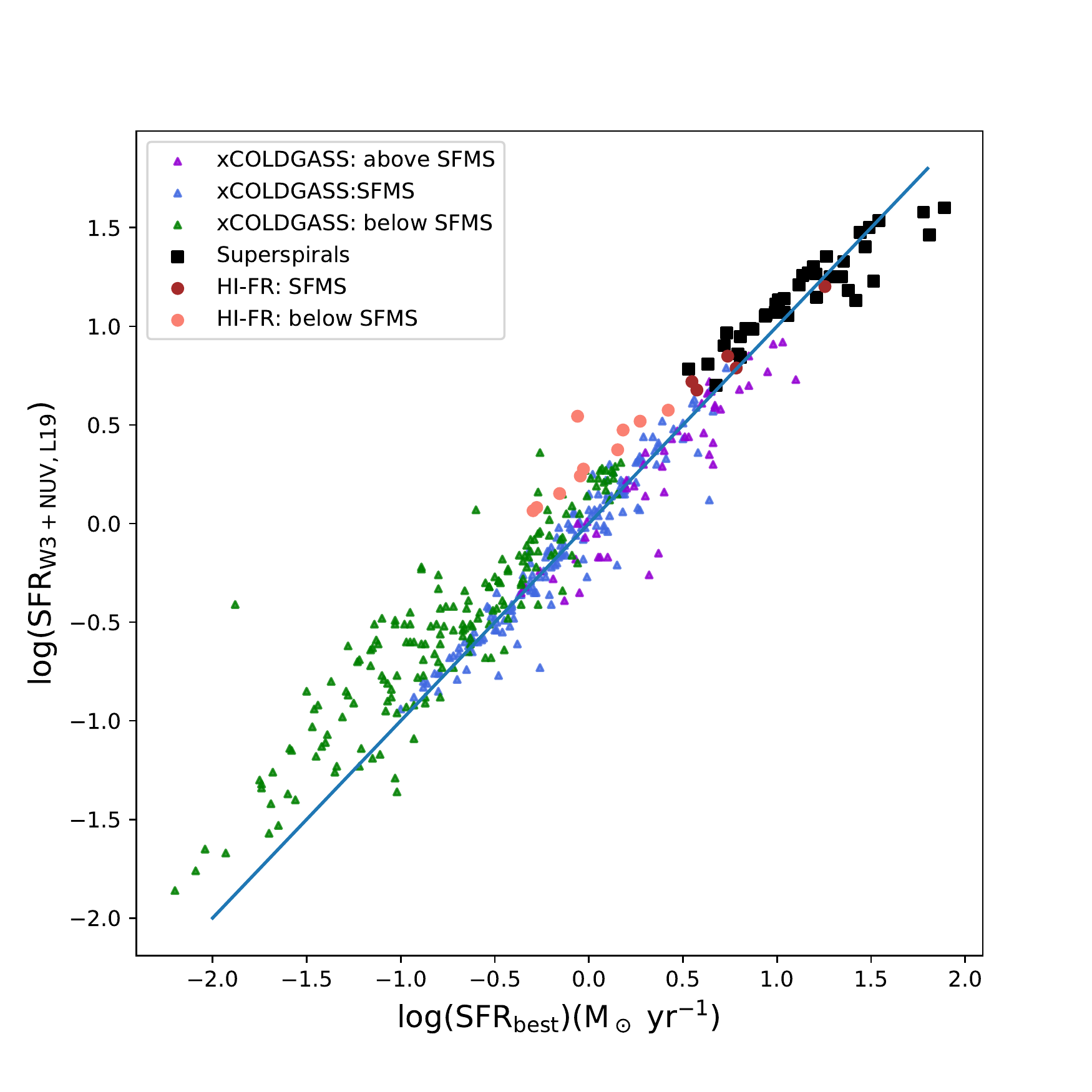}
\caption{Comparison of  \sfrbest\  to the prescription of \citet{leroy19}  (see eq.~\ref{eq:SFR-L19W3}).
The blue line is the unity line to guide the eye.}
\label{fig:compare_sfr_W3+NUV-L19-vs-SFRBEST}
\end{figure}

\begin{figure}
\centering
\includegraphics[width=8.cm,trim=0.cm 0.cm 0cm 0cm,clip]{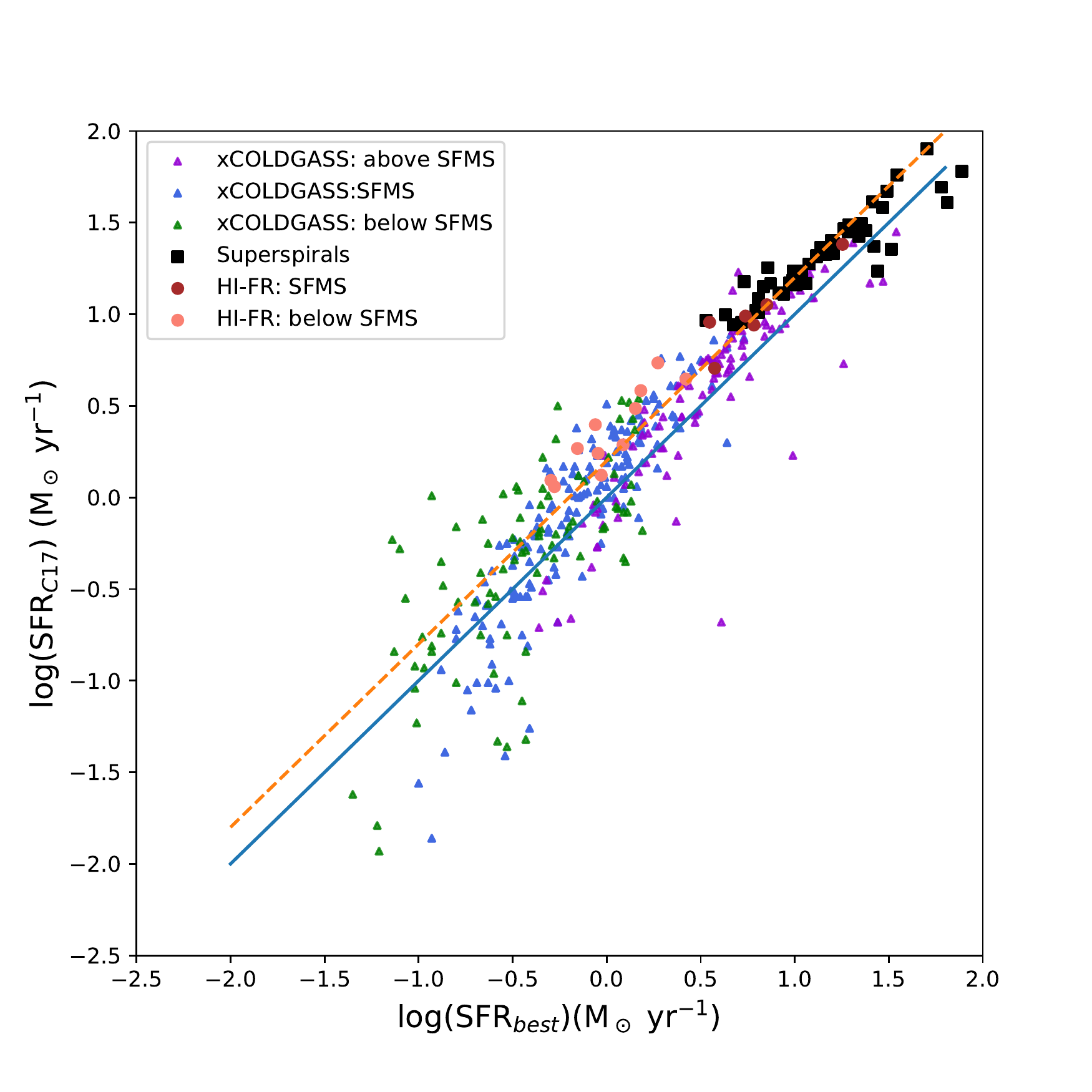}
\caption{Comparison of  \sfrbest\  to the prescription of \citet{cluver17} (see eq.~\ref{eq:SFR-W3_C17}). The blue line is the unity line to guide the eye, and the yellow dashed line  is offset by 0.2 dex, corresponding to the  mean value of log(SFR$_{W3,C17}$/SFR$_{\rm best})$.}
\label{fig:compare-SFRcluver17-vs-sfrbest}
\end{figure}

In Figs.~\ref{fig:compare_SFR_W4+NUV-L19-vs-SFRBEST}-\ref{fig:compare-SFRcluver17-vs-sfrbest} we show the results.
The comparison of \sfrbest\ with SFR$_{W4+NUV, L19}$ is excellent except for a few outliers. This is not too surprising since the coefficients of the prescriptions are very similar, the only difference being that the \citeauthor{leroy19} prescription is based on the total W3 and W4 luminosities, whereas the \citeauthor{janowiecki17} prescription is based on the W3 and W4 luminosities from dust only. The contribution from dust is higher for the W3 luminosity so that the comparison of \sfrbest\ and SFR$_{W3+NUV, L19}$ (Fig.~\ref{fig:compare_sfr_W3+NUV-L19-vs-SFRBEST}) presents a larger scatter.
We can also see a trend that galaxies with a more quiescent stellar population (as galaxies below the SFMS in the xCOLDGASS and  the FR-HI sample) have higher values of the SFR from the  \citeauthor{leroy19} prescription compared to \sfrbest. This is due to their higher $L_{W1}/L_{W3}$ values and therefore the higher stellar contribution in the W3 band.  But in general, also for SFR$_{W3+NUV, L19}$, the agreement between both prescriptions is good.


\begin{table}
\caption{\label{tab:comp-sfr} Comparison of different methods to calculate the SFR}
\resizebox{\columnwidth}{!}{%
\begin{tabular}{llll}
\noalign{\smallskip} \hline \noalign{\medskip}
Sample  & log($\rm \frac{SFR_{W4+NUV,L19}}{SFR_{\rm best}}$) & log($\rm \frac{SFR_{W3+NUV,L19}}{SFR_{\rm best}}$) & log($\rm \frac{SFR_{W3,C17}}{SFR_{\rm best}}$)  \\
 & mean (stdv)\tablefootmark{a} & mean (stdv)\tablefootmark{a} & mean (stdv)\tablefootmark{a}  \\
\noalign{\smallskip} \hline \noalign{\medskip}
SS & 0.03 (0.08) & 0.03 (0.14)& 0.17 (0.14) \\
FR-HI & 0.12 (0.12)& 0.22 (0.16) &  0.29 (0.12) \\
xCOLDGASS & -0.03 (0.09)  & -0.03 (0.13)& 0.05 (0.25)\\   
(SFMS)  &&&\\
xCOLDGASS & 0.02 (0.15) & 0.22 (0.22)& 0.06 (0.48)\\
(below SFMS)&&&\\
\noalign{\smallskip} \hline \noalign{\medskip}
   
\end{tabular}
}
\tablefoot{
\tablefoottext{a}{Mean value and standard deviation (in parenthesis).}
}
\end{table}

\FloatBarrier

The comparison with the \citet{cluver17} prescription also shows a good agreement, albeit with a constant offset of $\sim$ 0.2 dex. Towards lower SFRs there is a trend of lower values of  SFR$_{W3, C17}$ compared to \sfrbest\ which is most likely due to a larger contribution of dust-unobscured SF.

In Tab.~\ref{tab:comp-sfr} we list the mean values  and standard deviation of the ratio between the different tracers for the different subgroups. The standard deviation gives us an idea of the general uncertainty in the calculation of the SFR, and the differences in the mean values for the different sample an idea of the uncertainty when comparing the results between different groups. 
In general, we find a satisfactory agreement between the different tracers with roughly linear relations between them (see Figures). There are some differences in the mean values of the ratio between the different groups, with differences up to 0.20 - 0.25 dex between SF and quiscient subsamples, but less (up to $\sim$ 0.1 dex) between the SFMS samples. This means that  there could be  artifical differences  up to this order of magnitude in the mean SFR when comparing these subsamples.

\section{Comparison of different methods to calculate \mstar}
\label{app:sfr}

\begin{figure}
\centering
\includegraphics[width=8.cm,trim=0.cm 0.cm 0cm 0cm,clip]{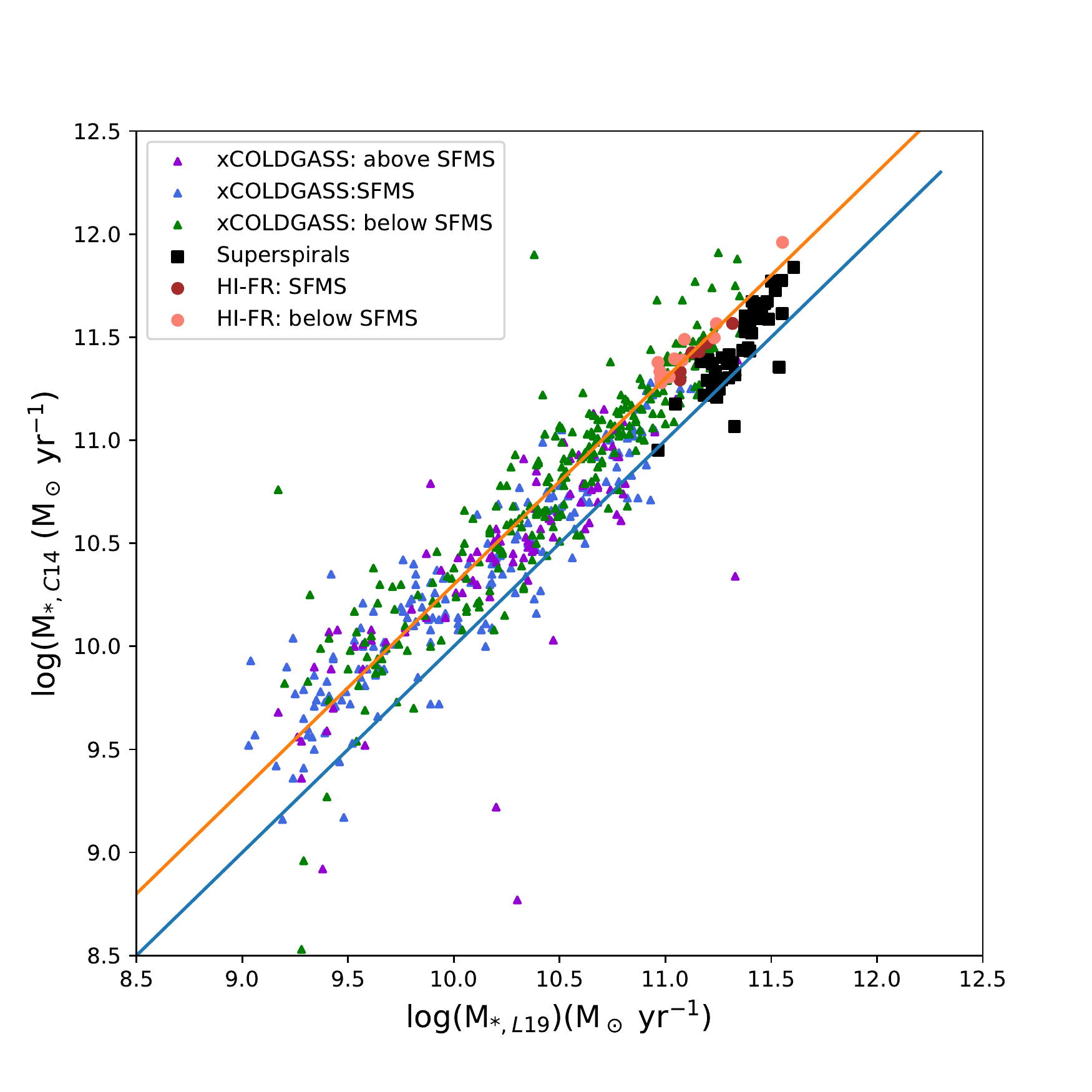}
\caption{ Comparison of the stellar mass derived from the methods of \citet{cluver14} and \citet{leroy19}. 
 The blue line is  the unity line to guide the eye and the yellow line is offset by 0.3 dex which corresponds to the mean value of log(\mstar$_{\rm C14}$/\mstar$_{L19})$ for the quiescient subsamples
 (see Tab.~\ref{tab:comp-mstar}). }
\label{fig:compare-mstarC14_L19}
   \end{figure}
\begin{figure}
\centering
\includegraphics[width=8.cm,trim=0.cm 0.cm 0cm 0cm,clip]{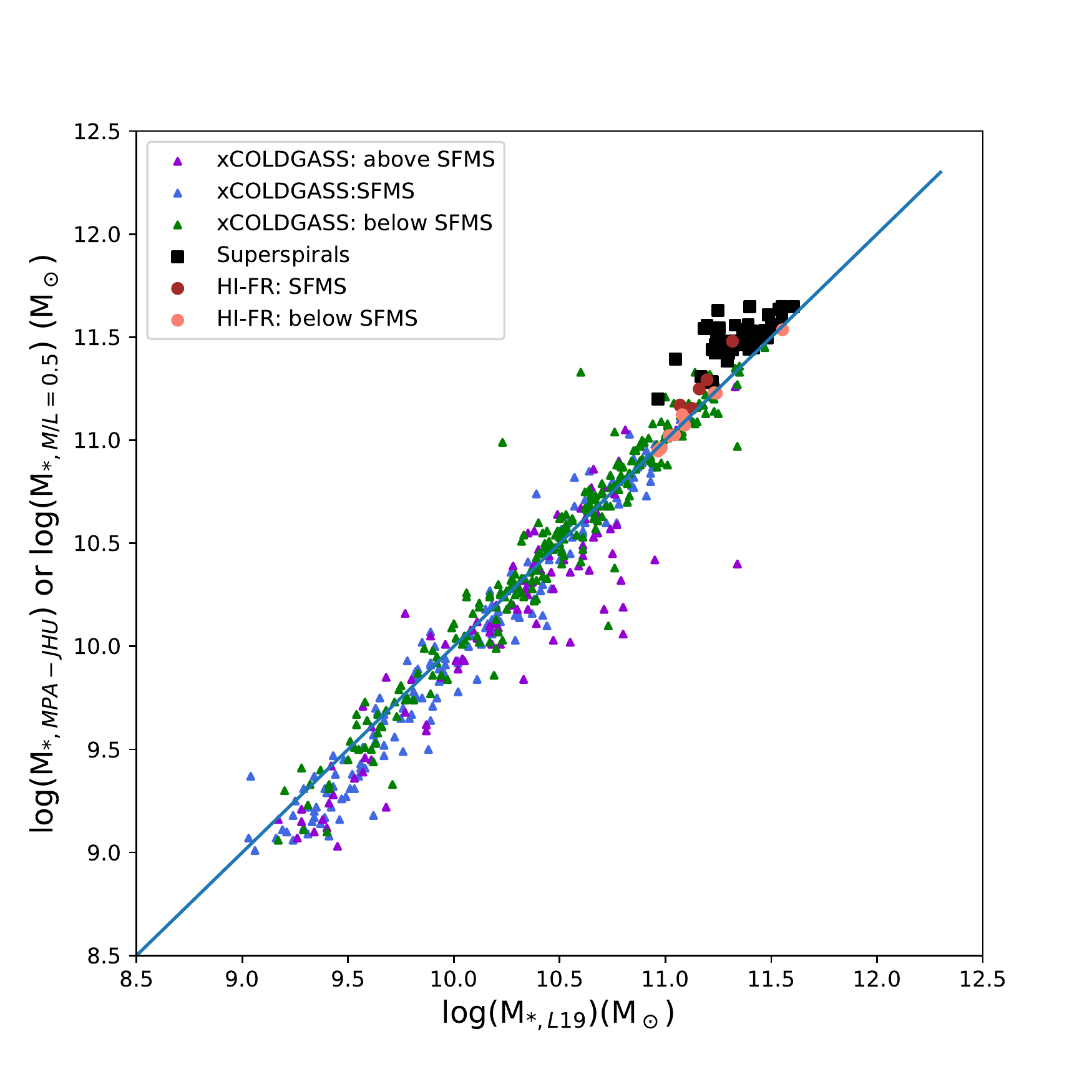}
\caption{ Comparison of the stellar mass derived from the method of  \citet{leroy19}, and MPA/JHU for the xCOLDGASS sample, respectively a constant 
 $\Upsilon_{\ast}$  = 0.5 for the SS+FR-HI sample. The blue line is  the unity line to guide the eye.
}
\label{fig:compare-mstar-L19-M-MPA/JHU}
   \end{figure}

We compared several prescriptions to calculate the stellar mass:

\begin{itemize}
\item  A constant mass-to-light ratio, $\Upsilon_{\ast}^{3.4}$ = 0.5. Whereas this is too simplistic for the entire sample, it is a reasonable assumption to test for the rather homogeneous  sample of massive spirals. 
\item The method of \citet{cluver14} who derive a color dependent $\Upsilon_{\ast}^{3.4}$ based on the analysis of a sample of galaxies with GAMA data for which the stellar mass was derived from an analysis of stellar populations. Their best-fit prescription for the entire sample (their eq. 2) is:

\begin{equation} 
\log(M_{\rm \ast, C14} /L_{\rm W1}  ) [M_\odot/L_{W1,\odot}]= -1.96 (W1-W2) -0.03,
 \label{eq:cluever14}
\end{equation}
where $(W1-W2)$ is the WISE color in mag.
\item Leroy et al. (2019) compared for a sample of $\sim$ 130.000 galaxies the stellar mass derived from  fitting the UV-to-mid-infrared spectral energy distribution (SED)  with CIGALE from \citet{salim18} to different observationally derived parameters (SFR, WISE luminosities and colours). The best correlation  for  $\Upsilon_{\ast}^{3.4}$ that they obtained was with  SFR$/\nu L_{\nu}$(W1) (their eq. 24, see also their Figs. 22 and 23):
 
 \begin{equation}
 \label{eq:leroy19}
\Upsilon_{\ast}^{3.4} [M_\odot/L_{W1,\odot}] = 
 \begin{cases}
    0.5, & \text{if}~ Q < a \\
    0.5 + b~\left( Q - a\right) , & \text{if}~a < Q <c \\
    0.2 , & \text{if}~Q>c
\end{cases}
 \end{equation}

where $L_{W1,\odot} = 1.6\cdot 10^{32}$ \ergs $= 0.042$ \lsun\ is the Solar luminosity in the W1 (3.4~\mi) band, $a =-11$, $b = -0.21$ and $c=-9.5$ (see Table 6 in Leroy et al. 2019). Q = SFR$/\nu L_{\nu}$(W1) with $\nu L_{\nu}$(W1) being the luminosity in the W1 in units of solar bolometric luminosity (\lsun). Given that $\nu L_{\nu}$(W1) is closely related to the stellar mass, Q is a quantity that is similar to the sSFR.
This prescription gives a high value ($0.5~M_\odot/L_{W1,\odot}^{-1}$)
for quiescient galaxies and a low value ($0.2~M_\odot L_{W1,\odot}^{-1}$) for actively star-forming objects. Applying this method to the SS sample, values for 
$\Upsilon_{\ast}^{3.4}$ between 0.25 and 0.5 were derived.

\end{itemize}

Fig.~\ref{fig:compare-mstarC14_L19} shows the comparision of the method of Cluver et al. (2014) and Leroy et al. (2019). A good correlation is visible, albeit with an difference of 0.1-0.3 dex between both methods. This offset is similar  for all subsamples except for SSs  for which
 M$_{\rm \ast, C14}$/M$_{\rm *, L19}$ is $\sim$ 0.1 dex lower than for  the star-forming galaxies in xCOLDGASS (see Tab.~\ref{tab:comp-mstar}).   This means that either the method of Leroy et al. overpredict the true stellar mass of super spirals, or Cluver et al. underpredicts it. The difference is, however, small. 

Fig.~\ref{fig:compare-mstar-L19-M-MPA/JHU} shows the comparison of the methods of Leroy et al. (2019) and the stellar masses from the MPA/JHU catalog for xCOLDGASS galaxies and a constant $\Upsilon_{\ast}^{3.6} = 0.5$ for the SS+FR-HI sample. The agreement between both methods is satisfactory  (see Tab.~\ref{tab:comp-mstar}). For the quiescient xCOLDGASS galaxies and for the FR-HI the agreement is perfect, whereas the mean value of M$_{\rm \ast, MPA/JHU}$ for SFMS xCOLDGASS galaxies  is slightly (0.08 dex) lower than the value from Leroy et al. (2019) and for SS galaxies the mean value for \mstar\ derived with a constant $\Upsilon_{\ast}^{3.4} = 0.5$  for SS galaxies is slightly higher (0.14 dex) than the value from \citet{leroy19}. Overall, the differences are small and close to the standard deviation of the ratios (see Tab.~\ref{tab:comp-mstar}).

\begin{table}
\caption{\label{tab:comp-mstar} Comparison of different methods to calculate the stellar mass}
\begin{tabular}{llll}
\noalign{\smallskip} \hline \noalign{\medskip}
Sample  & log($\rm \frac{\rm M_{\rm *,C14}}{M_{\rm *, L19}}$) &  log(\rm $\frac{M_{\rm *,MPA/JHU}}{M_{\rm *,L19}}$) & log(\rm $\frac{M_{*,\Upsilon_{*0.5}}}{M_{\rm *,L19}}$)  \\
 & mean (stdv)\tablefootmark{a} & mean (stdv)\tablefootmark{a} & mean (stdv)\tablefootmark{a}  \\
\noalign{\smallskip} \hline \noalign{\medskip}
SS & 0.11 (0.11) &  -- & 0.14 (0.10) \\
FR-HI & 0.31 (0.05)& -- & 0.03 (0.06) \\
xCOLDGASS & 0.21 (0.28)  & -0.08 (0.16)& -- \\   (SFMS) & & & \\ 
xCOLDGASS & 0.31 (0.22) & 0.01 (0.12)& -- \\
(below SFMS) & & & \\ 
\noalign{\smallskip} \hline \noalign{\medskip}
   
\end{tabular}
\tablefoot{
\tablefoottext{a}{Mean value and standard deviation (in parenthesis).}
}
\end{table}

\section{SED fitting of the SS galaxies with CIGALE}
\label{sect:app_cigale}

CIGALE \citep[Code Investigating GALaxy Emission;][]{boquien19}  is a python implemented code based on an energy balance principle, where the energy absorbed by dust from UV to near-infrared frequencies is re-emitted in the mid- and far-infrared. It has a Bayesian-like approach and has allowed us to model the SED of our SS+FR-HI galaxy sample from far-UV up to far-infrared wavelengths, and to estimate their physical properties, such as SFR and stellar mass. For reliability purposes, only objects with $\chi^{2}<1$ have been considered during the analysis. 

For the data input, we used the GALEX and WISE data presented in this paper (without the k-correction since CIGALE performs a k-correction in the fitting process), together with SDSS fluxes for the $u, g, r,i$ and $z$-band. For the GALEX and WISE data we added to the photometric errors a calibration error in quadrature (14.8 \% for GALEX, \citeauthor{gildepaz07} 2007, and 2.4\%, 2.8\%, 4.5\%, and 5.7\% for the W1, W2, W3, and W4 images, respectively, \citeauthor{jarrett11} 2011).

To perform the fits we used a series of modules that model the SF  history (SFH), stellar population, nebular emission, dust attenuation, and dust emission. The modules and parameters used in our fits follow those used by \citet{hunt19}, detailed in Table 1 of their article, with the exception of two parameters that have been slightly modified to model our sample better:

\begin {enumerate}
\item The SFH is modeled using a delayed + truncated parametrization \citep{ciesla17}, where $r_{\rm SFR}=SFR(t>t_{\rm trunc})/SFR(t_{\rm trunc})$ considers a reduction or increase in the SFR after the truncation time, $t_{\rm trunc}$. We allow the parameter set $r_{\text{SFR}}=(0.01,0.05,0.1,0.5,1.5,10)$.
\item We choose a modified starburst attenuation law \citep{calzetti00} that considers different attenuations for stellar populations of different ages. The baseline law is multiplied by $\lambda^{\delta}$, where we select the following values for the power-law slope, $\delta=(-1.0,-0.8,-0.6,-0.4,-0.2,0.0)$. 
\end{enumerate}

The mean values and standard deviations of the ratios between the value derived from CIGALE and from the prescriptions used here are given in  Tab.~\ref{tab:comp-cigale}

\begin{figure}
\centering
\includegraphics[width=8.cm,trim=0.cm 0.cm 0cm 0cm,clip]{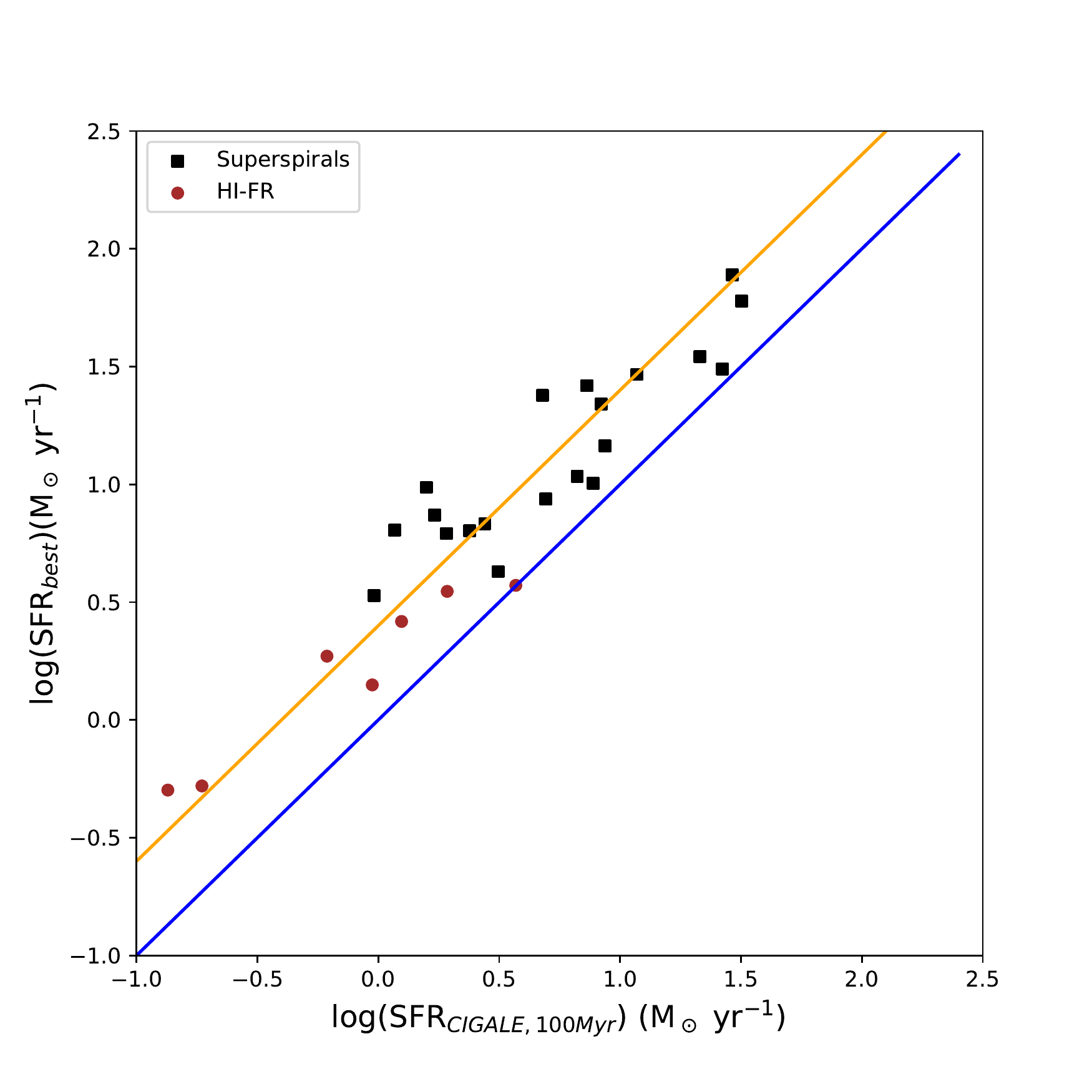}
\caption{ Comparison of  \sfrbest\  and SFR averaged over the past 100 Myr derived by CIGALE.    The blue line shows unity  to guide the eye and the yellow line is offset by 0.4 dex which corresponds to the mean value of log(SFR$_{\rm best}$/SFR$_{\rm CIGALE, 100 Myr}$).}
\label{fig:compare-SFRcigale-SFRbest}
   \end{figure}

Fig.~\ref{fig:compare-SFRcigale-SFRbest} gives a comparison between the SFR derived with CIGALE (averaged over the past 100 Myr) and \sfrbest\ for objects with a good fit  (reduced $\chi^2 <  1$). A good correlation is visible, albeit offset by 0.4 dex (which is the mean value of log(SFR$_{\rm CIGALE, 100 Myr}$/SFR$_{\rm best}$)).  The offset most likely  reflects the different definitions of both SFRs, as the SFR traced by UV+WISE data is not exactly the same as the SFR averaged over the past 100 Myr \citep[see][for a detailed discussion of the time-scales of SFRs derived from different tracers]{boquien14}.  The standard deviation of the correlation is 0.2 dex which gives an estimate for the uncertainty of the determination of the SFR.
\begin{figure}
\centering
\includegraphics[width=8.cm,trim=0.cm 0.cm 0cm 0cm,clip]{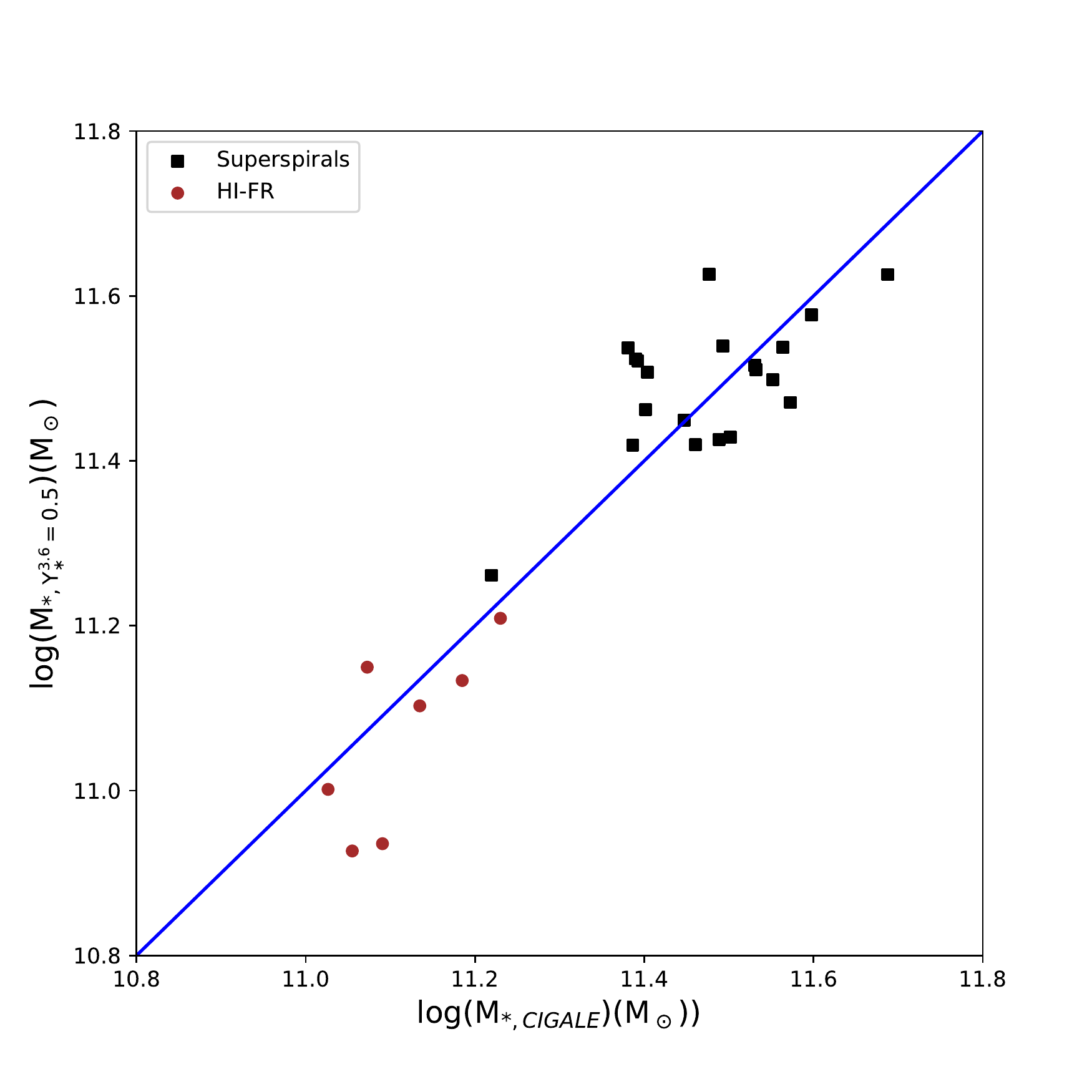}
\caption{ Comparison of  \mstar\  derived by CIGALE and \mstar\  derived from the W1 luminosity  assuming a $\Upsilon_{\ast}^{3.6} = 0.5$. Only objects with a good fit (reduced $\chi^2 <  1$) are taken into account. The blue line shows unity. }
\label{fig:compare-Mstar_cigale-constant_M_L}
   \end{figure}
\begin{figure}
\centering
\includegraphics[width=8.cm,trim=0.cm 0.cm 0cm 0cm,clip]{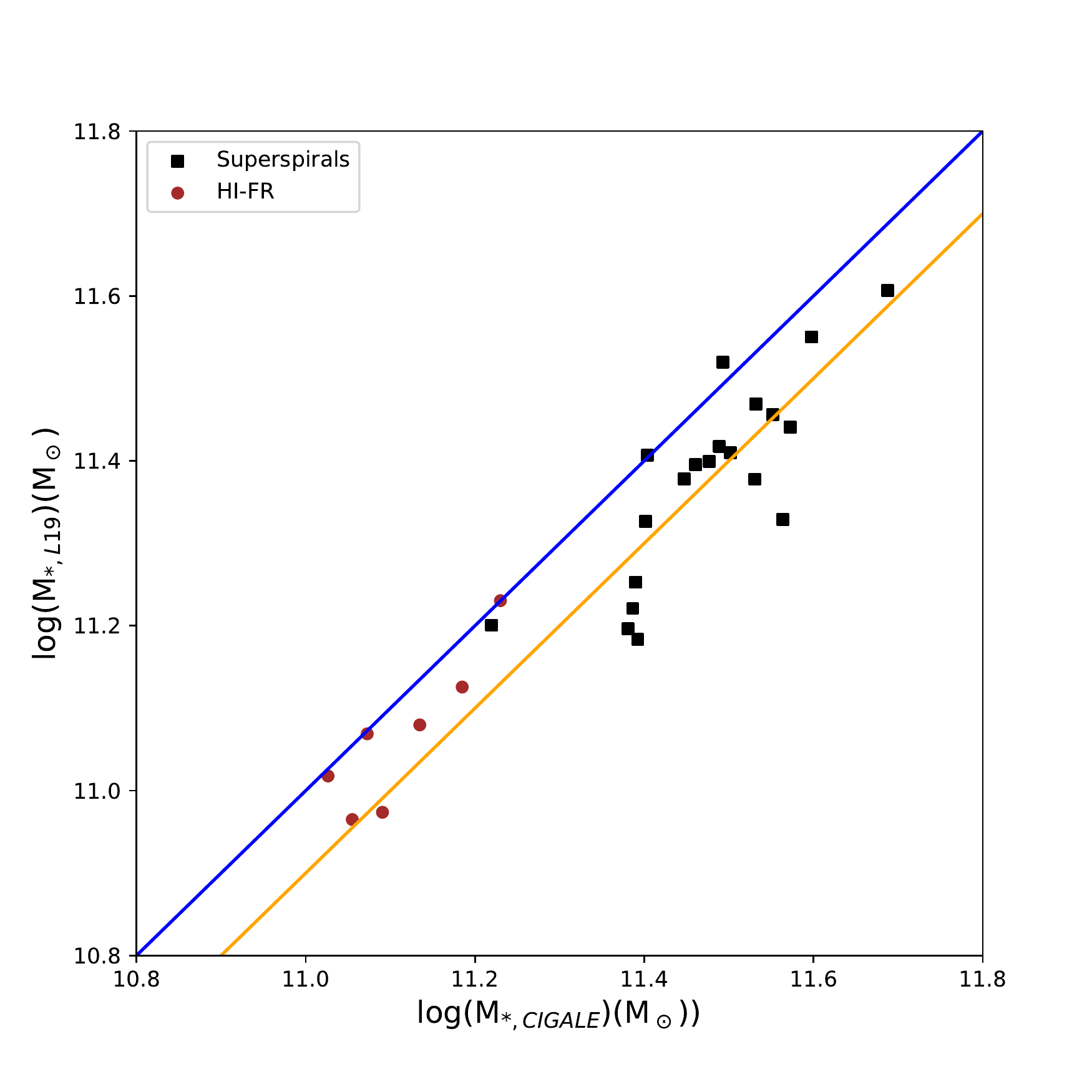}
\caption{ Comparison of  \mstar\  derived by CIGALE and \mstar\ following the prescription of \citet{leroy19}.  Only objects with a good fit (reduced $\chi^2 <  1$) are taken into account. The blue line shows unity and the orange line an offset of -0.1 dex.
}
\label{fig:compare-Mstar_cigale-L19}
   \end{figure}
  
  Fig.~\ref{fig:compare-Mstar_cigale-constant_M_L}  shows the comparison between the stellar mass derived from CIGALE and the value derived with a constant mass-to-light ratio, $\Upsilon_{\ast}^{3.4} = 0.5$ and  Fig.~\ref{fig:compare-Mstar_cigale-L19} the comparison of CIGALE with the values derived from  the prescription of Leroy et al. (2019)  (eq.~\ref{eq:leroy19}.) In both cases, good correlations exist. In the case of the Leroy et al. prescription there is a small, relatively constant offset between both measurement,  with CIGALE giving a slightly (by 0.1 dex) higher value for \mstar\ (see Tab.~\ref{tab:comp-cigale}).   We also compared the prescription by Cluver et al. (2014) to CIGALE (not shown)  and obtained a larger scatter (standard deviation 0.19). We conclude that both the calculation of \citet{leroy19} and a constant mass-to-light ratio $\Upsilon_{\ast}^{3.4} = 0.5$ give a good agreement with CIGALE. Taking the standard deviation as a reference, the uncertainty in the estimate of \mstar\ is 0.1-0.2 dex.

\begin{table}
\caption{\label{tab:comp-cigale} Comparison of SFR and \mstar\ from CIGALE and different methods}
\begin{tabular}{llll}
\noalign{\smallskip} \hline \noalign{\medskip}
Sample  & log($\rm \frac{\rm SFR_{\rm best}}{SFR_{\rm CIGALE, 100Myr}}$) &  log(\rm $\frac{M_{\rm *,L19}}{M_{\rm CIGALE}}$) & 
log(\rm $\frac{M_{*,\Upsilon_{\ast}^{3.4} = 0.5}}{M_{\rm CIGALE}} $)  \\
 & mean (stdv)\tablefootmark{a} & mean (stdv)\tablefootmark{a} & mean (stdv)\tablefootmark{a}  \\
\noalign{\smallskip} \hline \noalign{\medskip}
SS & 0.41 (0.21) &  -0.09 (0.07) & 0.03 (0.08) \\
FR-HI & 0.37 (0.20)& 0.01 (0.03) & 0.01 (0.05) \\
\noalign{\smallskip} \hline \noalign{\medskip}
   \end{tabular}
\end{table}
\end{appendix}
\end{document}